\pdfoutput=1
\documentclass[iop]{emulateapj}
\usepackage{graphicx}
\usepackage{natbib}
\usepackage{amsmath}
\usepackage[caption=false]{subfig} % mac
\usepackage{subfloat}
\makeatletter

\newcommand{\mocc}{m_{\rm O}^{\rm cc}}
\newcommand{\mfecc}{m_{\rm Fe}^{\rm cc}}
\newcommand{\mfeIa}{m_{\rm Fe}^{\rm Ia}}
\newcommand{\mxcc}{m_{\rm X}^{\rm cc}}
\newcommand{\mxIa}{m_{\rm X}^{\rm Ia}}
\newcommand{\KfeIa}{K_{\rm Fe}^{\rm Ia}}
\newcommand{\msicc}{m_{\rm Si}^{\rm cc}}
\newcommand{\msiIa}{m_{\rm Si}^{\rm Ia}}
\newcommand{\mscc}{m_{\rm S}^{\rm cc}}
\newcommand{\msIa}{m_{\rm S}^{\rm Ia}}
\newcommand{\Mo}{M_{\rm O}}
\newcommand{\Mfe}{M_{\rm Fe}}
\newcommand{\Mfecc}{M_{\rm Fe}^{\rm cc}}
\newcommand{\MfeIa}{M_{\rm Fe}^{\rm Ia}}
\newcommand{\Modot}{\dot{M}_{\rm O}}
\newcommand{\Mfedot}{\dot{M}_{\rm Fe}}
\newcommand{\MfeIadot}{\dot{M}_{\rm Fe}^{\rm Ia}}
\newcommand{\Moprod}{M_{\rm O,prod}}
\newcommand{\Mostars}{M_{\rm O,stars}}
\newcommand{\Moism}{M_{\rm O,ISM}}
\newcommand{\Moej}{M_{\rm O,ej}}
\newcommand{\Mform}{M_{\rm form}}
\newcommand{\mdotstar}{\dot{M}_*}
\newcommand{\mdotstaro}{\dot{M}_{*,0}}
\newcommand{\mdotstart}{\dot{M}_*(t)}
\newcommand{\bmdotstar}{\langle \mdotstar \rangle_{\rm Ia}}
\newcommand{\bmdotstart}{\langle \mdotstar(t) \rangle_{\rm Ia}}
\newcommand{\mdotstarone}{\dot{M}_{*,1}}
\newcommand{\mdotstartwo}{\dot{M}_{*,2}}
\newcommand{\mdotoutflow}{\dot{M}_{\rm outflow}}

\newcommand{\td}{t_D}
\newcommand{\tauIa}{\tau_{\rm Ia}}
\newcommand{\Yo}{Z_{\rm O}}
\newcommand{\Yfe}{Z_{\rm Fe}}
\newcommand{\Yx}{Z_{\rm X}}
\newcommand{\Yoinf}{Z_{\rm O,inf}}
\newcommand{\Yfeinf}{Z_{\rm Fe,inf}}

\newcommand{\Yfecc}{Z_{\rm Fe}^{\rm cc}}
\newcommand{\YfeIa}{Z_{\rm Fe}^{\rm Ia}}
\newcommand{\Yodot}{\dot{Z}_{\rm O}}

\newcommand{\YfedotIa}{\dot{Z}_{\rm Fe}^{\rm Ia}}
\newcommand{\Yoeq}{Z_{\rm O,eq}}
\newcommand{\Yomean}{\langle Z_{\rm O}\rangle}
\newcommand{\zo}{z_O}
\newcommand{\zomean}{\langle z_O(t)\rangle}

\newcommand{\Yoeqc}{Z_{\rm O,eqc}}
\newcommand{\Yfeeqc}{Z_{\rm Fe,eqc}}
\newcommand{\Yfeeqs}{Z_{\rm Fe,*}}
\newcommand{\Yfecceq}{Z^{\rm cc}_{\rm Fe,eq}}
\newcommand{\YfeIaeq}{Z^{\rm Ia}_{\rm Fe,eq}}
\newcommand{\Yfecceqtt}{Z^{\rm cc}_{\rm Fe,eq2}}
\newcommand{\YfeIaeqtt}{Z^{\rm Ia}_{\rm Fe,eq2}}
\newcommand{\YfeIaeqs}{Z^{\rm Ia}_{\rm Fe,*}}
\newcommand{\feh}{[{\rm Fe}/{\rm H}]}
\newcommand{\afe}{[\alpha/{\rm Fe}]}
\newcommand{\ofe}{[{\rm O}/{\rm Fe}]}
\newcommand{\xfe}{[{\rm X}/{\rm Fe}]}
\newcommand{\ohh}{[{\rm O}/{\rm H}]}
\newcommand{\Deltaofe}{\Delta\ofe}
\newcommand{\Deltaxfe}{\Delta\xfe}

\newcommand{\taustar}{\tau_*}
\newcommand{\taustaro}{\tau_{*,0}}
\newcommand{\tausfh}{\tau_{\rm sfh}}
\newcommand{\taudep}{\tau_{\rm dep}}
\newcommand{\taudepo}{\tau_{\rm dep,0}}
\newcommand{\taubarIaSfh}{\bar{\tau}_{[{\rm Ia,sfh}]}}
\newcommand{\taubarDepSfh}{\bar{\tau}_{[{\rm dep,sfh}]}}
\newcommand{\taubarDepIa}{\bar{\tau}_{[{\rm dep,Ia}]}}
\newcommand{\taubarXY}{\bar{\tau}_{[X,Y]}}
\newcommand{\mmdotstar}{\dot{M}_{*,1}}

\newcommand{\ttaustar}{\tau_{*,1}}
\newcommand{\ttausfh}{\tau_{\rm sfh1}}

\newcommand{\ttaubarIaSfh}{\bar{\tau}_{[{\rm Ia,sfh1}]}}
\newcommand{\ttaubarDepSfh}{\bar{\tau}_{[{\rm dep1,sfh1}]}}

\newcommand{\mmmdotstar}{\dot{M}_{*,2}}

\newcommand{\tttaustar}{\tau_{*,2}}
\newcommand{\tttausfh}{\tau_{\rm sfh2}}
\newcommand{\tttaudep}{\tau_{\rm dep2}}
\newcommand{\tttaubarIaSfh}{\bar{\tau}_{[{\rm Ia,sfh2}]}}
\newcommand{\tttaubarDepSfh}{\bar{\tau}_{[{\rm dep2,sfh2}]}}
\newcommand{\tttaubarDepIa}{\bar{\tau}_{[{\rm dep2,Ia}]}}

\newcommand{\Moc}{M_{\rm O,c}}

\newcommand{\fmet}{f_{\rm met}}
\newcommand{\etapr}{{\eta^\prime}}
\newcommand{\Mgpr}{{M_g^\prime}}
\newcommand{\Mopr}{{M_{\rm O}^\prime}}
\newcommand{\Mfepr}{{M_{\rm Fe}^\prime}}
\newcommand{\Yopr}{{Z_{\rm O}^\prime}}
\newcommand{\Yfepr}{{Z_{\rm Fe}^\prime}}
\newcommand{\rcc}{r_{\rm cc}}
\newcommand{\Funp}{F_{\rm unp}}

\newcommand{\Gyr}{\,{\rm Gyr}}
\newcommand{\Myr}{\,{\rm Myr}}
\newcommand{\yr}{\,{\rm yr}}

\makeatother
\begin{document}

\title{Equilibrium and Sudden Events in Chemical Evolution}
\author{David H.~Weinberg\altaffilmark{1},
Brett H.~Andrews\altaffilmark{2},
Jenna Freudenburg\altaffilmark{1}}

\altaffiltext{1}{Department of Astronomy, The Ohio State University, 
Columbus, OH 43210, dhw@astronomy.ohio-state.edu}
\altaffiltext{2}{Department of Physics and Astronomy, University of Pittsburgh,
Pittsburgh, PA}

%\begin{abstract}
%\end{abstract}


\begin{abstract}
We present new analytic solutions for one-zone (fully mixed) chemical evolution
models and explore their implications.  In contrast to existing analytic models,
we incorporate a realistic delay time distribution for Type Ia supernovae 
(SNIa) and can therefore track the separate evolution of $\alpha$-elements 
produced by core collapse supernovae (CCSNe) and iron peak elements synthesized
in both CCSNe and SNIa.
Our solutions allow constant, exponential, or linear-exponential 
($t e^{-t/\tausfh}$) star formation histories, or combinations thereof. 
In generic cases, $\alpha$ and iron abundances evolve to an equilibrium at
which element production is balanced by metal consumption and gas dilution,
instead of continuing to increase over time.
The equilibrium absolute abundances depend principally on supernova yields and 
the outflow mass loading parameter $\eta$, while the equilibrium abundance
ratio $\afe$ depends mainly on yields and secondarily on star formation history.
A stellar population can be metal-poor either because it has not yet evolved to 
equilibrium or because high outflow efficiency makes the equilibrium abundance 
itself low.  Systems with ongoing gas accretion develop metallicity 
distribution functions (MDFs) that are sharply peaked, while ``gas starved'' 
systems with rapidly declining star formation, such as the conventional 
``closed box'' model, have broadly peaked MDFs.
A burst of star formation that consumes a significant fraction of a system's
available gas and retains its metals can temporarily boost $\afe$ by 0.1-0.3 
dex, a possible origin for rare, $\alpha$-enhanced stars with intermediate 
age and/or high metallicity. 
Other sudden transitions in system properties can produce surprising behavior,
including backward evolution of a stellar population from high metallicity 
to low metallicity.
While one-zone models omit mixing processes that may play an important
role in chemical evolution, they provide a useful guide and flexible tool
for interpreting multi-element surveys of the Milky Way and its neighbors.
An Appendix provides a user's guide for calculating enrichment histories,
$\afe$ tracks, and MDFs in a wide variety of scenarios.
\end{abstract}

%%%%%%%%%%%%%%%%%%%%%%%%%%%%%% INTRODUCTION %%%%%%%%%%%%%%%%%%%%%%%%%%%%%%%%%
\section{Introduction}

The elemental abundances of stars provide essential clues to the
star formation and assembly history of the Milky Way and other galaxies.
One of the most useful ``clocks'' for tracing these histories is the
ratio of $\alpha$-elements, which are produced
in massive, short-lived stars that explode as core collapse supernovae
(CCSNe), to iron peak elements, which are produced by both CCSNe and
Type Ia supernovae (SNIa).  Old metal-poor stars in the stellar halo
and thick disk have enhanced $\afe$, while disk stars with near-solar
iron abundance ($\feh \approx 0$) typically have solar abundance ratios
as well ($\afe \approx 0$).  A standard theoretical account might be
that the enhanced $\afe$ characteristic of CCSN yields is driven towards
solar $\afe$ as SNIa iron enrichment becomes important, and thereafter
the iron abundance increases at fixed $\afe$ because of
continuing enrichment from both classes
of supernovae.  However, the behavior of simple one-zone chemical evolution
models is rather different: by the time that $\afe$ approaches zero,
the iron abundance $\feh$ has also approached an approximate ``equilibrium''
value in which new enrichment is balanced by dilution and 
depletion of existing metals,
so that a given model produces only a narrow range of $\feh$
when $\afe \approx 0$ (B. Andrews et al. 2016, in preparation,
hereafter AWSJ).

This paper focuses on the phenomenon of equilibrium abundances and on
the departures from equilibrium that can arise from bursts of star 
formation or other sudden changes.  
Over the past decade, theoretical 
discussions of the ``mass-metallicity'' relation ---
the correlation between the stellar mass and gas-phase oxygen abundance
of galaxies --- have concluded that the oxygen abundance of a galaxy's
interstellar medium (ISM) is controlled by an equilibrium among
fresh gas accretion, star formation, and outflows 
(e.g., \citealt{dalcanton2007,finlator2008,peeples2011,lilly2013}).
This account has early roots in the work of \cite{larson1972}, who
shows that, in the instantaneous recycling approximation, ISM abundances
in a galaxy with continuous infall approach an equilibrium determined
by nucleosynthetic yields.

It is less obvious that equilibrium is a useful concept for 
describing iron abundances, since SNIa provide enrichment over
long timescales not tied to the current star formation rate.
However, observational studies of the evolution of the SNIa rate
and the relative rates in star-forming and passive galaxies 
now indicate that a large fraction of SNIa explode within 1-3 Gyr
of the birth of their stellar progenitors 
\citep{scannapieco2005,maoz2012a,maoz2012b}, in agreement
with predictions of binary population synthesis models \citep{greggio2005}.
The dominance of relatively ``prompt'' SNIa makes equilibrium
approximations relevant even for iron abundances, though the
different timescale of CCSN and SNIa enrichment remains crucial
to understanding the evolution of abundance ratios.

Our discussion relies mainly on analytic
approximations, though we test their results against numerical calculations
that avoid the simplifying assumptions needed for analytic solutions.
Relative to existing analytic models of chemical evolution,
the distinctive feature of our approach is that we do not
assume instantaneous recycling for SNIa enrichment but instead
adopt an exponential form for the delay time distribution that
allows analytic solutions of the resulting differential equations
for interesting families of star formation histories.
We do assume instantaneous recycling for CCSN enrichment and
for the return of envelope material from evolved stars,
but incorporating realistic time evolution for SNIa allows
us to calculate evolution in the $\afe-\feh$ plane and to
make more realistic calculations of $\feh$ evolution and
metallicity distribution functions.
Our choices of parameters for our calculations are largely guided
by the discussion of AWSJ, who examine a broader range of models
and a broader range of chemical evolution results.
For definiteness, we focus on oxygen as our representative $\alpha$
element, but our results translate trivially to other elements
whose production is dominated by CCSNe.

The phenomena discussed here are necessarily incorporated into
most existing numerical chemical evolution models, as they have similar
physical ingredients.
However, the analytic discussion and the focus on equilibrium
provide physical insights into the behavior of these models,
and they suggest new ways to think about phenomena now being revealed
by large scale Galactic chemical evolution studies.
As surveys like SEGUE \citep{yanny2009}, 
APOGEE \citep{majewski2016}, Gaia-ESO \citep{gilmore2012}, 
and GALAH \citep{desilva2015} extend measurements
of $\alpha$ and iron abundances over much of the Milky Way,
flexible models that allow rapid explorations of a large parameter
space are a useful tool for developing interpretations.
They are also useful for modeling the stellar populations of
other galaxies and the relation of those populations to gas phase
abundances, and potentially for population synthesis modeling
of galaxy spectra.
Radial mixing of stellar populations and radial flows of enriched
gas are both likely to play important roles in the chemical evolution
of disk galaxies like the Milky Way 
(e.g., \citealt{schoenrich2009,bilitewski2012,pezzulli2016}), and
both processes violate the assumptions of one-zone models 
in which metals produced by stars are either ejected from the
system or retained locally.  However, fully understanding the behavior of
one-zone models is important for evaluating the empirical case for
mixing processes, and mixtures of one-zone models may provide a
useful approximate description of more complex scenarios.

Section~\ref{sec:context} gives a brief high-level overview of our models
in the broader context of chemical evolution calculations.
Section~\ref{sec:eq_equations} introduces our basic notation and
evolution equations for oxygen and iron mass, and 
\S\S\ref{sec:eq_csfr} and~\ref{sec:eq_expsfr} show how these
equations lead to equilibrium abundances for constant or exponentially
declining star formation histories.  Section~\ref{sec:evolution} derives
the full time evolution solutions for these cases and for
linear-exponential star formation histories, then discusses
metallicity distribution functions and the relation of our calculations
to traditional analytic models of chemical evolution.  This section
concludes with illustrations of behavior for a variety of model
parameters and tests against numerical solutions.
Section~\ref{sec:sudden} considers the impact of an instantaneous
burst of star formation on abundances and abundance ratios, 
then shows how to stitch together our previous analytic solutions
for cases where model parameters change suddenly from one set
of values to another.  These cases allow an interesting variety
of behaviors.  Section~\ref{sec:extensions} considers a variety
of extensions of our results, including more complex time histories
of SNIa or star formation, metal-enriched infall, and elements
that have both CCSN and SNIa contributions.
Section \ref{sec:conclusions} summarizes many of the insights
from our results in qualitative terms.
Some readers may prefer to start with the illustrations in 
\S\ref{sec:illustrations}, jump to the
qualitative conclusions in \S\ref{sec:conclusions},
then go back as needed to the analytic modeling that leads to them.
Tables~\ref{tbl:variables} and~\ref{tbl:results} provide a
guide to the notation used in the paper and the sections
where the principal analytic results appear.
Appendix~\ref{appx:userguide} provides a guide for readers who
want to use our results to compute enrichment histories,
$\afe-\feh$ tracks, and distribution functions of metallicity
or abundance ratios.

%%%%%%%%%%%%%%%%%%%%%%%%%%%%%%%%%%% MODEL %%%%%%%%%%%%%%%%%%%%%%%%%%%%%%%%%%%%
\section{Equilibrium Abundances}
\label{sec:equilibrium}

\subsection{Context}
\label{sec:context}

Before diving into notation and equations, it is useful to place
our calculations in the broader context of chemical evolution models.
Classic reviews of the subject include \cite{tinsley1980} and 
the monographs of \cite{pagel1997} and \cite{matteucci2001,matteucci2012}.
Key elements of any chemical evolution model include a gas
accretion history, a star formation law, an outflow prescription,
and nucleosynthetic yields as a function of time.  
In one-zone models \citep{schmidt1959,schmidt1963},
abundances in the gas phase are assumed to be homogeneous throughout
the model volume, and all stars form with the current abundances
of the ISM.  In our models we generally assume that the star formation
rate is proportional to the gas mass, with the star formation 
efficiency as a free but constant parameter.
This assumption, sometimes referred to in the literature as a
``linear Schmidt law'' (e.g., \citealt{recchi2008}), is arguably
the least desirable requirement of our solutions, since observations
imply that the star formation efficiency declines with decreasing
total (atomic + molecular) gas surface density
\citep{schmidt1959,kennicutt1998}.  
As in most models, we assume that the gas outflow rate is
a constant multiple of the star formation rate, specified
by a mass loading parameter.  We explicitly specify a model's
star formation history, which may be constant, exponentially declining,
or the product of a linear rise and an exponential decline.
With these assumptions, the gas accretion history follows implicitly
from the star formation history once the star formation efficiency
and mass loading are specified (see eq.~\ref{eqn:infall} below).
This choice to define the functional form of the star formation
rate rather than the gas accretion rate differs from most models,
but if the star formation rate is simply related to the gas supply
then these two histories usually track each other fairly closely.

Nearly all analytic models of chemical evolution rely on the 
instantaneous recycling approximation, that the elements synthesized
by newly formed stars are immediately returned and mixed into the
star-forming ISM.  We adopt this approximation for elements produced
by CCSNe, which come from stars with $M>8M_\odot$ and 
lifetimes less than 40 Myr.
The key insight that underlies much of this paper is that analytic
solutions for SNIa enrichment are possible if the delay time distribution
(DTD) is assumed to be an exponential in time (following a minimum delay),
or a sum of exponentials.  Even for CCSN products, instantaneous
recycling may be a poor approximation if the metals in SN ejecta
are locked up in a warm or hot phase of the ISM and return only
slowly to the cold, star-forming phase (see, e.g., \citealt{schoenrich2009}).
We ignore this possibility here, though we note that our solutions
could be easily adapted to this case if one assumed an exponential
form for this gas return.  Likewise, our DTD for SNIa enrichment
should be understood as representing the time for SNIa products
to return to the star-forming phase of the ISM, which may differ
from the time for the SNIa explosions themselves.
Our analytic solutions require assuming that nucleosynthetic
yields are independent of stellar metallicity.  
With the yield sources adopted by AWSJ (and described below),
metallicity independence is a good approximation for oxygen
and other $\alpha$-elements, and for iron.  However, the supernova
yields are uncertain, and population averaged yields could
in any case become metallicity dependent if the mass ranges of stars
that explode as CCSNe changes with metallicity.

Our standard solutions apply to models with parameters
that remain fixed throughout the time evolution, and that have
the smooth star formation histories described above.
However, in \S\ref{sec:sudden} we show how to combine solutions
to create models with a sharp transition in accretion history.
The most well known example of a (numerical) chemical evolution
model with a sharp transition in accretion history is the ``two infall''
scenario of \cite{chiappini1997}, which posits distinct infall
episodes for the formation of the Milky Way's halo-thick disk
and thin disk, respectively.  Some numerical models incorporate
accretion histories motivated by cosmological simulations
\cite{colavitti2008}, which typically transition from rapid
to slow accretion, somewhat analogous to the two infall model.

Models of disk galaxies are frequently built as a sequence of
annular zones, each of which evolves independently with its
own gas accretion history (e.g., \citealt{matteucci1989}).
Such a model could be created as a sequence of our solutions,
with parameters that depend on Galactocentric radius.  
However, there are processes that can redistribute stars and
metals from one annulus to another.  One of these is
radial mixing of stars \citep{sellwood2002,roskar2008,bird2012}, which is 
invoked to explain the observed dispersion between metallicity
and stellar age \citep{edvardsson1993,wielen1996} and could have
a large impact on metallicity gradients and distribution functions
\citep{schoenrich2009}.  These effects can be modeled crudely
by after-the-fact convolutions of a simple annular-zone model
(e.g., \citealt{hayden2015}).  A second mixing process arises because
gas accreted from the halo should have angular momentum below that of 
local circular orbits in the disk \citep{fraternali2008}, driving
radial flows that carry metals inward from the radius at 
which they are produced.  Numerical and analytic models that
include such radial gas flows are discussed by
\cite{spitoni2011}, \cite{bilitewski2012} and \cite{pezzulli2016}.
Large scale galactic winds could provide a third redistribution mechanism
that carries metals
from small radii to large radii, if some enriched gas is not ejected
entirely from the halo but instead returns as a galactic fountain.
We provide solutions with enriched infall of constant metallicity
(\S\ref{sec:enriched}), but we do not address this more
general case.

\subsection{Evolution equations}
\label{sec:eq_equations}

Our analytic models of chemical evolution assume that 
the star-forming ISM is always fully mixed,
that CCSNe and SNIa are the only sources of new metal production,
and that CCSNe redistribute their metals to the ISM instantaneously.
AGB stars are a source of some heavy elements, most notably nitrogen
and $s$-process neutron capture elements, but they make a negligible
contribution to the production of oxygen and other $\alpha$-elements.
They do return the metals they were born with, however, 
and to enable analytic solutions we assume that this return
of birth-metals happens instantaneously.
For the delay time distribution (DTD) of SNIa, we will usually assume
an exponential form \citep{schoenrich2009} with a minimum delay time $\td$,
\begin{equation}
\label{eqn:rIa}
R(t) = 
\begin{cases}
R_0 e^{-(t-\td)/\tauIa} & t \geq \td~, \\
0		      & t < \td~,
\end{cases}
\end{equation}
which proves analytically convenient.
Here $R(t)dt$ is the number of SNIa in the time interval 
$t \rightarrow t+dt$ per unit mass of stars formed at
time $t=0$; $R_0$ has units of $M_\odot^{-1}\,{\rm yr}^{-1}$.
With $\tauIa=1.5\Gyr$, equation~(\ref{eqn:rIa}) roughly tracks
the predictions of population synthesis models (\citealt{greggio2005};
see Fig.~6 of AWSJ for a comparison).
As discussed in \S\ref{sec:two-exp}, our analytic results generalize
easily to a DTD that is a sum of exponentials, which can be used
to approximate other forms such as a power-law DTD \citep{maoz2012a,maoz2012b}.
Note that population synthesis models and supernova statistics
both address the DTD for supernova explosions, but what matters for
chemical evolution is the time for SNIa elements to return to the
star-forming ISM, which may be longer if the supernova ejecta are
initially deposited in a warm or hot phase.

We define the dimensionless yield parameters:
\begin{itemize}
\item $\mocc$, the mass of newly produced oxgyen returned to the ISM by 
      CCSNe per unit mass of star formation,
\item $\mfecc$, the mass of newly produced iron returned to the ISM by CCSNe per 
      unit mass of star formation, and
\item $\mfeIa$, mass of newly produced iron returned to the ISM by SNIa per
      unit mass of star formation.
\end{itemize}
With the fiducial model assumptions of AWSJ,
a \cite{kroupa2001} initial mass function (IMF) and the CCSN yields
of \cite{chieffi2004} and \cite{limongi2006}, the core collapse quantities
are $\mocc = 0.015$ and $\mfecc=0.0012$:
for every $100 M_\odot$ of star formation with a Kroupa IMF (truncated outside
$0.1 - 100 M_\odot$),
CCSNe produce an average of $1.5M_\odot$ of oxygen and $0.12M_\odot$
of iron.  (These numbers are slightly metallicity-dependent, and
our values here are for $Z \approx 0.1 Z_\odot$.)
If the average mass of iron from an individual SNIa is $\KfeIa$, then 
\begin{equation}
\label{eqn:mfeIa}
\mfeIa \equiv \KfeIa \int_0^\infty R(t)dt = \KfeIa R_0 \tauIa,
\end{equation}
where the second equality holds for an exponential DTD.
(The notation $\KfeIa$ is chosen here only to reduce confusion
with variables used elsewhere in the paper.)
For the AWSJ fiducial model, with 
$R_0 = 2.2\times 10^{-3} M_\odot^{-1}\,\yr^{-1}$ (based on \citealt{maoz2012a})
and $\KfeIa = 0.77 M_\odot$ (based on the W70 model of \citealt{iwamoto1999}), 
the time-integrated yield is
$\mfeIa = 0.0017$.  Note that one can think of $\mocc$, $\mfecc$,
and $\mfeIa$ as having units of ``solar masses per solar mass.''

We define
\begin{itemize}
\item $\eta = \mdotoutflow/\mdotstar$, the mass loading factor, to be 
      the ratio of gas mass ejected from the ISM by stellar feedback to the
      gas mass being incorporated into stars, and
\item $r$, the recycling parameter,
      to be the fraction of mass formed into stars that is returned
      from the envelopes of CCSN progenitors and 
	  asymptotic giant branch (AGB) stars at its original metallicity.
\end{itemize}
In most theoretical models of galaxy formation and chemical evolution,
reproducing the observed stellar masses and metallicities of galaxies
requires substantial outflows, with time-averaged mass loading
factors $\eta \approx 1-10$.  The characteristic value of $\eta$ likely
varies with galaxy mass and evolutionary state, and in any given galaxy
it may vary with time and location.  
For a Kroupa IMF, the recycled fraction is $r(t) = 0.37$, 0.40, and 0.45
after 1, 2, and 10 Gyr, respectively. 
The quantities $\mocc$ and $\mfecc$ refer to the ``net yields'' of CCSNe,
while the ``absolute yields'' include the recycling of the progenitor's
original metals, which is encompassed here under $r$.
For our analytic solutions we must assume instantaneous recycling and 
a single value of $r$, and we adopt $r=0.4$ for a Kroupa IMF based on
comparing our analytic results to those of numerical calculations that
include time-dependent recycling (see \S\ref{sec:tests}).
While the timescale for envelope return from AGB
stars is comparable to the delay timescales for SNIa,
the supernovae are a primary source of new iron while the AGB
stars are merely returning the metals they were born with,
so it is useful to have full time evolution of SNIa even if
the AGB return is approximated as instantaneous.

We define the characteristic timescales
\begin{itemize}
\item $\taustar = M_g/\mdotstar$, the star formation efficiency (SFE) timescale,
      to be the ratio of the current ISM gas mass $M_g$ to the instantaneous 
      star formation rate $\mdotstar$, and
\item $\taudep = \taustar/(1+\eta-r)$, the gas depletion timescale, to be
      the net rate at which gas is being depleted by the combination of star
      formation and outflows.
\end{itemize}
The SFE timescale is more observationally accessible than the
gas depletion timescale because $\eta$
is usually difficult to determine.  The observations
of \cite{leroy2008} suggest a typical $\taustar \approx 2\Gyr$ for the
{\it molecular} ISM, over a wide range of star formation rate and gas
surface density, though the corresponding timescale for the 
molecular+atomic ISM will be longer.
With our instantaneous approximation, CCSNe and AGB stars return mass
to the ISM at a rate $r\mdotstar$, making the net depletion 
timescale $\taustar/(1+\eta-r)$ rather than $\taustar/(1+\eta)$.
We note that terminology in the literature is not universal, and that
some papers define the gas depletion timescale (or gas consumption timescale)
to be the quantity we call $\taustar$, rather than what we call $\taudep$.
Note also that because of recycling, the mass in stars and stellar
remnants at a given time is
\begin{equation}
\label{eqn:mstar_recycling}
M_*(t) = (1-r)\int_0^t \mdotstar(t') dt' ~.
\end{equation}

With these definitions and assumptions, the injection rate of oxygen
mass and iron mass into the ISM is
\begin{eqnarray}
\dot{M}_{\rm O,in} &=& \mocc\mdotstar + r\Yo\mdotstar~, \label{eqn:Oin} \\
\dot{M}_{\rm Fe,in} &=& \mfecc\mdotstar + \mfeIa\bmdotstar + r\Yfe\mdotstar~, 
   \label{eqn:Fein}
\end{eqnarray}
where $\Yo = \Mo/M_g$ and $\Yfe = \Mfe/M_g$ are the mass-weighted 
oxygen and iron abundances of the ISM and $\bmdotstar$
is the time-averaged star formation rate (SFR) weighted by the SNIa rate.
Specifically, as shown in Appendix~\ref{appx:SNIa}
\begin{equation}
\label{eqn:bmdotstar}
\bmdotstart \equiv {\int_0^t \mdotstar(t') R(t-t') dt' \over 
                   \int_0^\infty R(t') dt'}~.
\end{equation}
To compute the full time derivatives of oxygen and iron mass, we must
also take account of the consumption of ISM metals by star formation
and outflow, obtaining:
\begin{eqnarray}
\Modot &=& \mocc\mdotstar - (1+\eta-r)\mdotstar \Yo~, \label{eqn:Modot} \\
\Mfedot &=& \mfecc\mdotstar + \mfeIa\bmdotstar -(1+\eta-r)\mdotstar \Yfe ~.
  \label{eqn:Mfedot}
\end{eqnarray}
By approximating recycling as instantaneous, and occurring at
the current abundance $\Yo$ or $\Yfe$, we are able to include it
by simply using $1+\eta-r$ instead of $1+\eta$ in these equations.
In physical terms, the impact of recycling is the same as replacing
$\eta$ with an ``effective'' value $\eta-r$ (which may be negative)
because some gas and metals are immediately returned to the system.
With full time-dependent AGB recycling there would be a source term
that depends on the metallicity at earlier times, 
leading to integro-differential equations that would defy
analytic solution except in special circumstances.

The evolution of $M_g(t)$ is specified implicitly in our models
through a star formation history and the SFE timescale,
with $M_g(t) = \taustar\mdotstart$.
This in turn determines the rate at which abundances are
diluted by gas infall.
The time derivative of the gas supply is 
$\dot{M}_g = -(1+\eta-r)\mdotstar + \dot{M}_{\rm inf}$, where
$\dot{M}_{\rm inf}$ is the infall rate.
For constant $\taustar$, we can set $\dot{M}_g = \taustar \ddot{M}_*$ to
obtain
\begin{equation}
\label{eqn:infall}
\dot{M}_{\rm inf} = (1+\eta-r)\mdotstar + \taustar \ddot{M}_* ~.
\end{equation}

Our approximation of instantaneous recycling for CCSN products
implicitly assumes that the CCSN elements that are not ejected in outflow
become immediately available for star formation from the cold ISM.
It is possible that CCSN products are instead injected into a 
warm phase of the ISM and cool over time to join the star-forming
medium, as in, for example, the \cite{schoenrich2009} chemical evolution
model.  We ignore this complication here, but we note that one could
accommodate this scenario within our analytic framework if the 
return of CCSN elements were approximated as exponential in time (or
a sum of exponentials),
perhaps with a minimum delay.  In this case, one would simply
adapt our solutions for $\MfeIa$ and $\YfeIa$ to CCSN elements
with the appropriate $e-$folding and minimum delay timescales.
Similarly, one could incorporate time-dependent production of
AGB elements by approximating AGB enrichment as a delayed exponential.

\begin{table*}
\centering
\begin{tabular}{l|l|l|l}
Variable & Description & Section & Fiducial Value\\
\hline
$\eta$ & $=\dot{M}_{\rm outflow}/\mdotstar$, outflow efficiency & 
  \S\ref{sec:eq_equations} & 2.5 \\
$\eta_0$ & starting value of $\eta$ with time-dependent $\taustar$ &
  \S\ref{sec:timedeptaustar} & \\
$f_*$ & fraction of gas converted to stars in burst & \S\ref{sec:bursts} & \\
$f_g$ & $=M_g(t)/M_i$, gas fraction in closed/leaky box model & 
  \S\ref{sec:traditional} & \\
$\fmet$ & fraction of newly produced metals retained in burst & 
  \S\ref{sec:bursts} & \\
$\Funp$ & fraction of gas unprocessed in burst & \S\ref{sec:bursts} & \\
$\KfeIa$ & iron mass yield per SNIa & \S\ref{sec:eq_equations} & $0.77M_\odot$\\
$\mocc$ & IMF-integrated CCSN oxygen yield & \S\ref{sec:eq_equations} & 0.015 \\
$\mfecc$ & IMF-integrated CCSN iron yield & \S\ref{sec:eq_equations} & 0.0012 \\
$\mfeIa$ & IMF-integrated SNIa iron yield & \S\ref{sec:eq_equations} & 0.0017 \\
$M_g$ & gas mass & \S\ref{sec:eq_equations} & \\
$\Mo$ & oxygen mass & \S\ref{sec:eq_equations} & \\
$\Mfe$ & total iron mass & \S\ref{sec:eq_equations} & \\
$\MfeIa$ & iron mass from SNIa alone & \S\ref{sec:eq_equations} & \\
$\mdotstar$ & star-formation rate & \S\ref{sec:eq_equations} & \\
$M_*$ & mass of stars + stellar remnants & \S\ref{sec:eq_equations} & \\
$\Mform(t)$ & $= \int_0^t \mdotstar(t') dt' = M_*(t)/(1-r)$  
  & \S\ref{sec:oxygen_budget} & \\
$\bmdotstar$ & SFR averaged over SNIa DTD & \S\ref{sec:eq_equations} & \\
$\dot{M}_{\rm inf}$ & infall rate & \S\ref{sec:eq_equations} & \\
$R(t)$ & SNIa rate from population formed at $t=0$ & \S\ref{sec:eq_equations} 
  & \\
$r$ & mass recycling parameter (CCSN + AGB) & \S\ref{sec:eq_equations} & 0.4 \\
$\rcc$ & mass recycling parameter (CCSN only) & \S\ref{sec:bursts} & 0.2 \\
$\taustar$ & $=M_g/\mdotstar$, star formation efficiency (SFE) timescale &
  \S\ref{sec:eq_equations} & 1 Gyr \\
$\taustaro$ & starting value of $\taustar$ with time-dependent $\taustar$ &
  \S\ref{sec:timedeptaustar}\\
$\taudep$ & $=\taustar/(1+\eta-r)$, gas depletion timescale &
  \S\ref{sec:eq_equations} & 0.323 Gyr \\
$\taudepo$ & starting value of $\taudep$ with time-dependent $\taustar$ &
  \S\ref{sec:timedeptaustar} & \\
$\tauIa$ & $e$-folding timescale of SNIa DTD & \S\ref{sec:eq_equations} & 
  1.5 Gyr\\
$\td$ & minimum delay time for SNIa & \S\ref{sec:eq_equations} & 0.15 Gyr\\
$\Delta t$ & $=t-\td$, shifted time variable & \S\ref{sec:eq_csfr} & \\
$t_c$ & transition time in sudden-change models & \S\ref{sec:changes} & \\
$\tausfh$ & star formation history timescale, $\mdotstar \propto e^{-t/\tausfh}$
  & \S\ref{sec:eq_expsfr} & 6 Gyr \\
$\taubarXY$ & $=1/(\tau_X^{-1}-\tau_Y^{-1})$, harmonic difference timescale &
  \S\ref{sec:eq_expsfr}\\
$\taubarIaSfh$ & $=1/(\tauIa^{-1}-\tausfh^{-1})$ & \S\ref{sec:eq_expsfr} & 
  2 Gyr\\
$\taubarDepSfh$ & $=1/(\taudep^{-1}-\tausfh^{-1})$ & \S\ref{sec:eq_expsfr} & 
  0.341 Gyr\\
$\taubarDepIa$ & $=1/(\taudep^{-1}-\tauIa^{-1})$ & \S\ref{sec:ev_csfr} & 
  0.412 Gyr\\
$\Yo$ & $=\Mo/M_g$, oxygen abundance & \S\ref{sec:eq_equations} & \\
$\Yfe$ & $=\Mfe/M_g$, iron abundance & \S\ref{sec:eq_equations} & \\
$\Yfecc $ & $=\Mfecc/M_g$, CCSN iron abundance & \S\ref{sec:ev_csfr} & \\
$\YfeIa $ & $=\MfeIa/M_g$, SNIa iron abundance & \S\ref{sec:ev_csfr} & \\
$\Yoeq$ & equilibrium oxygen abundance & \S\ref{sec:eq_expsfr} & 0.51\% \\
$\Yfecceq$ & equilibrium CCSN iron abundance & \S\ref{sec:eq_expsfr} & 0.041\%\\
$\YfeIaeq$ & equilibrium SNIa iron abundance & \S\ref{sec:eq_expsfr} & 0.079\%\\
$\Yoeqc$ & equilibrium oxygen abundance for constant SFR & \S\ref{sec:eq_csfr} 
 & \\
$\Yfeeqc$ & equilibrium iron abundance for constant SFR & \S\ref{sec:eq_csfr}
 & \\
$\Yoinf$ & oxygen abundance of enriched infall & \S\ref{sec:enriched} & \\
$\Yfeinf$ & iron abundance of enriched infall & \S\ref{sec:enriched} & \\
$Z_{{\rm O},\odot}$ & solar oxygen abundance by mass & \S\ref{sec:ev_csfr} 
  & 0.56\% \\
$Z_{{\rm Fe},\odot}$ & solar iron abundance by mass & \S\ref{sec:ev_csfr} 
  & 0.12\% \\
$z_O$ & $= \Yo/\Yoeq$, scaled oxygen abundance & \S\ref{sec:mdf} & \\
\hline
\end{tabular}
\caption{Variables used in the paper}
\tablecomments{Variables used frequently in the paper are listed along with 
the section where they first appear.
Values of parameters in our fiducial model are listed in the right column.
In \S\ref{sec:bursts}, primes are used to denote post-burst quantities.
Numbered subscripts are used in \S\ref{sec:changes} to denote separate
contributions to oxygen and iron evolution from different phases of
the model.  Frequent abbreviations in the paper
are CCSN (core collapse supernova), DTD (delay time distribution for SNIa), 
SFE (star formation efficiency), SNIa (Type Ia supernova), 
and SFH (star formation history).}
\label{tbl:variables}
\end{table*}

\begin{table*}
\centering
\begin{tabular}{lr}
Result & Location \\
\hline
Equilibrium abundances & 
  \S\ref{sec:eq_expsfr}, eqs.~\ref{eqn:Yoeq}, \ref{eqn:Yfecceq}, 
  \ref{eqn:YfeIaeq} (or eqs.~\ref{eqn:Yoeq2}--\ref{eqn:YfeIaeq2}) \\
Evolution of simple models:\\
\ \ \ \ Constant SFR & 
           \S\ref{sec:ev_csfr}, eqs.~\ref{eqn:YoEvol}, \ref{eqn:YfeccEvol}, 
	   \ref{eqn:YfeIaEvol} \\
\ \ \ \ Exponential SFH & 
           \S\ref{sec:ev_expsfr}, eqs.~\ref{eqn:YoEvolExp}, 
	   \ref{eqn:YfeccEvolExp}, \ref{eqn:YfeIaEvolExp} \\
\ \ \ \ Lin-Exp SFH & 
           \S\ref{sec:ev_linexpsfr}, eqs.~\ref{eqn:YoEvolLinExp}, 
           \ref{eqn:YfeccEvolLinExp}, \ref{eqn:YfeIaEvolLinExp} \\
Metallicity distribution functions & 
  \S\ref{sec:mdf}, eqs.~\ref{eqn:mdf1}--\ref{eqn:mdf2} \\
Impact of a star formation burst &
  \S\ref{sec:bursts}, eqs.~\ref{eqn:Mgpr}--\ref{eqn:Mfepr}, 
  \ref{eqn:PostBurstRatio}, Fig.~\ref{fig:burst} \\
Models with sudden parameter changes & 
  \S\ref{sec:changes}, Fig.~\ref{fig:change1} \\
Two-exponential SNIa DTD &
  \S\ref{sec:two-exp} \\
Enriched infall &
  \S\ref{sec:enriched} \\
Stellar vs. ISM oxygen abundance & 
  \S\ref{sec:oxygen_budget}, eq.~\ref{eqn:YomeanExp} \\
Evolution of $\xfe$ &
  \S\ref{sec:other_elements}, eq.~\ref{eqn:Deltaxfe} \\
Complex SFH &
  \S\ref{sec:sfhsum}, eq.~\ref{eqn:sfhsum} \\
\hline
\end{tabular}
\caption{Principal Analytic Results}
\tablecomments{Reference guide for finding the analytic results
presented in the paper.  For illustrations of the behavior of 
the simple models, including MDFs, see \S\ref{sec:illustrations}.
For a qualitative summary of findings, see \S\ref{sec:conclusions}.
The two-exponential SNIa DTD considered in \S\ref{sec:two-exp} can
be used to approximate a power-law DTD (e.g., $t^{-1.1}$).
The results in \S\ref{sec:sfhsum} allow generalization to an
arbitrary sum of exponential and linear-exponential star formation
histories.
}
\label{tbl:results}
\end{table*}

\subsection{Constant SFR}
\label{sec:eq_csfr}

First consider the case of a constant $\mdotstar$, 
for which (see Appendix~\ref{appx:SNIa})
\begin{equation}
\label{eqn:bmdotstar_csfr}
\bmdotstar = \mdotstar \left[1-e^{-(t-\td)/\tauIa}\right]
\end{equation}
at times $t \geq \td$ and $\bmdotstar = 0$ for $t < \td$.
Also assume that the mass loading
factor $\eta$ and SFE timescale $\taustar$ are themselves
not changing in time.  In this case, a constant SFR corresponds to
constant $M_g = \taustar \mdotstar$, 
requiring that accretion and recycling from evolved
stars balance the rate at which the ISM loses mass to star formation
and outflow.  For $t-\td > \tauIa$, $\bmdotstar \approx \mdotstar$,
and the abundance ratios can approach an equilibrium in which
$\Modot = \Mfedot = 0$.  Solving 
equations~(\ref{eqn:Modot})-(\ref{eqn:Mfedot})
with this condition implies
\begin{eqnarray}
\Yoeqc &=& \mocc / (1+\eta-r) \label{eqn:YoeqCsfr}\\
\Yfeeqc &=& (\mfecc+\mfeIa) / (1+\eta-r) ~, \label{eqn:YfeeqCsfr}
\end{eqnarray}
where the ``eqc'' subscript denotes equilibrium for a constant
star formation rate.
As discussed in \S\ref{sec:ev_csfr},
the timescale to approach
the equilibrium oxygen abundance is $\taudep$, while the timescale
to approach the equilibrium iron abundance depends on both $\taudep$
and $\tauIa$.
The equilibrium abundances depend on the nucleosynthetic yields
and the outflow rate, but they are independent of the gas depletion
timescale.  Furthermore, the equilibrium abundance ratio 
\begin{equation}
\label{eqn:eqratio_Csfr}
{\Yoeqc \over \Yfeeqc} = 
{\mocc \over \mfecc + \mfeIa}, \quad \hbox{constant SFR},
\end{equation}
is independent of the outflow rate.

\subsection{Exponentially declining SFR}
\label{sec:eq_expsfr}

Next consider a case in which $\taustar$ and $\eta$ are constant 
but $\mdotstar = M_g/\taustar$
declines exponentially on a timescale $\tausfh$ because accretion
and recycling of gas does not keep pace with depletion of gas
by star formation and outflow.  (Note the distinction between
$\taustar$ and $\tausfh$, where the latter refers to the exponential
decline of the star formation history.)
It is useful to rewrite equation~(\ref{eqn:Modot}) in the form
\begin{equation}
\label{eqn:oxygen2}
\Modot = \mocc {M_g \over \taustar} 
       -(1+\eta-r)\left({M_g \over \taustar}\right) 
	   \left({\Mo \over M_g}\right)~.
\end{equation}
The equilibrium oxygen abundance 
corresponds to constant $\Mo/M_g$ and thus
$\Modot/\Mo = \dot{M}_g/M_g = -1/\tausfh$.
Applying this condition to equation~(\ref{eqn:oxygen2}) 
and solving for $\Mo/M_g$ yields
\begin{equation}
\label{eqn:Yoeq}
\Yoeq = {\mocc \over 1+\eta-r-\taustar/\tausfh}~.
\end{equation}
Note that the smallest physical value for $\tausfh$ is 
$\taudep = \tausfh/(1+\eta-r)$,
since even with no accretion the gas supply only diminishes on the depletion
timescale.  Equation~(\ref{eqn:Yoeq}) diverges as $\tausfh$
approaches this critical value, but we will show in 
\S\ref{sec:ev_expsfr} that the timescale to reach equilibrium
also diverges for $\tausfh \rightarrow \taudep$.
Conversely, for $\tausfh \rightarrow \infty$ and thus a nearly constant SFR,
equation~(\ref{eqn:Yoeq}) yields equation~(\ref{eqn:YoeqCsfr})
as a special case.

In similar fashion, we can rewrite the iron mass evolution equation as
\begin{eqnarray}
\label{eqn:iron2}
\Mfedot &=& \mfecc {M_g \over \taustar} + 
 \mfeIa\left({\bmdotstar \over \mdotstar}\right)\left({M_g \over \taustar}\right) \nonumber \\
 && -(1+\eta-r)
    \left({M_g \over \taustar}\right) \left({\Mfe \over M_g}\right)~. 
\end{eqnarray}
Setting $\Mfedot/\Mfe = -1/\tausfh$ yields 
\begin{equation}
\label{eqn:Yfestar}
\Yfeeqs = \Yfecceq + \YfeIaeqs
\end{equation}
with
\begin{equation}
\label{eqn:Yfecceq}
\Yfecceq = {\mfecc \over 1 + \eta -r -\taustar/\tausfh}
\end{equation}
and
\begin{equation}
\label{eqn:YfeIaeqs}
\YfeIaeqs = {\mfeIa \over 1+\eta- r - \taustar/\tausfh} 
            {\bmdotstart \over \mdotstart}~.
\end{equation}
Here we have separated the iron contributions from CCSNe and SNIa,
and we have used subscript $*$ instead of eq because 
$\bmdotstart/\mdotstart$ itself evolves with time,
so equations~(\ref{eqn:Yfestar}) and~(\ref{eqn:YfeIaeqs})
do not yet represent a stationary equilibrium.
%\begin{equation}
%\label{eqn:FeeqExpsfr}
%\Yfeeqs = {\mfecc +\mfeIa{\bmdotstar\over \mdotstar} 
%          \over 1+\eta-\taustar/\tausfh}~.
%\end{equation}
Oxygen and iron abundances are both enhanced relative to the
constant SFR case because the gas supply is declining over time and
providing less dilution of supernova enrichment.  
The final factor in equation~(\ref{eqn:YfeIaeqs}) further boosts
the SNIa iron abundance because the time-averaged SFR $\bmdotstar$ is higher
than the instantaneous SFR, thus raising the rate of SNIa relative to CCSNe.
For the abundance ratio, the first effect cancels while
the second does not, yielding
\begin{equation}
\label{eqn:eqratio_expsfr}
{\Yoeq \over\Yfeeqs}
= {\mocc \over \mfecc + \mfeIa {\bmdotstar\over \mdotstar}}~.
\end{equation}

For an exponential SNIa DTD and an exponential SFH, 
Appendix~\ref{appx:SNIa} shows that
\begin{equation}
\label{eqn:mdotstar_ratio}
{\bmdotstart \over \mdotstar(t)} =
  \left({\taubarIaSfh \over \tauIa}\right)\,e^{\td/\tausfh} 
  \left[1-e^{-\Delta t/\taubarIaSfh}\right] 
\end{equation} 
at times $t > t_D$, where 
\begin{equation}
\label{eqn:Deltat}
\Delta t \equiv t - t_D
\end{equation}
is the time interval since the commencement of SNIa explosions.
%\begin{equation}
%\label{eqn:bmdotstar_exp}
%\bmdotstart = \left({\taubarIaSfh \over \tauIa}\right)
%  \left[1-e^{-\Delta t/\taubarIaSfh}\right] \mdotstar(t-\td) 
%\end{equation}
%at times $t > \td$. 
Here we have introduced the notation
\begin{equation}
\label{eqn:taubar}
\taubarXY \equiv \left({1 \over \tau_X} - {1 \over \tau_Y}\right)^{-1}
\end{equation}
for a ``harmonic difference timescale'' that
arises frequently in our solutions from integrating the
products of exponentials of opposite sign.
The limiting cases are
\begin{eqnarray}
\label{eqn:taubarlimits}
\taubarXY &\approx & \tau_X, \qquad \tau_X \ll \tau_Y \\
\taubarXY &\approx & -\tau_Y, \qquad \tau_Y \ll \tau_X~.
\end{eqnarray}
If one timescale is much smaller than the other, then it controls
the rate of evolution, but 
if the two timescales are close then $\taubarXY$ can
be much longer than either one individually.
While the sign of $\taubarXY$ can be positive or negative,
it typically appears in multiplicative combinations like that of
equation~(\ref{eqn:mdotstar_ratio}), which yield a positive-definite value
and suppress divergences for $\tau_X \approx \tau_Y$.
In equation~(\ref{eqn:mdotstar_ratio})
we have the identification $\tau_X=\tauIa$ and $\tau_Y=\tausfh$.  
When $\Delta t \ll |\taubarIaSfh|$, which is always the case if 
$\tauIa$ and $\tausfh$ are sufficiently close, then Taylor-expanding
equation~(\ref{eqn:mdotstar_ratio}) yields
\begin{equation}
\label{eqn:mdotstar_ratio_approx}
{\bmdotstart \over \mdotstar(t)} \approx
  \left(\Delta t \over \tauIa\right) e^{t_D/\tausfh}, \qquad
  \Delta t \ll |\taubarIaSfh|~.
\end{equation}

The iron abundance and oxygen-to-iron ratio typically approach
the values of equation~(\ref{eqn:Yfestar}) and~(\ref{eqn:eqratio_expsfr})
on a gas depletion timescale.  On the (usually longer) timescale
$\taubarIaSfh$, $\YfeIaeqs$ approaches the equilibrium value
\begin{equation}
\label{eqn:YfeIaeq}
\YfeIaeq = \left({\taubarIaSfh \over \tauIa}\right)\,e^{\td/\tausfh} 
           {\mfeIa \over 1+\eta-r-\taustar/\tausfh}~.
\end{equation}
For a slowly declining star formation history with $\tauIa \ll \tausfh$,
the harmonic difference timescale is
$\taubarIaSfh \approx \tauIa$, and $\YfeIaeq$ is only slightly
larger than $\mfeIa/(1+\eta-r-\taustar/\tausfh)$.
Conversely, if $\tauIa$ and $\tausfh$ are 
close then the final equilibrium $\YfeIaeq$ can be large,
but the timescale $|\taubarIaSfh|$ for reaching it becomes long.
If $\tausfh$ is smaller than $\tauIa$ then $\bmdotstart/\mdotstart$
grows indefinitely (eq.~\ref{eqn:mdotstar_ratio}), and the
iron abundance never approaches an equilibrium
because the gas supply is falling off
more rapidly than the SNIa enrichment from previously formed stars.
This case corresponds to negative values of 
$\taubarIaSfh$ and thus of $\YfeIaeq$, but in 
our equations $\YfeIaeq$ appears in multiplicative combinations
that ensure a positive iron abundance.

In equations~(\ref{eqn:Yoeq}), (\ref{eqn:Yfecceq}),
(\ref{eqn:YfeIaeqs}), and~(\ref{eqn:YfeIaeq}),
one can readily identify the
impact of yields, outflow mass loading, recycling,
and the declining gas supply implied by a declining
star formation history.  For notational compactness
and connection to subsequent results, it is useful to recognize
that by substituting $\tau_*/\taudep = 1+\eta-r$ one can also
express the equilibrium abundances in the form:
\begin{eqnarray}
\Yoeq &=& \mocc\times \taubarDepSfh/\taustar~, \label{eqn:Yoeq2} \\
\Yfecceq &=& \mfecc\times \taubarDepSfh/\taustar~, \label{eqn:Yfecceq2} \\
\YfeIaeq &=& \mfeIa \times (\taubarDepSfh/\taustar)(\taubarIaSfh/\tauIa) 
              e^{\td/\tausfh}. \label{eqn:YfeIaeq2}
\end{eqnarray}

%%%%%%
%\begin{figure*}
%\centerline{\includegraphics[width=19cm]
%{pcaa_chemevol/sims/dec_sfr/yale_visit/plots/cl04_scatter_pc1_bin.pdf}}
%\caption{PCAA on bins of $\Delta$[Fe/H] = 0.01: PC1 eigenvector component of
%  each element as a function of [Fe/H].}
%\end{figure*}
%%%%%%

\section{Full Time Evolution}
\label{sec:evolution}

In the following subsections, we present analytic solutions to the
one-zone chemical evolution equations under three different assumptions
about the star formation history, with specific discussions of
metallicity distribution functions and the relation to traditional
analytic chemical evolution models in \S\S\ref{sec:mdf} 
and~\ref{sec:traditional}.  We present illustrative 
results in \S\ref{sec:illustrations} , and some readers may
prefer to start with those illustrations and refer back to the
equations as needed.  In \S\ref{sec:tests} we compare the analytic models to
numerical results that include
the full time dependence of AGB recycling.
Some analytic results for other cases appear in \S\S\ref{sec:sudden}
and~\ref{sec:extensions}.
We often refer to a ``fiducial'' model, chosen to have physically
reasonable parameter values that yield near-solar equilibrium 
abundances.  The values of these parameters, including our adopted
supernova yields and solar abundances, are listed in 
Table~\ref{tbl:variables}.
Table~\ref{tbl:results} provides a ``finding chart'' for our
principal analytic results.

%%%%%%
\begin{figure}
\centerline{\includegraphics[width=2.8truein]{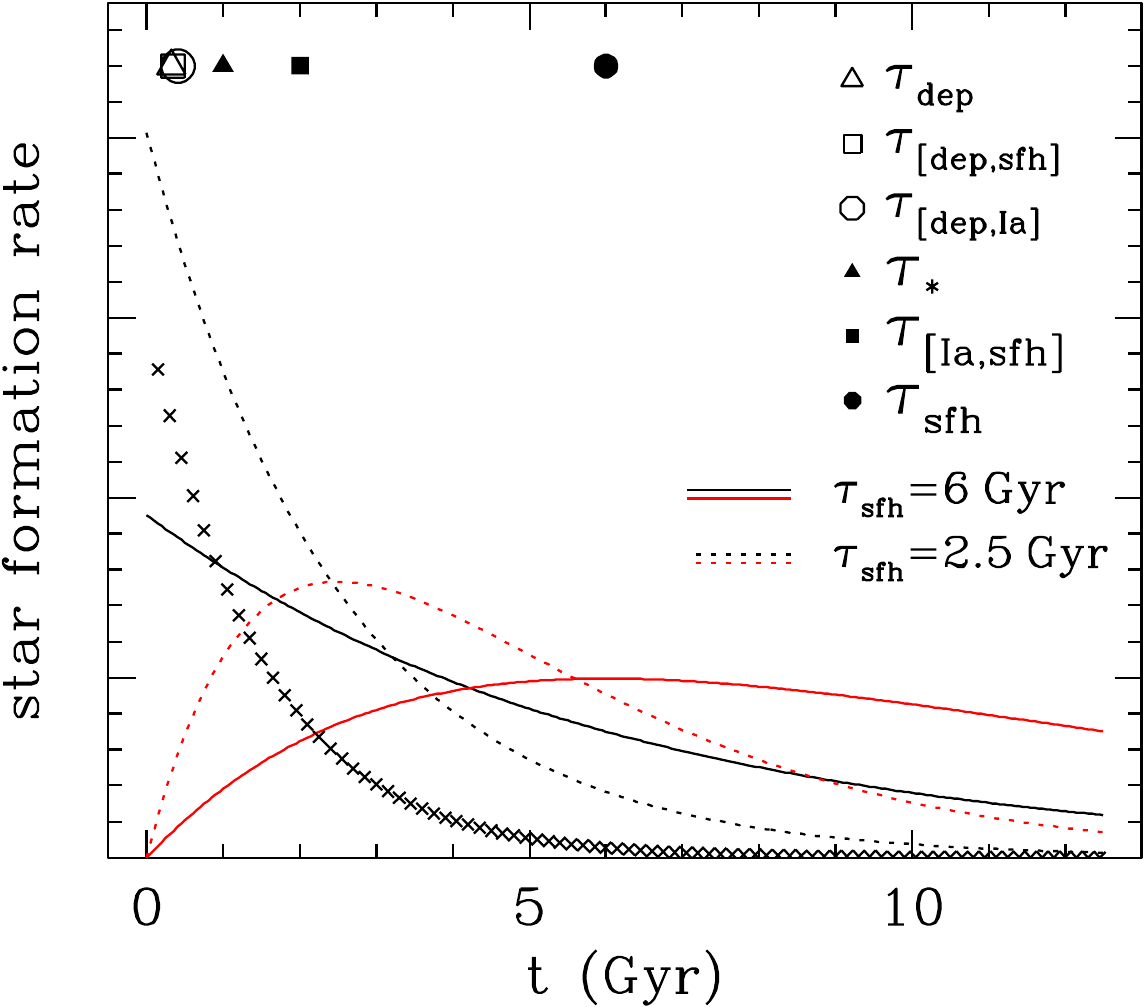}}
\caption{Star formation histories and timescales.  The solid black curve
shows the SFH of our fiducial model, an exponential with $\tausfh=6\Gyr$.
The dotted black curve shows a $\tausfh=2.5\Gyr$ exponential, used in 
some of our examples.  Red solid and dotted curves show linear-exponential
SFH cases with the same two timescales.  SFR units are arbitrary, with
all histories normalized to the same integral over $12.5\Gyr$.
Crosses show the exponential SNIa DTD with $\tauIa=1.5\Gyr$ and
minimum delay time $\td=0.15\Gyr$ assumed in most of our calculations.
Points near the top axis show quantities for the fiducial model:
the depletion, SFE, and SFH timescales ($\taudep$, $\taustar$, $\tausfh$)
and the three harmonic difference timescales
$\taubarDepSfh$, $\taubarDepIa$, $\taubarIaSfh$.
}
\label{fig:timescales}
\end{figure}
%%%%%%

To set the scene for thinking about evolutionary behavior, 
Figure~\ref{fig:timescales} compares star formation histories for
several of the cases that we consider in our examples of 
\S\ref{sec:illustrations}, and it shows the timescales that govern
chemical evolution in our fiducial model.  The fiducial model
adopts a $6\Gyr$ exponential SFH (solid black curve), SFE timescale
$\taustar=1\Gyr$, and outflow efficiency $\eta=2.5$.  
The solid red curve shows a linear-exponential SFH with the 
same $\tausfh$, which rises to a broad plateau over the range
$t \approx 4-11\Gyr$.  Dotted curves show exponential and linear-exponential
curves with a shorter $e$-folding timescale $\tausfh=2.5\Gyr$.
Production of CCSN elements directly tracks the SFH, while production
of iron from SNIa is given by the convolution of the SFH with our
fiducial DTD, an exponential with $\tauIa=1.5\Gyr$, shown by crosses.
We will see below that the evolution of CCSN abundances is governed
by the harmonic difference timescale $\taubarDepSfh$, while the
evolution of SNIa iron depends additionally on $\taubarDepIa$ and
$\taubarIaSfh$.  Because of the high $\eta$ adopted in the fiducial model, 
the depletion time $\taudep=0.323\Gyr$ is much shorter than $\tausfh$ or
$\tauIa$.  The first two of these timescales are therefore very
close to $\taudep$ itself.  The longest timescale in the fiducial
model is $\taubarIaSfh=2\Gyr$, so the iron abundance approaches equilibrium
on a timescale that is much longer than that of oxygen but still
short compared to the SFH timescale and the age of the Galaxy.

While a constant SFR is just the limiting case of an exponential SFH
with $\tausfh \rightarrow \infty$, it is useful to begin with this
analytically and physically simpler case.

\subsection{Constant SFR}
\label{sec:ev_csfr}

For a constant SFR, and constant $\taustar$ and $\eta$, it is
straightforward to solve for the full evolution of the 
oxygen abundance.  It is helpful 
to first rewrite 
equation~(\ref{eqn:Modot}) in the form
\begin{equation}
\Yodot + {\Yo \over\taudep} = {\mocc \over \taustar} ~,
\end{equation}
using the substitutions $\mdotstar = M_g/\taustar$ and $\Yo = \Mo/M_g$.
We have used the fact that 
constant SFR and $\taustar$ imply that $\dot{M}_g = 0$ and
thus $\Yodot = \Modot/M_g$.
The general solution to a differential equation of the form
\begin{equation}
\label{eqn:diffeq}
\dot y + p(t)y = f(t)
\end{equation} 
is 
\begin{equation}
\label{eqn:diffeqsol}
y(t) = {1 \over \mu(t)}\left[\int_0^t \mu(t')f(t')dt' + C\right], \quad
\mu(t) = e^{\int p(t) dt}~,
\end{equation}
where in this case we have simply $p(t) = \taudep^{-1}$,
$\mu(t) = e^{t/\taudep}$, and $f(t) = \mocc\taustar^{-1}$.  
Setting the integration constant $C=0$ corresponds to setting
$\Yo=0$ at $t=0$.  With this initial condition,
the solution is
\begin{equation}
\label{eqn:YoEvol}
\Yo(t) = {\mocc \over (1+\eta-r)}\left[1-e^{-t/\taudep}\right],
\end{equation}
which approaches the equilibrium abundance $\Yoeqc$ on
a gas depletion timescale $\taudep = \taustar/(1+\eta-r)$.
\cite{larson1972} derives an equivalent equation for the
case $\eta=r=0$.
(For further discussion of the relation to previous analytic
results see \S\ref{sec:traditional}.)
Starting from a non-zero abundance adds a term
$Z_{O,{\rm init}} e^{-t/\taudep}$ to equation~(\ref{eqn:YoEvol}).

For iron, it is useful to consider the evolution of the core collapse
and SNIa contributions separately, with the total iron abundance
being simply the sum of the two.  The core collapse contribution
follows the same evolution as the oxygen abundance:
\begin{equation}
\label{eqn:YfeccEvol}
\Yfecc(t) = {\mfecc \over (1+\eta-r)}\left[1-e^{-t/\taudep}\right].
\end{equation}
At times $t < \td$, there is no SNIa contribution, and the 
abundance ratio is just determined by the CCSN yields,
$\Yo/\Yfe = \mocc/\mfecc$.

For the evolution of the SNIa iron contribution,
a similar set of substitutions in equation~(\ref{eqn:Mfedot}) yields
\begin{equation}
\label{eqn:YfedotIa}
\YfedotIa + {\YfeIa \over \taudep} = 
  {\mfeIa \over \taustar}{\bmdotstar \over \mdotstar} 
  = {\mfeIa \over \taustar}\left[1- e^{-\Delta t/\tauIa}\right]
  ~,
\end{equation}
where the second equality uses $\bmdotstar$ for an exponential
SNIa DTD with constant SFR and $\Delta t = t-\td$ 
(see equation~\ref{eqn:bmdotstar_csfr}).
With a slightly tedious but straightforward
calculation, one can use the same method to solve this differential
equation with the boundary condition $\YfeIa=0$ at $\Delta t = 0$.
The result can be expressed in the form
\begin{equation}
\label{eqn:YfeIaEvol}
\begin{split}
\YfeIa(t) = & {\mfeIa \over 1+\eta-r}\Big[1-e^{-\Delta t/\taudep}-  \\
%  & \left(1-{\taudep \over\tauIa}\right)^{-1}
   & {\taubarDepIa \over \taudep}
  \left( e^{-\Delta t/\tauIa} - e^{-\Delta t/\taudep}\right) \Big] ~,
\end{split}
\end{equation}
with $\taubarDepIa$ defined by equation~(\ref{eqn:taubar}),
subject to the condition $\taudep \neq \tauIa$.
%In the limit of short $\tauIa$, when SNIa enrichment is effectively
%instantaneous except for the time delay $\td$, 
%equation~(\ref{eqn:YfeIaEvol}) approaches the core collapse 
%evolution~(\ref{eqn:YfeccEvol}) with the substitutions
%$\mfecc \rightarrow \mfeIa$ and $t \rightarrow \Delta t$.
%In the opposite limit of $\tauIa \gg \taudep$, the factor in $[~]$
%becomes $[1-e^{-\Delta t/\tauIa}]$ instead of $[1-e^{-t/\taudep}]$.
%These limits make intuitive sense, but in realistic cases we
%expect $\tauIa$ and $\taudep$ to be of similar order.

It is evident from equation~(\ref{eqn:YfeIaEvol}) that the equilibrium
iron abundance is reached only when $\Delta t \gg \taudep$ {\it and}
$\Delta t \gg \tauIa$, i.e., the timescale to reach equilibrium
is controlled by the longer of the gas depletion timescale and the SNIa
timescale.  In the limiting cases of $\tauIa \ll \taudep$ or
$\taudep\ll \tauIa$, the factor $\taubarDepIa/\taudep$ goes to 0 or 1
respectively, and equation~(\ref{eqn:YfeIaEvol}) yields the pleasingly
intuitive results
\begin{eqnarray}
\YfeIa(t) &=& {\mfeIa \over 1+\eta-r}\left[1-e^{-\Delta t/\taudep}\right],
  \quad \tauIa \ll \taudep, \label{eqn:YfeIaEvolLimit1} \\
\YfeIa(t) &=& {\mfeIa \over 1+\eta-r}\left[1-e^{-\Delta t/\tauIa}\right],
  \quad \taudep \ll \tauIa.\label{eqn:YfeIaEvolLimit2} 
\end{eqnarray}
In the former case, SNIa enrichment is effectively instantaneous
after $\td$, so the evolution is simply a shifted version of the
core collapse evolution~(\ref{eqn:YfeccEvol}) with the SNIa yield
in place of the CCSN yield.  
In the latter case, evolution to equilibrium is controlled instead
by the SNIa timescale.

For many realistic situations we
expect $\tauIa$ and $\taudep$ to be of similar order.
Equation~(\ref{eqn:YfeIaEvol}) looks
it could diverge for $\taudep \approx \tauIa$,
but rearranging and Taylor expanding the exponentials shows that 
when $\Delta t \ll |\taubarDepIa|$ the solution approaches
\begin{equation}
\label{eqn:YfeIaEvolConverge}
\YfeIa(t) \approx {\mfeIa \over 1+\eta-r}\left[1-e^{-\Delta t/\taudep}-
  {\Delta t \over \taudep}e^{-\Delta t/\taudep}\right]~,
\end{equation}
so there is no divergence.

At early times when $\Delta t \ll \taudep$ and $\Delta t \ll \tauIa$,
Taylor expanding all of the exponentials in equation~(\ref{eqn:YfeIaEvol})
shows that $\YfeIa(t)=0$ to first order in $\Delta t$.
Expanding to second order yields
\begin{equation}
\label{eqn:YfeIaEvol2ndOrder}
\YfeIa(t) \approx {\mfeIa \over 1+\eta-r} \cdot 
   {1\over 2}{\Delta t\over \tauIa}{\Delta t \over \taudep},
   \qquad \Delta t \ll \tauIa,~\Delta t \ll \taudep~.
\end{equation}
In this sense, SNIa enrichment ``turns on slowly'' starting at $t=\td$,
so the ``knee'' in an $\ofe-\feh$ diagram is a smooth bend rather
than a discontinuous change of slope (see Fig.~\ref{fig:ofe1} below).

%For comparison to other solutions derived below, it is worth noting
%that equation~(\ref{eqn:YfeIaEvol}) can be expressed in the alternative form
%\begin{equation}
%\label{eqn:YfeIaEvol2}
%\begin{split}
%\YfeIa(t) = & {\mfeIa \over 1+\eta}\Big[1-e^{-\Delta t/\taudep}\Big(1+  \\
%  & {\taubarDepIa \over \taudep}\left(e^{\Delta t/\taubarDepIa}-1\right)\Big)
%  \Big]~,
%\end{split}
%\end{equation}
%where $\taubarDepIa$ is defined according to equation~(\ref{eqn:taubar}).
%{\it Well, maybe it is worth noting, and maybe it isn't.  Leaving it
%in for now to decide later whether it's useful.}

Since core collapse iron follows the same evolution as core collapse oxygen,
the drop in $\ofe$ relative to the early-time CCSN plateau is determined
by the ratio of SNIa iron to core collapse iron:
\begin{equation}
\label{eqn:Deltaofe}
\Deltaofe = -\log_{10}\left[1+\YfeIa(t)/\Yfecc(t)\right]~,
\end{equation}
where we have defined
\begin{equation}
\Deltaofe \equiv \ofe-\ofe_{\rm plateau}
\end{equation}
and
\begin{equation}
\label{eqn:plateau}
\ofe_{\rm plateau}=\log_{10}
  \left[\frac{\mocc/\mfecc}{Z_{\rm O,\odot}/Z_{\rm Fe,\odot}}\right]~.
\end{equation}
The late time result, in equilibrium, is 
\begin{equation}
\Deltaofe = -\log_{10}\left[1+\mfeIa/\mfecc\right], \quad t \gg \taudep~.
\end{equation}
The {\it change} in $\ofe$ depends only on
the iron yields, not the oxygen yields; for our adopted yields
$\mfecc=0.0012$ and $\mfeIa=0.0017$ the drop is 0.38 dex
for a constant SFR.
Relative to the low metallicity plateau, the evolution 
of $[{\rm X}/{\rm Fe}]$ 
should be the same for any element X whose production is 
dominated by CCSNe, provided the IMF stays constant in time
and the element's yield is independent of metallicity.
Departures from this behavior could be a useful diagnostic for
element production mechanisms or IMF changes.
We return to this point in \S\ref{sec:other_elements}.

For small $\Delta t$ (specifically $\YfeIa/\Yfecc \ll 1$,
$\Delta t \ll \taudep$, and $\Delta t \ll \tauIa$), 
we can use a Taylor expansion of
equation~(\ref{eqn:Deltaofe})
together with equations~(\ref{eqn:YfeccEvol}) 
and~(\ref{eqn:YfeIaEvol2ndOrder}) 
to approximate the evolution of $\ofe$ near the knee, obtaining
\begin{equation}
\label{eqn:ofeKnee2ndOrder}
\Deltaofe \approx 
  -{\mfeIa/\mfecc \over 2 \ln 10} {(\Delta t)^2 \over \tauIa\taudep}
  \times \left(1-e^{-t/\taudep}\right)^{-1}~,
\end{equation}
with $t \approx \td$.
Thus, the initial turndown of the $\ofe$ track is quadratic in $\Delta t$.
The full expression based on combining equations~(\ref{eqn:YfeIaEvol})
and~(\ref{eqn:YfeccEvol}) is not especially illuminating, but in
the limiting cases represented by 
equations~(\ref{eqn:YfeIaEvolLimit1}) and~(\ref{eqn:YfeIaEvolLimit2}) 
we obtain, respectively,
\begin{equation}
\label{eqn:ofeEvolLimit1}
\Deltaofe \approx 
  -\log_{\rm 10}\left[1+{\mfeIa\over \mfecc}
    {\left(1-e^{-\Delta t/\taudep}\right) \over
     \left(1-e^{- t/\taudep}\right)}\right]~
\end{equation}
for $\tauIa \ll \taudep$ and 
\begin{equation}
\Deltaofe \approx 
  -\log_{\rm 10}\left[1+{\mfeIa\over \mfecc}
    {\left(1-e^{-\Delta t/\tauIa}\right) \over
     \left(1-e^{-t/\taudep}\right)}\right]
\end{equation}
for $\taudep \ll \tauIa$.
In the former case, the abundance drop approaches its equilibrium
value by the time $\Delta t \approx t$ (i.e., $t \gg \td$),
while in the latter case the approach to
equilibrium depends mainly on $\Delta t/\tauIa$.

\subsection{Exponentially declining SFR}
\label{sec:ev_expsfr}

When considering time-dependent SFR, we assume that the time dependence
is driven by a time-dependent gas supply $M_g(t)$, with constant $\taustar$.
(See \S\ref{sec:timedeptaustar} for a time-dependent $\taustar$ case.)
It is most straightforward to solve for the metal mass evolution itself,
then divide by $M_g(t)$ to get the evolving abundance.
In the case of oxygen, substituting $\mdotstar(t)=M_g(t)/\taustar$ and
$\Yo=\Mo(t)/M_g(t)$ in equation~(\ref{eqn:Modot}) yields
\begin{equation}
\label{eqn:ModotTimedep}
\Modot + {\Mo\over \taudep} = \mocc{M_g(t)\over \taustar}~.
\end{equation}
The solution proceeds like the above solution for $\Yo(t)$ with
constant SFR, but now the driving function $f(t)$ in 
equation~(\ref{eqn:diffeq}) is $\mocc M_g(t)/\taustar$
instead of simply $\mocc/\taustar$.
For an exponential star formation history we have
$M_g(t)=M_{g0}e^{-t/\tausfh}$, and
a short calculation yields the solution (assuming $\tausfh \neq \taudep$)
\begin{equation}
\label{eqn:YoEvolExp}
\Yo(t) = 
% {\mocc \over 1+\eta-\taustar/\tausfh}
  \Yoeq
  \left[1-e^{-t/\taubarDepSfh}\right]~,
\end{equation}
with $\Yoeq$ given by equation~(\ref{eqn:Yoeq2}) and
\begin{equation}
\label{eqn:taubar2}
\taubarDepSfh \equiv \left(\taudep^{-1}-\tausfh^{-1}\right)^{-1}~.
\end{equation}
This solution has the same form as equation~(\ref{eqn:YoEvol}) for
constant SFR, but the equilibrium abundance is now the higher one
from equation~(\ref{eqn:Yoeq}), and the timescale for approaching
equilibrium is $\taubarDepSfh$ rather than $\taudep$.
Provided there is continuing gas accretion, we expect $\tausfh > \taudep$,
and if $\tausfh \gg \taudep$ then $\taubarDepSfh \approx \taudep$.
However, if $\tausfh \approx \taudep$, as expected for a very low gas
accretion rate, then the equilibrium abundance itself is high
because $\taustar/\tausfh\approx 1+\eta-r$, and the timescale for
reaching that abundance is long ($\taubarDepSfh \gg \taudep$)
because the enrichment rate is evolving on the same timescale as
the gas supply itself.  

If gas is being swept from the system
by ram pressure stripping, or by some other process not associated
with star formation, then $\tausfh$ can in principle be shorter
than $\taudep$.  In this case $\taubarDepSfh$ is negative, but
equation~(\ref{eqn:YoEvolExp}) still yields a positive result
because both the pre-factor and the factor in $[~]$ change sign.

For iron, it is again useful to separate the evolution of the 
CCSN contribution and the SNIa contribution.
The former has the same behavior as oxygen,
\begin{equation}
\label{eqn:YfeccEvolExp}
\Yfecc(t) = 
  % {\mfecc \over 1+\eta-\taustar/\tausfh}
  \Yfecceq
  \left[1-e^{-t/\taubarDepSfh}\right]~.
\end{equation}
The solution for SNIa is somewhat tedious because there are
three exponential timescales involved, $\taudep$, $\tauIa$, and
$\tausfh$, and they enter in several different combinations.
The end result is
\begin{equation}
\label{eqn:YfeIaEvolExp}
\begin{split}
\YfeIa(t) = & 
% {\mfeIa \over 1+\eta} 
% {\taubarIaSfh \taubarDepSfh \over \tauIa\taudep} e^{\td/\tausfh} ~\times \\
  \YfeIaeq 
  \Big[1-e^{-\Delta t/\taubarDepSfh} \\
  & - {\taubarDepIa \over \taubarDepSfh} 
  \left( e^{-\Delta t/\taubarIaSfh} - e^{-\Delta t/\taubarDepSfh}\right) \Big] 
  ~.
\end{split}
\end{equation}
In the limit that $\tausfh$ is much larger than $\tauIa$ and 
$\taudep$, the harmonic difference timescales simplify to
$\taubarDepSfh \approx \taudep$ and $\taubarIaSfh \approx \tauIa$,
making equation~(\ref{eqn:YfeIaEvolExp}) equivalent to
equation~(\ref{eqn:YfeIaEvol}).  This is as expected, since an
infinite exponential timescale is equivalent to a constant SFR.
In the alternative limit that $\tauIa$ and $t_D$ are much shorter
than $\Delta t$, $\taudep$, and $\tausfh$, the ratio
$\taubarDepIa/\taubarDepSfh \rightarrow 0$, and the 
solution~(\ref{eqn:YfeIaEvolExp}) approaches that for 
$\Yfecc(t)$ in equation~(\ref{eqn:YfeccEvolExp}), except
for the difference in yield.
%[A short calculation shows that 
%$(1+\eta)^{-1}\taubarDepSfh/\taudep = (1+\eta-\taustar/\tausfh)^{-1}.$]
In this limit, SNIa enrichment is effectively instantaneous,
so the distinction from CCSN enrichment goes away.
Finally, in the limit that $\taudep$ is much smaller than
either $\tauIa$ or $\tausfh$, equation~(\ref{eqn:YfeIaEvolExp}) approaches
\begin{equation}
\label{eqn:YfeIaEvolExpApprox}
\YfeIa(t) = 
  \YfeIaeq
  \left[1-e^{-\Delta t/\taubarIaSfh}\right]~,
\end{equation}
with the form of equation~(\ref{eqn:YfeccEvolExp}) but
the timescale $\taubarIaSfh$.
While it is not obvious by inspection, equation~(\ref{eqn:YfeIaEvolExp})
yields a positive iron abundance even when $\taubarIaSfh$, and
thus $\YfeIaeq$, are negative.

\subsection{Linear-Exponential SFR}
\label{sec:ev_linexpsfr}

For the early stages of evolution, it is useful to consider cases
in which the gas supply is growing from a small initial value
rather than starting at a high value and either staying constant
or declining with time.  For example, the functional form
$t e^{-t/\tausfh}$ provides a better match to the predicted star 
formation histories of galaxies in semi-analytic models and
cosmological simulations
than $e^{-t/\tausfh}$ \citep{lee2010,simha2014}, and at times $t \ll \tausfh$
this form is just linear in time.  

For a linear gas supply 
$M_g(t) \propto t$, our previous solution methods for oxygen 
abundance yield the result
\begin{equation}
\label{eqn:YoEvolLinear}
\Yo(t) = {\mocc \over 1+\eta-r}
  \left[1-{\taudep \over t}\left(1- e^{-t/\taudep}\right)\right]~,
\end{equation}
which has the same equilibrium abundance as the constant SFR
solution (eq.~\ref{eqn:YoEvol}) but a different evolutionary behavior.  
At times $t \ll \taudep$, a 2nd-order Taylor expansion shows
that the abundance from equation~(\ref{eqn:YoEvolLinear}) is
just half that for a constant SFR model with the same parameters,
because for a given $M_g(t)=\taustar\mdotstar$ the stellar
mass formed (and hence CCSN enrichment produced)
is just half that formed for a constant gas supply.
For oxygen, it is also straightforward to solve the evolution equations
for $\mdotstart \propto t e^{-t/\tausfh}$, with the result
\begin{equation}
\label{eqn:YoEvolLinExp}
\Yo(t) = \Yoeq
  \left[1-{\taubarDepSfh\over t}\left(1-e^{-t/\taubarDepSfh}\right)\right]~,
\end{equation}
where $\Yoeq$ is the equilibrium abundance~(\ref{eqn:Yoeq2}) for
the exponentially declining SFH.
In the limit of $\tausfh \gg \taudep$ this approaches the
result for pure linear evolution as expected.
In general, the oxygen abundance for a linear-exponential SFH
is half that of an exponential SFH at early times and
approaches the same final equilibrium at a slower pace.
For both cases, early time evolution is sensitive to the 
assumption that CCSN recycling is instantaneous, and it 
would be altered if the CCSN products take time to rejoin
the star-forming phase of the ISM.

%In particular,
%for $t \ll \taudep$, equation~(\ref{eqn:YoEvolLinear}) vanishes
%to first order in $t$, making the early evolution
%\begin{equation}
%\label{eqn:YoEvolLinearEarly}
%\Yo(t) \approx {\mocc \over 1+\eta-r} \cdot 
%  {1\over 2}\left({t\over\taudep}\right)^2, \qquad t \ll \taudep
%\end{equation}
%in constrast to the linear growth that arises for constant SFR,
%\begin{equation}
%\label{eqn:YoEvolCsfrEarly}
%\Yo(t) \approx {\mocc \over 1+\eta} \cdot 
%  \left({t\over\taudep}\right)~.
%\end{equation}
%In both cases the oxgyen mass is growing linearly at
%early times {\it check this statement}, 
%but if $M_g(t)$ is itself growing linearly then
%the ratio $\Mo/M_g$ only grows quadratically.

For iron, the full solution is cumbersome, but this case is of
sufficient physical interest to make the result worth reporting.
The CCSN contribution follows the same behavior as oxygen, of course,
\begin{equation}
\label{eqn:YfeccEvolLinExp}
%\begin{split}
%\Yfecc(t) = & {\mfecc \over 1+\eta - \taustar/\tausfh} \times \\
\Yfecc(t) = \Yfecceq 
  \left[1-{\taubarDepSfh\over t}\left(1-e^{-t/\taubarDepSfh}\right)\right]~.
%\end{split}
\end{equation}
The solution for SNIa iron involves all three combinations of the
depletion, SNIa, and SFH timescales:
\begin{equation}
\label{eqn:YfeIaEvolLinExp}
\begin{split}
\YfeIa(t) =& 
 %  {\mfeIa \over 1+\eta} 
 %  {\taubarIaSfh \taubarDepSfh \over \tauIa\taudep}
 %            e^{t_D/\tausfh} {\taubarIaSfh \over t} \times \\
   \YfeIaeq {\taubarIaSfh \over t} \times \\
   & \bigg[ {\Delta t \over \taubarIaSfh} + 
           {\taubarDepIa \over \taubarDepSfh} e^{-\Delta t/\taubarIaSfh} + \\
   & 	   \left(1+{\taubarDepSfh\over\taubarIaSfh}
                 -{\taubarDepIa\over\taubarDepSfh}\right) 
	   e^{-\Delta t/\taubarDepSfh} - \\
   &       \left(1+{\taubarDepSfh\over\taubarIaSfh}\right) \bigg]~.
\end{split}
\end{equation}
The equilibrium abundances in
equations~(\ref{eqn:YfeccEvolLinExp}) and~(\ref{eqn:YfeIaEvolLinExp})
again match those of the exponential cases, given in 
equations~(\ref{eqn:Yfecceq2}) and~(\ref{eqn:YfeIaeq2}).
One can again take the limit of small $t_D$ and $\tauIa$ and find
that the SNIa iron evolution follows that of CCSN iron,
i.e., equation~(\ref{eqn:YfeIaEvolLinExp}) approaches
equation~(\ref{eqn:YfeccEvolLinExp}) with the substitution
$\mfecc \rightarrow \mfeIa$.
It is also illuminating to consider the limit in which $\taudep$
is much smaller than both $\tauIa$ and $\tausfh$, in which case
equation~(\ref{eqn:YfeIaEvolLinExp}) becomes
\begin{equation}
\label{eqn:YfeIaEvolLinExpApprox}
\begin{split}
\YfeIa(t) \approx \YfeIaeq 
  & \Big[{\Delta t \over t}-{\taudep\over t}\left(1-e^{-t/\taudep}\right) - \\
  & {\taubarIaSfh\over t}\left(1-e^{-t/\taubarIaSfh}\right)\Big]~,
\end{split}
\end{equation}
which is similar in form to equation~(\ref{eqn:YfeccEvolLinExp}) but
with the additional dependence on the timescale $\taubarIaSfh$.

\subsection{Metallicity distribution functions}
\label{sec:mdf}

In many cases we are interested in the distribution of stars as
a function of metallicity in addition to the evolution of abundances
over time.  If the abundance $Z(t)$ is monotonically increasing,
then the number of stars born as a function of metallicity will
be 
\begin{equation}
\label{eqn:mdfbasic}
{dN \over dZ} = {dN/dt \over dZ/dt} \propto {\mdotstar \over \dot{Z}}~.
\end{equation}
For the simplest cases, this distribution can be obtained analytically.

The oxygen abundance evolution for constant SFR can be expressed in 
the form
\begin{equation}
\label{eqn:YoEvolCsfrScaled}
\zo(t) \equiv \Yo(t)/\Yoeq = 1-e^{-t/\taudep}~,
\end{equation}
where $\Yoeq = \Yoeqc = \mocc/(1+\eta-r)$.
From this we obtain $\dot{\zo} \propto (1-\zo)$, and since $\mdotstar$ is
constant we get
\begin{equation}
\label{eqn:mdf1}
{dN \over d\zo} \propto (1-\zo)^{-1} ~,
\end{equation}
truncated at the value of $\zo(t)$ 
given by equation~(\ref{eqn:YoEvolCsfrScaled}).

For the exponentially declining SFR (eq.~\ref{eqn:YoEvolExp}), the
result is similar, but now the equilibrium abundance is that of
equation~(\ref{eqn:Yoeq2}) and we must multiply by
$\mdotstar(t) \propto e^{-t/\tausfh} \propto (1-\zo)^{\taubarDepSfh/\tausfh}$,
where we have used equation~(\ref{eqn:YoEvolExp}) to find
$t = -\taubarDepSfh\ln(1-\zo)$.
The result is
\begin{equation}
\label{eqn:mdf2}
{dN \over d\zo} \propto (1-\zo)^{-1+\taubarDepSfh/\tausfh} ~.
\end{equation}
In the limit of $\tausfh \gg \taudep$ this approaches the constant SFR
result as expected, but in general the exponentially declining SFR
produces a distribution that is less sharply peaked at the equilibrium 
abundance,
in part because it takes longer to reach equilibrium, but mostly because
a smaller fraction of stars are produced at late times.
Noting that $\taubarDepSfh/\tausfh = (\tausfh/\taudep-1)^{-1},$ one
can see that $\tausfh=2\taudep$ is a critical case for which
$dN/d\zo$ is constant up to the cutoff at $\zo(t)$.
Note that metallicity distribution functions
(MDFs) are frequently plotted as histograms
in logarithmic bins of metallicity (e.g., bins of $\feh$), which
show $dN/d\ln Z \propto Z\,dN/dZ$.
Analytic expressions for the population mean metallicity corresponding
to equations~(\ref{eqn:mdf1}) and~(\ref{eqn:mdf2}) appear
in \S\ref{sec:oxygen_budget} below.

\cite{qian2012} present analytic forms for the MDF with instantaneous
recycling and enrichment using essentially the same methods
that we have used here for oxygen, though with quite different notation.
Unfortunately, the other cases we have considered do not allow
an analytic expression for $t$ or $d\zo/dt$ in terms of $\zo$,
so there is no closed form for the oxygen MDF in the linear 
or linear-exponential cases, or for the iron MDF once SNIa enrichment
becomes important.  However, it is easy to tabulate the analytic
solution for $Z(t)$ with equal time steps, then 
sum $\mdotstart$ in the desired bins of $Z$ to compute the MDF
(see Appendix~\ref{appx:userguide}).

\subsection{Relation to Traditional Analytic Models}
\label{sec:traditional}

Conventional analytic models of chemical evolution adopt the
instantaneous recycling approximation for all elements, 
both CCSN and SNIa products, so they should be
comparable to our results for oxygen evolution.
However, the connection to traditional analytic models
has some illuminating subtleties.
Binney \& Merrifield (\citeyear{binney1998}, \S 5.3) review
the ``closed box'' \citep{talbot1971}, ``leaky box'' \citep{hartwick1976},
and ``accreting box'' \citep{larson1972}
models (see also the classic and still
illuminating chemical evolution review of \citealt{tinsley1980}).
The leaky box allows outflow but no inflow, and the closed box
is the limiting case with $\eta=0$.
For constant $\taustar$, which is assumed in our models,
the star formation rate from an initial gas mass $M_i$ is
\begin{equation}
\label{eqn:leakysfr}
\mdotstart = {M_g(t) \over \taustar} = 
  {1\over\taustar}\left[M_i - (1+\eta)\int_0^t dt'\,\mdotstar(t')\right]~,
\end{equation}
where we have set the recycling fraction $r=0$.
The solution to this equation is
\begin{equation}
\label{eqn:leakyexp}
\mdotstart = {M_i\over \taustar} e^{-t/\taudep}~,
\end{equation}
an exponentially declining star formation history with
$\tausfh = \taudep$.  This is precisely the case that is not
allowed in our analytic expressions because $\taubarDepSfh$ 
diverges, but we can consider the limit in which $\tausfh$
approaches $\taudep$ from above, so that $t/\taubarDepSfh \rightarrow 0$.
Taylor-expanding equation~(\ref{eqn:YoEvolExp}) with
expression~(\ref{eqn:Yoeq2}) for $\Yoeq$ yields
\begin{equation}
\label{eqn:YoEvolExpApprox}
\begin{split}
\Yo(t) &= \Yoeq \left[1-e^{-t/\taubarDepSfh}\right] \\
  & \approx \mocc {\taubarDepSfh\over\taustar}\times {t\over \taubarDepSfh}
  = \mocc {t \over\taustar}~.
\end{split}
\end{equation}
Defining $f_g(t) = M_g(t)/M_i = e^{-t/\taudep}$ and substituting
$\taustar = (1+\eta)\taudep$ leads to the conventional expression
for the leaky box model,
\begin{equation}
\label{eqn:leaky}
\Yo(t) = - {\mocc \over 1+\eta} \ln f_g(t)~,
\end{equation}
where the quantity $\mocc/(1+\eta)$ is usually
referred to as the ``effective yield.''

Next consider our expression~(\ref{eqn:mdf2}) for the oxygen MDF.
In the leaky box limit $\tausfh \rightarrow \taudep$ and 
$\taubarDepSfh \rightarrow \infty$,
\begin{equation}
\label{eqn:ylimit}
z_O = {\Yo \over \Yoeq} = {\Yo \over \mocc}{\taustar \over \taubarDepSfh}
 = {\Yo (1+\eta) \over \mocc} {\taudep \over \taubarDepSfh} \ll 1~,
\end{equation}
i.e., $\Yo$ is always $\ll \Yoeq$ because the latter diverges.
Defining $q = \taubarDepSfh/\tausfh \approx \taubarDepSfh/\taudep$
allows us to approximate
\begin{equation}
\label{eqn:mdfapprox}
{dN \over d\Yo} \propto (1-y)^{-1+\taubarDepSfh/\tausfh} \approx
  \left(1-{\Yo(1+\eta)/\mocc \over q}\right)^q~.
\end{equation}
Applying the $n\rightarrow \infty$ limit
$(1-x/n)^n \approx e^{-x}$ yields the final result
\begin{equation}
\label{eqn:leakymdf}
{dN \over d\Yo} \propto \exp\left[-{\Yo \over \mocc/(1+\eta)}\right]~,
\end{equation}
an exponentially declining MDF with effective yield $\mocc/(1+\eta)$.
This again matches the conventional leaky box result.
Although expression~(\ref{eqn:YoEvolExp}) for $\Yo(t)$ applies
for the constant $\taustar$ assumed in our modeling,
the expressions~(\ref{eqn:leaky}) and~(\ref{eqn:leakymdf})
apply to a no-inflow box regardless of the star formation history.

For constant $\taustar$, the 
``accreting box'' model with constant gas mass is identical
to our constant SFR model with $\eta=r=0$.
At time $t$, the total mass is
\begin{equation}
\label{eqn:accbox_mass}
M_t(t) = M_g + \int_0^\infty \mdotstar(t')dt' = M_g(1+t/\taustar)~.	
\end{equation}
Equation~(5.5.8) of \cite{binney1998} is, in our notation,
\begin{equation}
\label{eqn:accbox_y}
\Yo = \mocc\left[1-\exp\left(1-{M_t\over M_g}\right)\right]~,
\end{equation}
which in combination with equation~(\ref{eqn:accbox_mass}) 
yields $\Yo=\mocc(1-e^{-t/\taustar})$,
equivalent to our equation~(\ref{eqn:YoEvol}) for $\eta=r=0$.
The closed box and leaky box scenarios are fuel-starved,
and they have falling $dN/dZ$ because they form a small fraction
of their stars at late times.
The accreting box scenario, by contrast, has a rising $dN/dZ$.

While the notation and focus is different, our oxygen results overlap
those of classic analytic models based on the instantaneous recycling
approximation, such as the work of \cite{lynden-bell1975},
\cite{pagel1975}, \cite{tinsley1975}, and \cite{tinsley1978}.
Many of these papers focus on the metallicity distribution function
and the relation between metallicity and gas fraction, while here
we focus on time evolution.  More similar in formulation is the work of,
e.g., \cite{recchi2008} and \cite{spitoni2015}, who build on analytic
models from \cite{matteucci2001} to examine a variety of enriched
infall scenarios, of which only the simplest case is considered
here (see \S\ref{sec:enriched}).  In these papers, our assumption
of constant SFE timescale $\taustar$ is described as a
``linear Schmidt law.''
Qian \& Wasserburg's (\citeyear{qian2012})
formulation is also similar to ours and yields similar results
for oxygen evolution and the corresponding MDF.
Our analytic results for iron evolution with an exponential SNIa DTD
are, to our knowledge, new, and they underlie most of the key
findings of this paper.

\subsection{Illustrations}
\label{sec:illustrations}

We now illustrate the behavior of our analytic solutions 
for a variety of parameter choices.
Unless otherwise specified, all calculations
in this section adopt yield parameters
$\mocc=0.015$, $\mfecc=0.0012$, $\mfeIa=0.0017$ and recycling
fraction $r=0.4$.  These population-averaged yields correspond
to those computed by AWSJ for a \cite{kroupa2001} IMF with 
mass limits $0.1-100M_\odot$ and 
the SN yields of \cite{chieffi2004} and 
\citeauthor{iwamoto1999} (\citeyear{iwamoto1999}, the W70 model).
We discuss the choice of recycling fraction further 
in \S\ref{sec:tests}.
As a fiducial case for comparison to others we adopt
outflow parameter $\eta=2.5$, SFE timescale $\taustar=1\Gyr$,
an exponentially declining SFH with $\tausfh=6\Gyr$,
and SNIa parameters $\tauIa=1.5\Gyr$ and $\td=0.15\Gyr$.
These are the same parameters as in the fiducial model of AWSJ.
We use the solar photospheric abundance scale of \cite{lodders2003}, 
corresponding to solar mass fractions of 0.0056 for oxygen and 0.0012 
for iron (8.69 and 7.47 on the conventional $12+\log(X/H)$
number density scale).
Note that there are significant uncertainties in supernova yields,
photospheric solar abundances, and corrections for diffusion to
connect photospheric abundances to proto-solar abundances,
which can easily affect predicted values of 
$\ofe$ and $\feh$ at the 0.1-dex level or larger,
though potential corrections typically take the form
of constant offsets in these logarithmic quantities.

%%%%%%
\begin{figure}
\centerline{\includegraphics[width=2.5truein]{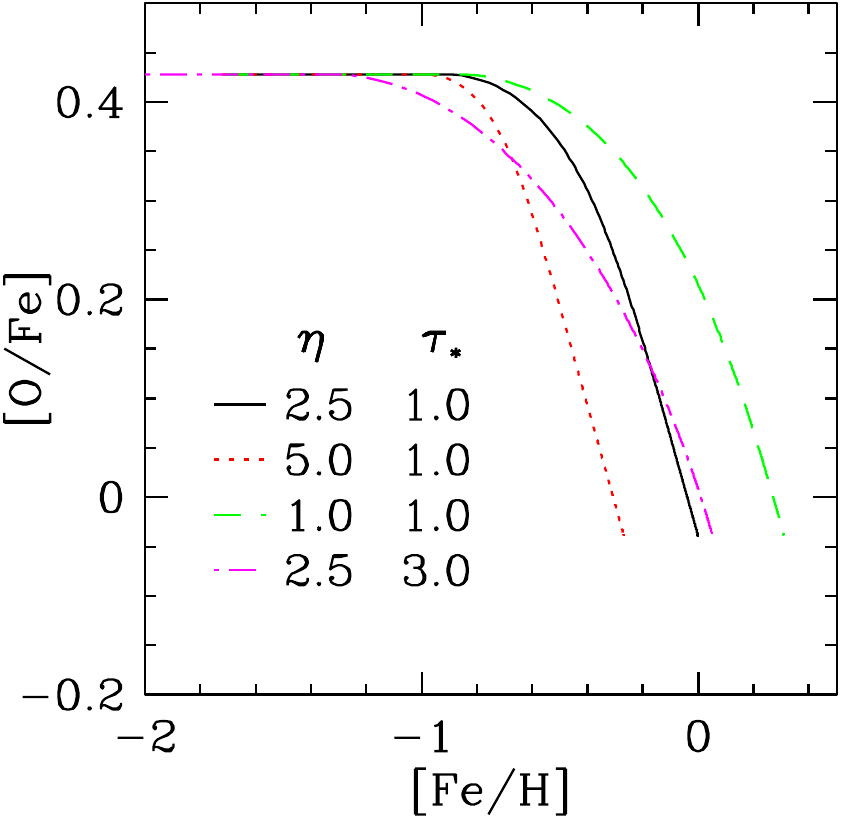}}
\caption{Tracks in the $\ofe-\feh$ plane for our fiducial model,
which has $\eta=2.5$, $\taustar=1\Gyr$, $\tausfh=6\Gyr$,
$\tauIa=1.5\Gyr$, and $\td=0.15\Gyr$ (black curve), and for
alternative models that have a higher outflow rate
($\eta=5$, red dotted curve, lower equilibrium abundance), a lower
outflow rate ($\eta=1.0$, green dashed curve, higher 
equilibrium abundance),
or a longer SFE timescale ($\taustar=3\Gyr$, magenta dot-dashed curve, 
with a ``knee'' at lower $\feh$).
}
\label{fig:ofe1}
\end{figure}
%%%%%%

%%%%%%
\begin{figure*}
\centerline{\includegraphics[width=6.5truein]{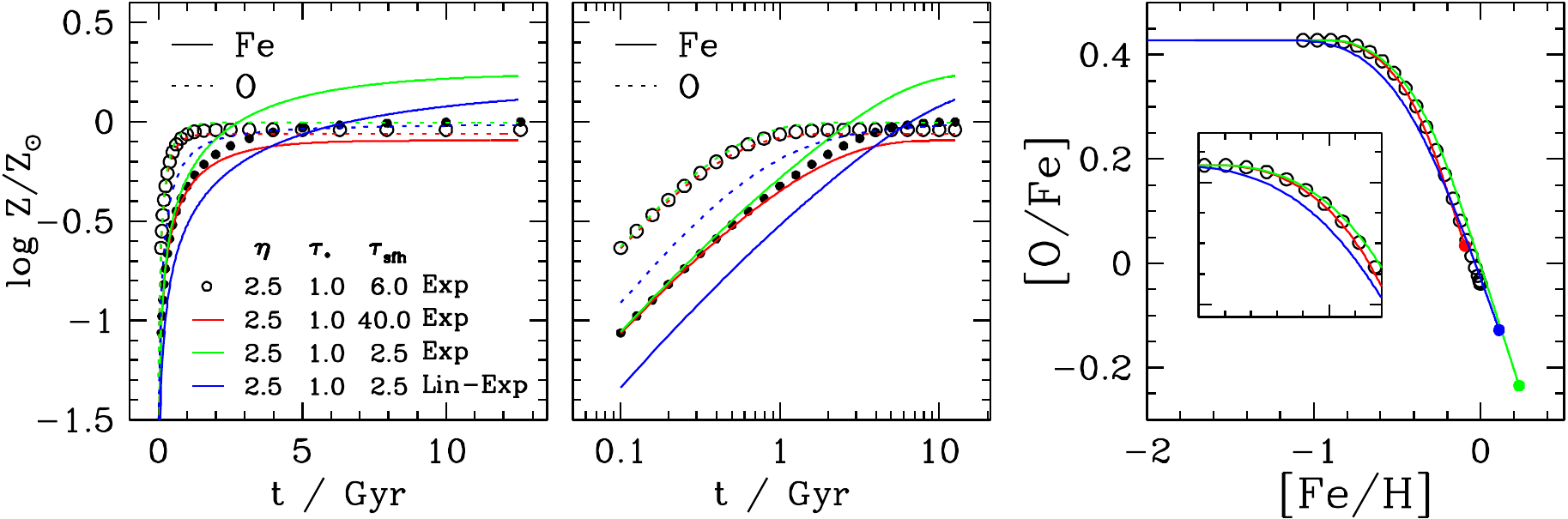}}
\caption{Abundance evolution (left two panels, with linear time axis on
the left and logarithmic time axis in the middle) and $\ofe-\feh$ tracks
(right panel) for the fiducial model and three variants.
In the left and middle panels, dotted and solid curves show oxygen and iron
abundances, respectively; the fiducial model is represented by open 
and filled circles for visual clarity.  
In the right panel, the fiducial model is represented by
open circles, and filled circles on the other three tracks mark
the abundances reached at $t=12.5\Gyr$.  
The alternative models have $\tausfh=40\Gyr$, approximating
a constant SFR (red curves), a rapidly declining SFH with
$\tausfh=2.5\Gyr$ (green curves), and a model with $\tausfh=2.5\Gyr$
but a linear-exponential SFH (blue curves).
The inset in the right panel presents a magnified view of
the knee in the $\ofe-\feh$ tracks.
All models have $\tauIa=1.5\Gyr$ and $\td=0.15\Gyr$.
}
\label{fig:history1}
\end{figure*}
%\end{figure*}
%%%%%%

%%%%%%
\begin{figure*}
\centerline{\includegraphics[width=6.5truein]{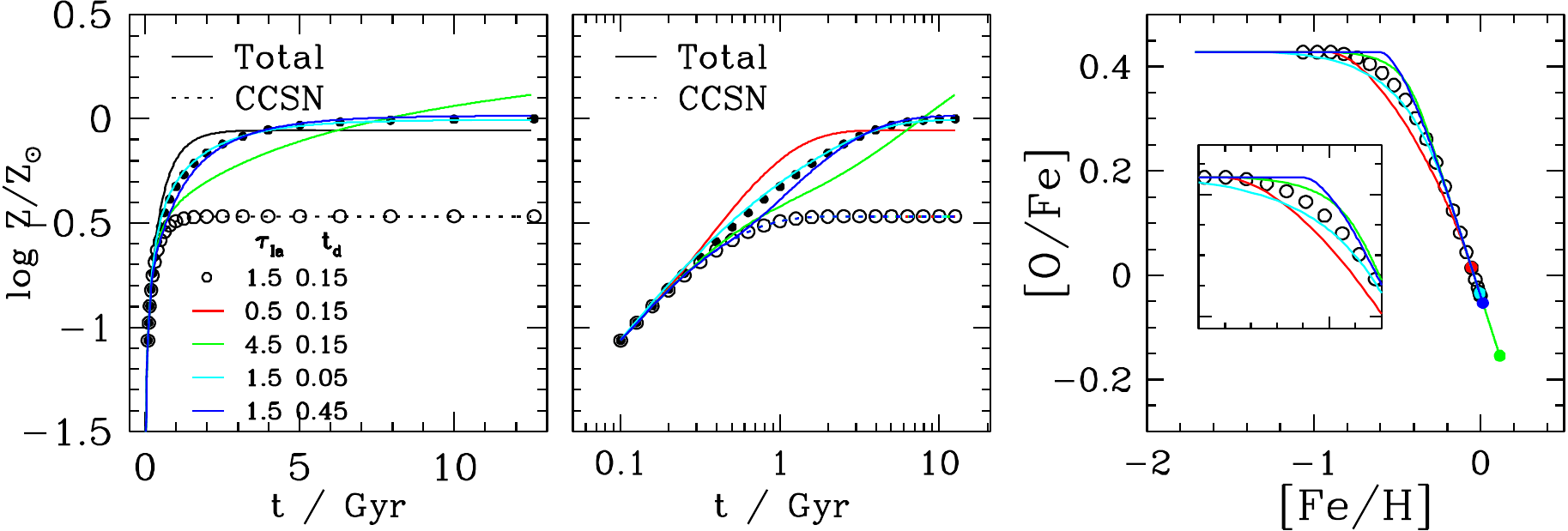}}
\caption{Similar to Fig.~\ref{fig:history1}, but comparing 
the fiducial model (circles) to models with different
SNIa delay timescales ($\tauIa=0.5\Gyr$, red curves; 
$\tauIa=4.5\Gyr,$ green curves) or different SNIa minimum
delay times ($\td=0.05\Gyr$, cyan curves; $\td=0.45\Gyr$, blue curves).
In the left and middle panels, the open circles and dotted curve show
the iron abundance from CCSNe only, which is the same in all
six models, while filled circles and solid curves show the
total iron abundance including SNIa.
}
\label{fig:history2}
\end{figure*}
%\end{figure*}
%%%%%%

Figure~\ref{fig:ofe1} reproduces one of the key results from
AWSJ (cf.\ their figure 3): the end-point of $\ofe-\feh$ tracks
is determined principally by the value of $\eta$, while the
location of the knee in $\ofe$ vs. $\feh$ is determined
principally by the SFE timescale.  With $\eta=2.5$ we obtain
approximately solar equilibrium abundances for an exponential
SFH with $\tausfh=6\Gyr$, in agreement with AWSJ.
Raising (lowering) $\eta$ lowers (raises)
the equilibrium value of $\feh$ without changing the initial
or final values of $\ofe$.  Our adopted CCSN yields produce
$\ofe = +0.43$ on the low-metallicity plateau, and these models
end with $\ofe \approx -0.05$.
Tripling $\taustar$ (i.e., lowering the SFE by a factor of three)
shifts the location of the knee from $\feh\approx -0.8$ to
$\feh\approx -1.2$ because CCSN have produced less enrichment
by the time SNIa enrichment starts to drive down $\ofe$.
Figure~\ref{fig:ofe1} shows good qualitative and quantitative
agreement with the full numerical results from AWSJ.
In AWSJ the equilibrium $\ofe$ depends slightly (at the 0.05 dex
level) on $\eta$, in contrast to the results here.
In tests with the numerical code we find that this
difference arises from the metallicity dependence of CCSN oxygen
yields (ignored in our calculation here), which are reduced at the 
lower equilibrium metallicity of the $\eta=5$ model.

In Figure~\ref{fig:history1} we investigate the dependence of chemical
enrichment histories on the star formation history.
The left and middle panels compare the oxygen and iron evolution
for our fiducial model parameters to several alternative models,
with time plotted linearly on the left and logarithmically in the middle.
The logarithmic time axis provides a better view of early-time
evolution, but it can give a misleading visual impression of
slow approach to equilibrium by compressing the late-time evolution.
The right panel shows the corresponding $\ofe-\feh$ tracks.
(See Fig.~\ref{fig:timescales}) for the star formation histories themselves.)
Starting with the fiducial model, we see that oxygen rises to
its equilibrium abundance quickly, on a timescale 
$\taudep=\taustar/(1+\eta-r) = 0.323\Gyr$, while the iron
evolution is controlled by the longer SNIa timescale ($\tauIa=1.5\Gyr$).
Changing to a nearly constant SFR (red curves, with $\tausfh=40\Gyr$)
does not change the early-time
behavior, but it slightly lowers the equilibrium oxygen abundance
as expected from equation~(\ref{eqn:Yoeq2}).
The reduction in the equilibrium iron abundance is larger than for oxygen
because at late times a constant SFR yields $\bmdotstar/\mdotstar = 1$ 
while the $\tausfh=6\Gyr$ model has
$\bmdotstar/\mdotstar \approx \taubarIaSfh/\tauIa = 2$.
As a result, the long-$\tausfh$ model has a higher equilibrium
$\ofe$, as evident in the right panel.

Green curves show a model with a much more rapid cutoff in star formation,
$\tausfh=2.5\Gyr$.  Oxygen evolution is almost unchanged relative to
the fiducial model, except for a slight increase in the equilibrium
abundance.  However, moving $\tausfh$ closer to $\tauIa$ increases
the harmonic difference timescale $\taubarIaSfh$ from $2\Gyr$ (fiducial model)
to $3.75\Gyr$.  The equilibrium iron abundance is therefore
substantially higher (eq.~\ref{eqn:YfeIaeq2}), but the approach to
this equilibrium is much slower.  
Physically, the high abundance arises because the SNIa iron ejecta
are deposited into a rapidly declining gas supply.
At $t=12.5\Gyr$, this model has
$\feh=0.25$ and $\ofe=-0.25$.  Blue curves show the $t e^{-t/\tausfh}$
model, with the same $2.5\Gyr$ timescale.
The oxygen abundance is a factor of two lower at early times
when the SFR is linearly growing rather than constant,
but it approaches the same equilibrium abundance at late times.
The iron abundance similarly starts a factor of two below
that of the exponential model, and the approach to equilibrium
is now very slow, approximately $\propto (1-\taubarIaSfh/t)$ 
at late times (see eq.~\ref{eqn:YfeIaEvolLinExpApprox}).
In contrast to the other models plotted, this model remains
significantly below the equilibrium abundance at $t=12.5\Gyr$,
though it does asymptotically approach the abundance of
the exponential SFH model if extended further in time.
Except for the endpoints, the $\ofe-\feh$ tracks of all of these 
models are similar, though the knee for the linear-exponential
model is shifted to lower $\feh$ because of the slower
early enrichment.  The model differences here parallel those
found in the numerical calculations of AWSJ (their fig.~3).

Figure~\ref{fig:history2} focuses on the role of the 
SNIa timescales $\tauIa$ and $\td$, with other parameters
fixed to those of the fiducial model.
In the left and middle panels, open circles show the CCSN iron
contribution, which is unaffected by the SNIa parameters.
In the fiducial model, the total iron (filled circles)
begins to depart significantly from the CCSN iron at
$t \approx 1\Gyr$, eventually settling to an equilibrium
that is 0.5-dex higher.  Decreasing or increasing $\tauIa$
by a factor of three (red and green curves, respectively)
shifts the $\ofe-\feh$ knee towards lower or higher $\feh$
as expected.  For $\tauIa = 4.5\Gyr$, $\taubarIaSfh$ becomes
large (18 Gyr, vs. 2 Gyr for the fiducial model), producing
a slow approach to a high equilibrium iron abundance
as in the short-$\tausfh$ model of Figure~\ref{fig:history1}.
For $\tauIa=0.5\Gyr$ the approach to equilbrium is rapid;
even though $\tauIa \approx \taudep$ in this case, making
$\taubarDepIa$ long, this does not lead to slow evolution
(see eq.~\ref{eqn:YfeIaEvolConverge} and associated discussion).
Changing $\td$ by a factor of three with fixed $\tauIa$ does
change the onset of the knee in $\ofe-\feh$, but the curves
have nearly converged by the time they have dropped $\sim 0.1$-dex
below the plateau, and changes to $\tauIa$ have a larger overall
impact on the tracks.  
We note, however, that the minimum delay time has a larger
impact for a $t^{-1.1}$ DTD because the SNIa rate in the
power-law model diverges at early times (see \S\ref{sec:two-exp}).

\begin{figure*}
\centerline{\includegraphics[width=6.5truein]{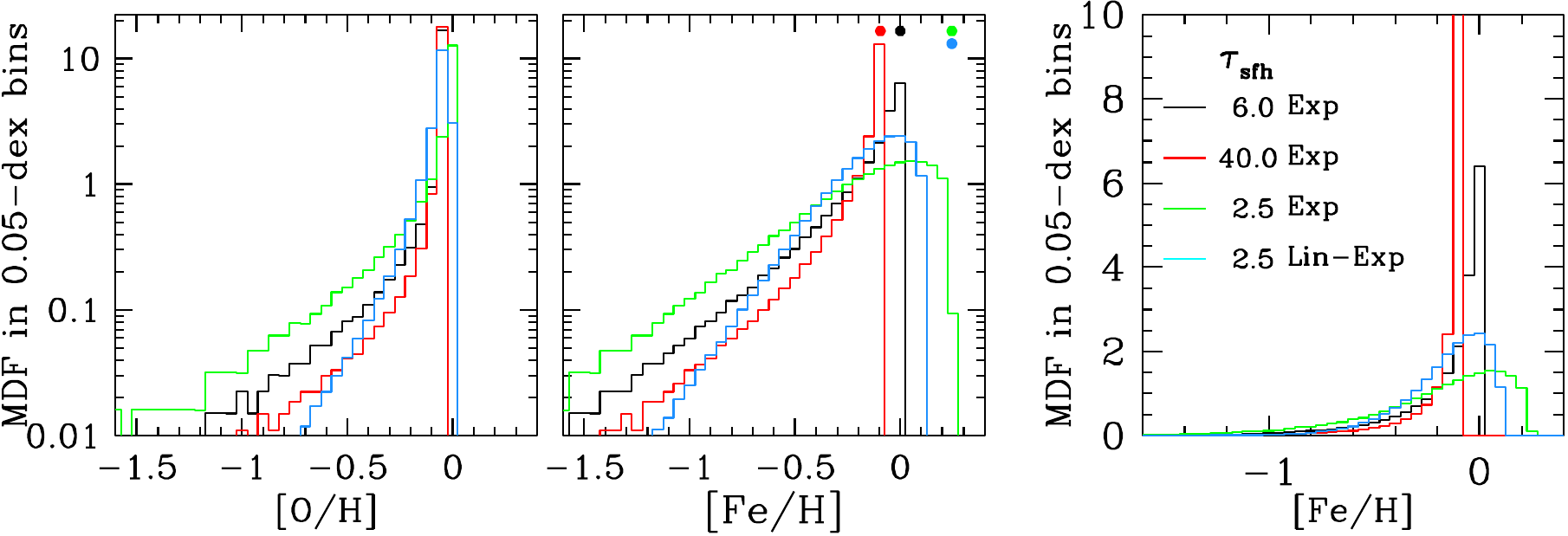}}
\caption{Metallicity distribution functions (MDFs), computed as
the fraction of stars formed in 0.05-dex bins of $\ohh$ or $\feh$
by $t=12.5\Gyr$ and normalized to unit integral, for the four
models shown in Fig.~\ref{fig:history1}.  
The left and middle panels
show the MDFs for oxygen and iron, respectively, while the right
panel repeats the iron MDF with a linear vertical axis scale.
Note that histograms binned in $\ohh$ or $\feh$ approximate
$Z\,dN/dZ$ rather than $dN/dZ$.
In the middle plot, colored dots mark the equilibrium $\feh$
of the four models.
All models have $\eta=2.5$, $\taustar=1\Gyr$, $\tauIa=1.5\Gyr$,
$\td=0.15\Gyr$.
}
\label{fig:mdf1}
\end{figure*}

Figure~\ref{fig:mdf1} plots metallicity distribution functions
for the four models shown in Figure~\ref{fig:history1}.  The upper left
panel shows the distribution of $\ohh$, which for exponential
SFH models follows equation~(\ref{eqn:mdf2}), while the other panels
show distributions of $\feh$ with a logarithmic or linear $y$-axis.
Here the MDF is the fraction of stars born in bins of 0.05-dex
in metallicity, proportional to $dN/d\log Z \propto Z\,dN/dZ$.
%The fiducial model is shown as a black histogram; other curves
%are computed as histograms but plotted as continuous for visual
%clarity.  
The MDF of the fiducial model rises linearly at early times
and is sharply peaked at the equilibrium abundance, for both
oxygen and iron.  
(For $t \ll \taudep$, the SFR is nearly constant, making
$Z \propto t$ and $Z\,dN/dZ \propto Z$.)
The peak for iron is less sharp because the timescale for SNIa
enrichment is longer, but $\tauIa$
is still short compared to $\tausfh=6\Gyr$.
For the $\tausfh=40\Gyr$ model, the equilibrium abundances are lower
and the peaks are even sharper.

%%%%%%
%\end{figure*}
%%%%%%

%%%%%%
\begin{figure*}
\centerline{\includegraphics[width=6.5truein]{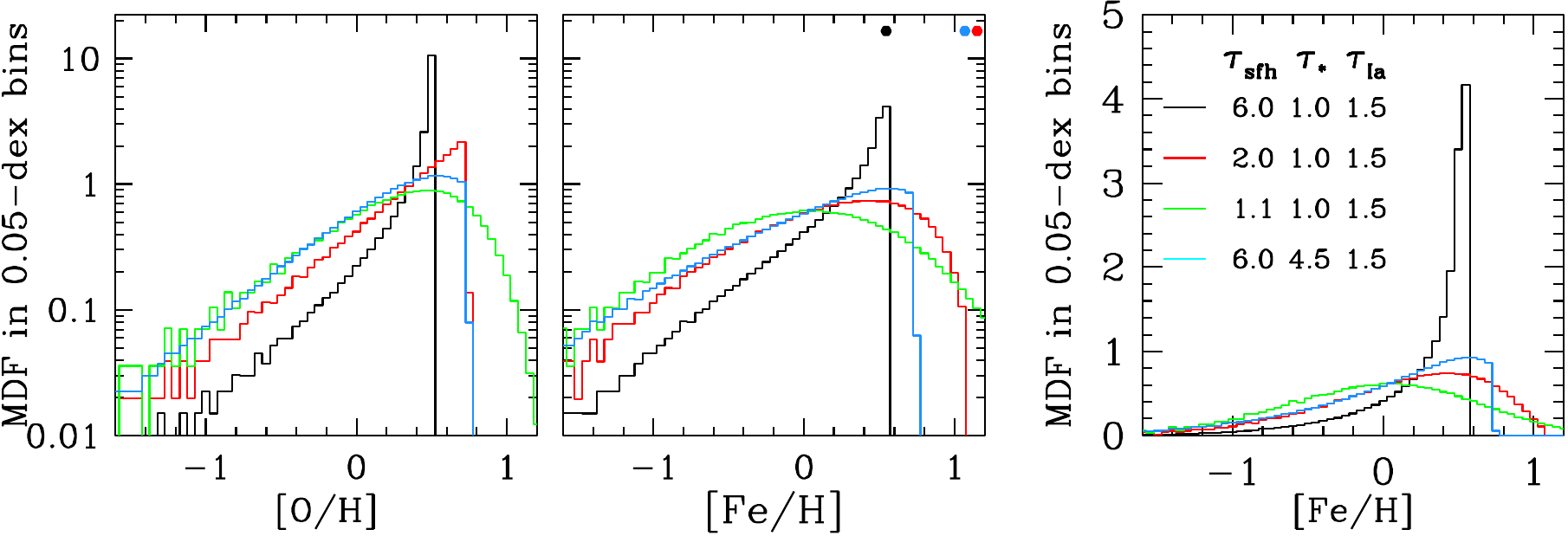}}
\caption{MDFs as in Fig.~\ref{fig:mdf1} for an alternative
set of models, all with outflow rate $\eta$ and recycling
fraction $r$ set to zero.  Black histograms show a model with
the same $\taustar=1\Gyr$ and $\tausfh=6\Gyr$ as our fiducial model, 
while other curves show models with shorter SFH timescales
($\tausfh=2\Gyr$, red histograms; $\tausfh=1.1\Gyr$, green histograms)
or a longer SFE timescale ($\taustar=4.5\Gyr$, blue histograms).
All models have $\tauIa=1.5\Gyr$ and $\td=0.15\Gyr$.
}
\label{fig:mdf2}
\end{figure*}
%%%%%%

Shortening $\tausfh$ to 2.5 Gyr makes only moderate difference 
to the oxygen MDF, as this timescale is still much larger than
$\taudep = \taustar/(1+\eta-r) = 0.323\Gyr$.  However, the peak
of the iron MDF changes substantially in this case, in part
because of the higher equilibrium abundance, but also because
SNIa can still provide significant iron enrichment after the
SFR has declined, producing a much softer cutoff
of the MDF.  The linear-exponential model with the same $\tausfh$
has the same equilibrium iron abundance as the exponential model, but,
as shown previously in Figure~\ref{fig:history1}, it does not
reach this abundance by $t=12.5\Gyr$.  Its MDF therefore turns over
significantly before that of the exponential model.
Both the oxygen and iron MDFs rise quadratically at early
times because $Z\propto t$, $\mdotstar \propto t$, and
thus $Z\,dN/dZ \propto t^2 \propto Z^2$.

Figure~\ref{fig:mdf2} shows MDFs for cases that illustrate a variety
of behaviors, where we have set $\eta=r=0$ for simplicity of interpretation.
With $\taustar=1\Gyr$ and $\tausfh=6\Gyr$, the oxygen and iron
MDFs are similar to those of our fiducial model except for the $\sim 0.4$-dex
increase in equilibrium abundance that results from eliminating outflows.
Setting $\tausfh=2\Gyr = 2\taustar$ corresponds to the critical
case of constant $dN/dz_O$ in equation~(\ref{eqn:mdf2}), producing
an oxygen MDF that rises linearly until a sharp cutoff at
$\ohh = 0.5$, the enrichment level achieved by $t=12.5\Gyr$.
The iron MDF of this model exhibits a smooth turnover instead of a sharp 
cutoff because of the longer timescale of SNIa,
which allows them to produce significant iron even after
the star formation rate has fallen substantially.
A model with $\tausfh=3\Gyr = 2\tauIa$ (not shown) produces an
iron MDF that is almost linearly rising to a sharp cutoff,
similar to the oxygen MDF for $\tausfh=2\taustar$.

Reducing $\tausfh$ to $1.1\Gyr = 1.1\taustar$ corresponds
nearly to a conventional closed box model, which has $\tausfh=\taustar$
as discussed in \S\ref{sec:traditional}.  The form of the oxygen MDF
is very close to the closed box form $dN/d\ln\Yo \propto \Yo\exp(-\Yo/\mocc)$,
where the turnover scale corresponds to $\ohh = 0.43$ for 
our adopted values of $\mocc$ and the solar oxygen abundance.
In contrast to the other cases we have examined, the iron MDF
in this model is nearly symmetric in $\feh$, and extremely broad.
The substantial change of behavior relative to the previous
model arises because $\tausfh$ has crossed the critical
threshold $\tausfh = \tauIa$.  For this model (and any model
with $\tausfh < \tauIa$) the equilibrium 
abundance (eq.~\ref{eqn:YfeIaeq2}) is negative, and it no
longer represents a late time asymptotic value.
Instead, the iron abundance
grows exponentially at late times on the timescale $|\taubarIaSfh|$
because the exponentially declining SNIa enrichment is deposited 
into a gas supply that is declining exponentially on a still
shorter timescale.

The final model in Figure~\ref{fig:mdf2} has 
$\tausfh=6\Gyr$ but a low star formation efficiency,
with $\taustar = 4.5\Gyr$.  
With $\tausfh/\taustar=1.33$, the oxygen MDF of this model
is close to that of the closed box model except that it
cuts off sharply at $\ohh=0.5$, the abundance reached at $t=12.5\Gyr$.
Because the depletion timescale is now much longer than $\tauIa$,
the iron MDF is similar to the oxygen MDF, as SNIa are 
approximately ``instantaneous'' compared to the chemical
evolution timescale.

The cases in Figures~\ref{fig:mdf1} and~\ref{fig:mdf2} illustrate
several general points about the MDFs of one-zone models.
When the depletion time is short and the SFH timescale
is long, the MDF is usually sharply peaked near the equilibrium
abundance, with a strong negative skewness, as in our fiducial model.  
Many plausible chemical evolution models fall in this regime of parameter space.
A rapidly declining star formation history (short $\tausfh$) can
produce a gentler cutoff and less asymmetry of the MDF.
The transition between regimes arises at $\tausfh \approx 2\taudep$
for oxygen and $\tausfh \approx 2\tauIa$ for iron.
A linear-exponential SFH model approaches equilibrium more slowly
than an exponential SFH model, so its abundances can remain 
significantly below equilibrium even at $t \approx 12.5\Gyr$.
Inefficient star formation, with long $\taudep$, can
change the shape of MDFs by approaching the regime of 
$\tausfh \approx \taudep$ and by truncating evolution before
equilibrium is reached.

In disk galaxies like the Milky Way, radial migration of stars
and flows of enriched gas may also play critical roles in shaping
MDFs (e.g., \citealt{schoenrich2009,bilitewski2012,pezzulli2016}).
A solid understanding of the MDFs of one-zone models is
essential background for interpreting constraints on these 
more complex processes.

\subsection{Numerical Tests}
\label{sec:tests}

%%%%%%
\begin{figure}
\centerline{\includegraphics[width=3.3truein]{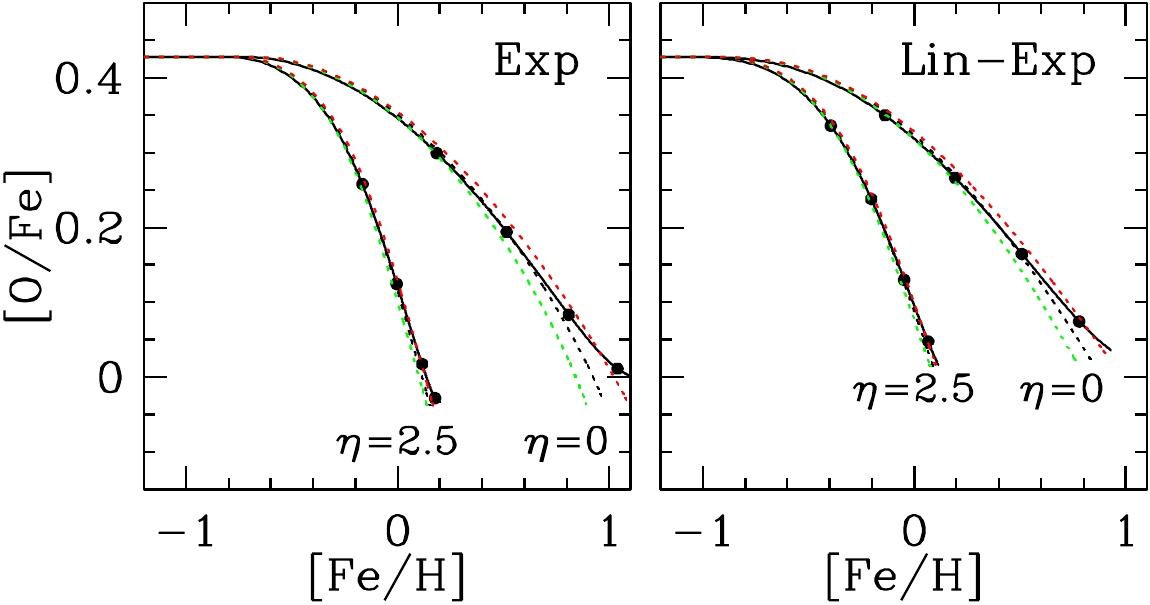}}
\caption{Tracks in $\ofe-\feh$ for a numerical calculation with
continuous AGB recycling (solid curves) and the analytic results
with recycling parameters of $r=0.3$, 0.4, and 0.5 (green, black, and
red dotted curves), assuming an exponential (left panel) or
linear-exponential (right panel) star formation history.
In each panel the left set of curves has the parameters of the
fiducial model and the right set of curves has the same parameters
but no outflow.  Points on the solid curves mark $t=1$, 2, 4, and 8 Gyr;
curves end at $t=12.5\Gyr$.
}
\label{fig:recycling}
\end{figure}

To check our analytic solutions and test the impact of our limiting
assumptions, we have written a code that numerically integrates the
equations for the evolution of oxygen and iron mass for an arbitrary
star formation history and mass outflow history.  Using this code,
we have numerically confirmed our analytic solutions for the 
time-evolution of the oxygen and iron abundances with a constant,
exponential, or linear-exponential SFH.
We have also confirmed analytic results that appear later in 
the paper for evolution with sudden parameter changes (\S\ref{sec:changes}),
with a time-dependent $\taustar$ (\S\ref{sec:timedeptaustar}),
or with a star formation history that is the sum of exponentials
or linear-exponentials (\S\ref{sec:sfhsum}).
With this numerical code we can also relax two of the assumptions
needed to allow our analytic solutions: instantaneous return of
metals incorporated into stars, and an exponential SNIa DTD.

Figure~\ref{fig:recycling} compares $\ofe-\feh$ tracks from our
analytic calculations assuming instantaneous recycling of birth metals
to numerical results that incorporate continuous recycling.
Specifically, we adopt our usual Kroupa IMF and
assume that a star of mass $M$ returns a mass
$M-M_{\rm rem}$ of gas to the ISM at its birth metallicity after a time
$t=10\Gyr(M/M_\odot)^{-3.5}$, where the remnant mass is $1.44M_\odot$
for stars with $M>8M_\odot$ and $0.394M_\odot+0.109M$ for $M<8M_\odot$
based on \cite{kalirai2008}.
While this simple lifetime formula becomes inaccurate at high masses,
recycling is fast there in any case, so for our purposes we care mainly
about the behavior in the range $M \sim 0.8-2M_\odot$.
Stars above $8M_\odot$ also return newly synthesized oxygen and 
iron with our standard net yields, as in our analytic calculations.

In each panel of Figure~\ref{fig:recycling}
(exponential SFH on the left and linear-exponential
on the right, with $\tausfh=6\Gyr$),
the left set of curves shows results for our fiducial
outflow rate $\eta=2.5$, which yields near-solar equilibrium abundances.
The parameter combination that appears in our analytic solutions
is $1+\eta-r$, so with $\eta=2.5$ setting $r=0.3$, 0.4, or 0.5
makes little difference (three dotted curves).
Since recycling is a small effect in any case, it is unsurprising
that agreement with the continuous recycling
result (solid curve) is nearly perfect.  
The right set of curves sets $\eta=0$
to maximize the impact of recycling.  The instantaneous curve
with $r=0.4$ (equal to the Kroupa IMF recycling fraction after $2\Gyr$)
agrees well with the continuous recycling calculation up to $t=4\Gyr$,
but at later times the full calculation yields higher $\feh$ and
slightly higher $\ofe$ because of the greater return of metals from
stars formed at earlier times.  We conclude that our approximation
of instantaneous return of birth metals is generally accurate
for models that produce near-solar abundances but can fail at
the $0.1-0.2$ dex level for models with low outflow efficiency
and declining star formation histories.

We investigate the effects of a power-law SNIa DTD in \S\ref{sec:two-exp}
below, where we show that this form can be well approximated by a
sum of two exponentials, which remains soluble by our techniques.

\section{Sudden Events}
\label{sec:sudden}

\subsection{Bursts of Star Formation}
\label{sec:bursts}

For a constant SFR, reaching an equilibrium abundance ratio
$\ofe$ requires $\bmdotstar = \mdotstar$ so that oxygen and
iron are being added to the system at an equal rate.
For declining SFR, equilibrium requires $\bmdotstart/\mdotstart$
to reach a constant ratio.
A burst of star formation --- i.e., an increase of $\mdotstar$
well above its recent trend --- will boost $\mdotstar$ relative
to $\bmdotstar$ and therefore increase CCSN enrichment relative
to SNIa enrichment, driving the system to higher $\ofe$.
The ``system'' here could be an entire dwarf galaxy if the
ISM is well mixed throughout.  However, it could also be a single
molecular cloud, or the kpc-scale star-forming region 
around a spiral arm.  The defining element of the system is the
scale over which newly produced metals are retained 
and mixed into the gas supply.
\cite{gilmore1991} highlighted the potential influence of
bursty star formation histories on the $\afe$ ratios of dwarf galaxies,
though they emphasized the effect of depressed $\afe$ during
quiescent phases (see the black curve in Fig.~\ref{fig:change1} below)
rather than enhanced $\afe$ in the bursts themselves.

In this section we consider the impact of an instantaneous burst that
converts a fraction $f_*$ of the system's remaining gas into stars.
(There are some subtleties to the meaning of ``instantaneous,''
as discussed below.)
This burst of star formation could be induced by a rapid 
influx of gas, in which case the abundances of the pre-existing
gas would be diluted before the burst takes place.
For a simple, well defined model, starting from gas mass $M_g$ and
abundances $\Yo$ and $\Yfe$, we
(1) convert a fraction $f_*$ of the initial gas supply into stars,
(2) eject an amount of gas $\etapr f_* M_g$ in an 
    outflow at the pre-burst metallicity, and
(3) retain a fraction $\fmet$ of the CCSN ejecta, including their
      newly produced metals, their original metals, and their original
      hydrogen and helium.
We use primes to distinguish post-burst from pre-burst quantities;
thus $\etapr$ may differ from the outflow parameter $\eta$
that characterized the preceding evolution.

We require $(1+\etapr)f_* \leq 1$ so that we do not 
use up more than 100\% of the original gas supply.
Of the gas that does not participate in star formation,
the fraction ejected is 
$\etapr f_* M_g / [(1-f_*)M_g] = \etapr f_*/(1-f_*)$.
If the SN ejecta are well mixed into the ISM {\it before} 
the outflow occurs, then we expect $\fmet = 1 - \etapr f_*/(1-f_*)$.
However, in two extreme limits, all CCSN ejecta could escape
without entraining much gas, yielding $\fmet=0$ regardless of $\etapr$, 
or the metals could be captured in dense gas around the supernovae
while stellar winds or radiation pressure eject the lower density
gas of the system, yielding $\fmet=1$.
Note that even for $\fmet=1$, the pre-burst metals swept up 
in the outflow are still ejected.

For notational convenience, we define
\begin{equation}
\label{eqn:Funp}
\Funp = 1 - (1+\etapr)f_*~,
\end{equation}
the fraction of the pre-existing
gas supply that is ``unprocessed,''
neither formed into stars nor ejected in the associated outflow.
The relation between post-burst and pre-burst gas mass, oxygen mass,
and iron mass is:
\begin{eqnarray}
\Mgpr &=& \Funp M_g + \rcc M_g f_* \fmet
  \label{eqn:Mgpr} \\
\Mopr &=& \Funp \Mo + \rcc\Mo f_*\fmet + \mocc M_g f_*\fmet 
  \label{eqn:Mopr} \\
\Mfepr &=& \Funp\Mfe + \rcc\Mfe f_*\fmet + \mfecc M_g f_*\fmet .
   \label{eqn:Mfepr} 
\end{eqnarray}
In each equation, the first term represents the initial gas
that remains after star formation and outflow, and the
second term represents the return of gas from CCSN ejecta.  
The quantity $\rcc$ is analogous to our previously used recycling
parameter $r$ but includes only the contribution from massive stars
that produce CCSNe.  For a Kroupa IMF and all stars with $M \geq 8M_\odot$
exploding as CCSNe, $\rcc = 0.20$.
The final terms in equations~(\ref{eqn:Mopr}) and~(\ref{eqn:Mfepr})
represent the oxygen and iron newly synthesized by CCSNe.
The factor $\fmet$ appears in the second terms of
these equations because we assume that all CCSN ejecta are
retained with efficiency $\fmet$, not just the newly synthesized
metals within those ejecta.

In the limit that $\fmet=0$ the abundances are unchanged,
because the outflow carries gas at the pre-burst metallicity.
In the opposite limit that $\Funp \rightarrow 0$, we obtain
\begin{equation}
\label{eqn:FunpLimit}
\Yopr = {\Mopr \over \Mgpr} \approx {\Yo} 
        {\rcc + \mocc\Yo^{-1} \over \rcc} = \Yo + {\mocc \over \rcc}
	~,
\end{equation}
with an analogous result for iron.  
Here we have used the substitution $M_g = \Yo^{-1}\Mo = \Yfe^{-1}\Mfe$.
Note that $\mocc/\rcc$ is the mass
fraction of newly synthesized oxygen in CCSN ejecta.

The most interesting impact is on the abundance ratio of the ISM
following the burst:
\begin{equation}
\label{eqn:PostBurstRatio}
{\Yopr \over \Yfepr} = {\Yo\over\Yfe} 
  {\left[\Funp+\rcc f_*\fmet + \mocc\Yo^{-1} f_*\fmet\right] \over
   \left[\Funp+\rcc f_*\fmet + \mfecc\Yfe^{-1} f_*\fmet\right] }~.
\end{equation}
If the pre-existing metals (first two terms) dominate over the 
newly produced metals, then the abundance ratio is essentially unchanged.
However, if the final terms dominate then 
$\Yopr/\Yfepr \rightarrow \mocc/\mfecc$, returning the abundance
ratio to the CCSN plateau.

%%%%%%
\begin{figure}
\centerline{\includegraphics[width=3.5truein]{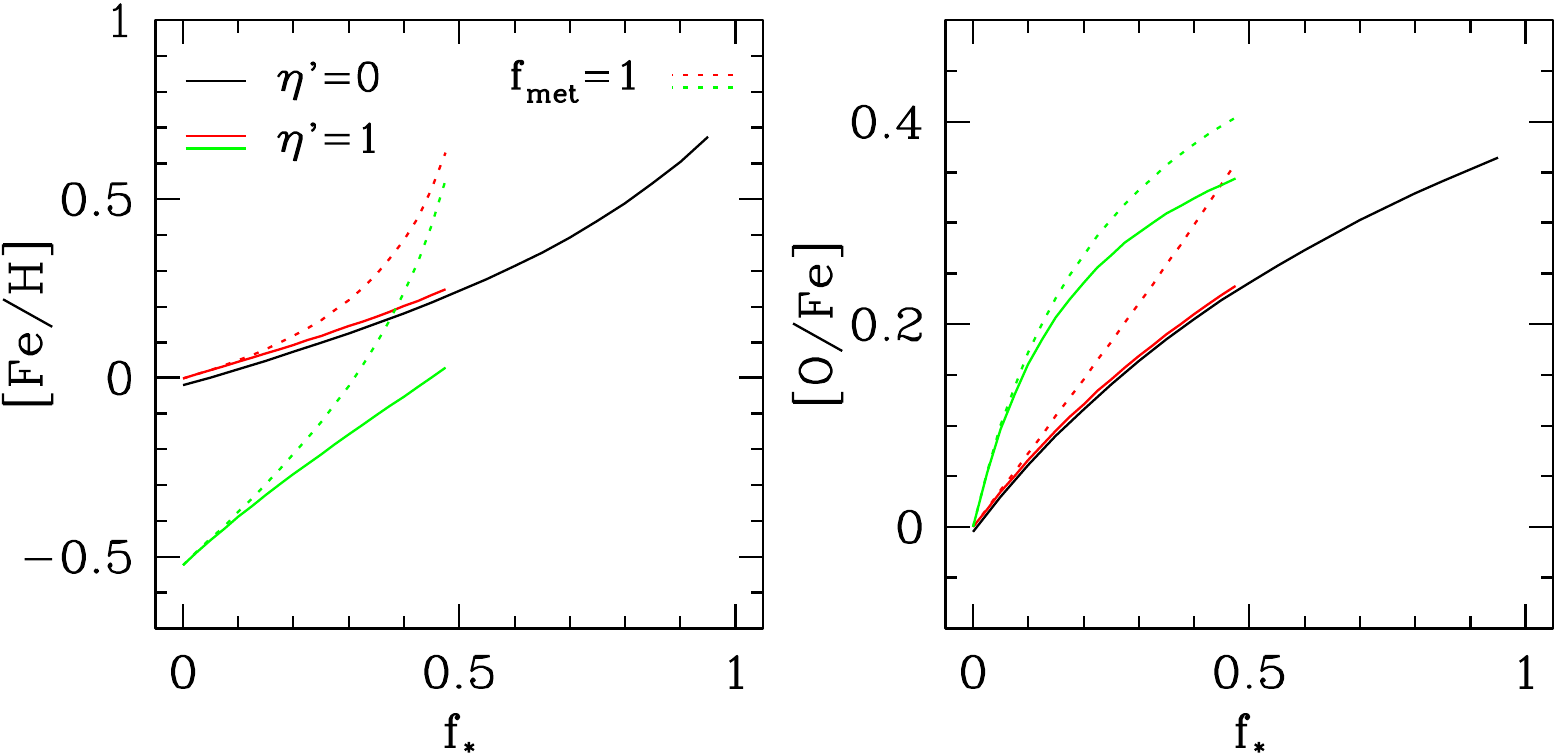}}
\caption{Post-burst iron abundance (left) or $\ofe$ (right) as a
function of the fraction $f_*$ of the initial gas mass that is
converted to stars during the burst.
Black curves show the case in which all gas and metals are retained.
Red and green curves show cases with an outflow of mass-loading
parameter $\etapr=1$, starting from
solar or $0.3\times$solar abundances, respectively.
In these cases the maximum $f_*$ is $1/(1+\etapr)=0.5$.
Solid curves assume that metals and recycled gas from
CCSN are ejected with the same efficiency as pre-existing gas,
while dotted curves show the case in which all CCSN ejecta are retained.
Black curves have been shifted slightly downward, as they would
otherwise be obscured by the solid red curve.
}
\label{fig:burst}
\end{figure}

Figure~\ref{fig:burst} shows abundance changes as a function of $f_*$.
Black curves show an example in which all gas and metals are 
retained by the system, $\etapr=0$ and $\fmet=1$.
Starting from solar abundances, converting 50\% of the gas
into stars boosts $\feh$ by 0.15 dex and $\ofe$ by 0.2 dex.
As the conversion efficiency approaches 100\%, the abundance
ratio approaches the CCSN plateau value $\mocc/\mfecc$,
corresponding to $\ofe=+0.43$ for our adopted yields and solar
abundance scale.  
Solid red curves show a case with outflow parameter
$\etapr=1$ and $\fmet=1-\etapr f_*/(1-f_*)$, i.e., SN metals
are ejected with the same efficiency as the overall outflow.
In this situation the change of abundances is determined by
$f_*$ independent of $\etapr$, so the red curve lies on top 
of the black curve, but it extends only to the maximum
allowed $f_*=1/(1+\etapr)=0.5$, where the $\ofe$ enhancement is 0.25 dex.
If all metals are retained ($\fmet=1$, dotted red curves) then the 
impact on abundances rises sharply as $f_*$ approaches 0.5.
Green curves show cases with $\etapr=1$ starting from abundances
of $0.3\times$ solar.  Here the burst has a much larger impact,
with a $0.35-0.45$ dex boost in $\ofe$ as $f_*$ approaches 0.5,
because the newly produced metals are more important compared
to the pre-existing metals.  

The takeaway message from Figure~\ref{fig:burst} is that
once a population has evolved to roughly solar $\ofe$, a 
burst of star formation can readily boost $\ofe$ by 
$\sim 0.1-0.3$ dex if it consumes a significant fraction 
of the available gas and the metals produced by the CCSNe
are retained.  
Gas with low $\feh$, perhaps because of dilution
by a recent accretion event that triggers the burst,
is more susceptible to such a boost.
The quantities in Figure~\ref{fig:burst}
represent the post-burst gas phase abundances after 
CCSN enrichment, so if the star formation is truly
instantaneous then all stars in the burst are born with
the pre-burst abundances.  However, if the star formation extends
over a $\sim 40\Myr$ time span comparable to the lives of
$8 M_\odot$ stars, and metals are efficiently mixed, then
stars will be born with the whole range of abundances
from the initial to the final values.  For our calculation
here to be reasonably accurate, the timescale of the ``burst''
need only be short compared to the $\sim 1\Gyr$ timescale
on which SNIa enrichment becomes important.

Individual molecular clouds are usually thought to form
stars quite inefficiently, with $f_*$ of a few percent \citep{murray2011}.
Furthermore, CCSN ejecta may frequently escape their parent
molecular clouds, making $\fmet$ low.
However, the occasional molecular cloud that forms stars with unusually
high efficiency and traps its supernova ejecta in dense surroundings
could produce some stars with significantly enhanced $\ofe$.
This ``cloud burst'' phenomenon is a possible explanation for the
rare population of $\alpha$-enhanced stars with intermediate
ages found in the SDSS-III APOGEE survey \citep{martig2014,chiappini2015}
and in local samples \citep{haywood2015}.
While low star formation efficiency and metal loss may keep these effects
small on the scale of individual molecular clouds, they could
have a larger impact on the $\sim\hbox{kpc}$ scale of a spiral arm passage,
where the CCSN products of many molecular clouds may be trapped
and mixed much more rapidly than SNIa enrichment occurs.

Boosting of CCSN abundances by starbursts could play a significant
role in the chemical evolution of dwarf galaxies, which frequently
show evidence of bursty star formation histories \citep{weisz2014}.
However, the time resolution of inferred star formation histories
is usually too coarse to determine whether the starbursts are
short compared to $\tauIa$.
The impact on the galaxy depends critically on what
happens to the eventual SNIa metals associated with a burst.
If these are retained by the galaxy's star-forming gas,
or return to it after fountaining into the halo, then
the CCSN metals produced in a burst may only ``catch up''
with the SNIa metals from a previous burst, and some inter-burst
stars could even be born with substantial deficits of $\alpha$ elements
(as emphasized by \citealt{gilmore1991}).
Investigation of these issues requires more detailed models
of dwarf galaxy evolution, but the distribution of $\afe$ ratios
may set interesting constraints on timescales of starbursts
and the physics of dwarf galaxy outflows.

The scatter in $\afe$ at fixed $\feh$ is hard to
constrain precisely because of the difficulty of subtracting
observational errors.  However, recent studies suggest that
this scatter is $< 0.05$ dex rms once one separates the
high-$\alpha$ (``thick disk'') and low-$\alpha$ (``thin disk'')
sequences
(\citealt{ramirez2013}; Bertran de Lis et al. 2016, submitted;
P.\ Kempski et al., in preparation).
Reversing the arguments above, one can use this
small observed scatter to set limits
on the stochasticity of star formation and metal mixing
in the Milky Way and other galaxies.

\subsection{Sudden Changes of Model Parameters}
\label{sec:changes}

The methods employed in \S\ref{sec:evolution} can be extended
to calculate evolution that incorporates a sudden change in
model parameters at a time $t_c$.  We consider cases in which
the star formation history is $\mdotstart = \mmdotstar e^{-t/\ttausfh}$
for $t \leq t_c$, with SFE timescale and outflow parameters
$\ttaustar$ and $\eta_1$, and
$\mdotstart = \mmmdotstar e^{-t_2/\tttausfh}$ for $t > t_c$,
with corresponding parameters $\tttaustar$ and $\eta_2$,
where $t_2 \equiv t -t_c$ represents time since the transition.
We assume that the yields and SNIa timescale $\tauIa$ stay
constant, though it will be obvious from the solutions below
how one would incorporate changes in these parameters as well.

It is useful to imagine that stars formed prior to $t_c$ produce
distinct isotopes from those formed after $t_c$.  With this
artificial assumption, it is obvious that one can separately
calculate the evolution of the elements produced by the two
separate phases of star formation, then add them together
to get the total elemental abundance.  Mathematically, this
split works because the differential equations for 
the evolution of oxygen and iron mass are linear.

Before $t_c$, the time evolution is simply given by the previous
results from \S\ref{sec:ev_expsfr} with the appropriate parameters.
At time $t_c$ the oxygen mass is
\begin{equation}
\label{eqn:Moc}
\Moc = \Yo(t_c)M_g(t_c)~,
\end{equation}
with $\Yo(t_c)$ evaluated from equation~(\ref{eqn:YoEvolExp})
and $M_g(t_c) = \ttaustar\mmdotstar e^{-t/\ttausfh}$.
After $t_c$, this abundance evolves according to 
equation~(\ref{eqn:ModotTimedep}) but with the source term
on the right hand side set to zero, with the solution
\begin{equation}
\label{eqn:Mo1}
M_{\rm O,1}(t_2) = \Moc e^{-t_2/\tttaudep}~.
\end{equation}
If we assume that $M_g(t)$ is continuous across $t_c$,
then the post-$t_c$ star formation history requires
$M_g(t_2) = M_g(t_c) e^{-t_2/\tttausfh}$.  In combination with
equation~(\ref{eqn:Moc}), this implies
\begin{equation}
\label{eqn:Yo1}
\begin{split}
Z_{\rm O,1}(t_2) &= \Yo(t_c) e^{-t_2/\tttaudep} / e^{-t_2/\tttausfh} \\
  &= \Yo(t_c) e^{-t_2/\tttaubarDepSfh}~.
\end{split}
\end{equation}
Comparing~(\ref{eqn:Mo1}) and~(\ref{eqn:Yo1}) we see that while
the pre-$t_c$ oxygen mass evolves on the timescale $\tttaudep$,
the corresponding abundance evolves on the longer timescale
$\tttaubarDepSfh$.

For stars forming after $t_c$, the oxygen evolution equation
is identical to the original exponential SFR case but with
time variable $t_2$ rather than $t$.  The solution is therefore
equation~(\ref{eqn:YoEvolExp}) with the substitution $t \rightarrow t_2$
and post-$t_c$ values of the parameters:
\begin{equation}
\label{eqn:YoEvolExp2}
Z_{\rm O,2}(t_2) = 
           Z_{\rm O,eq2} \left[1-e^{-t_2/\tttaubarDepSfh}\right]~.
\end{equation}
The full solution for $t > t_c$ is
\begin{equation}
\label{eqn:YoEvolChange}
\Yo(t_2) = Z_{\rm O,1}(t_2)+Z_{\rm O,2}(t_2).
\end{equation}
This result makes intuitive sense: the oxygen abundance evolves 
from its original value at $t_c$ to the post-$t_c$ equilibrium
abundance on the timescale $\tttaubarDepSfh$.

For iron we must consider three separate contributions.  The
first is iron produced before $t_c$, by either CCSNe or SNIa.
After $t_c$ this component evolves exactly like pre-$t_c$ oxygen, so
\begin{equation}
\label{eqn:Yfe1}
Z_{\rm Fe,1}(t_2) = \Yfe(t_c) e^{-t_2/\tttaubarDepSfh}~.
\end{equation}
The second is iron produced by stars that form after $t_c$.
This follows the usual evolution for an exponential star formation
history but with time variable $t_2$, thus following 
equations~(\ref{eqn:YfeccEvolExp}) and~(\ref{eqn:YfeIaEvolExp})
with $t \rightarrow t_2$, $\Delta t \rightarrow \Delta t_2 = t_2 - t_D$,
and post-$t_c$ values for all of the other parameters:
\begin{equation}
\label{eqn:YfeEvolExp2}
\begin{split}
Z_{\rm Fe,2}(t_2) = &
  \Yfecceqtt
  \left[1-e^{-t/\tttaubarDepSfh}\right] + \\
  & \YfeIaeqtt
  \Big[1-e^{-\Delta t_2/\tttaubarDepSfh} 
    - {\tttaubarDepIa \over \tttaubarDepSfh} \times \\
  & \left( e^{-\Delta t_2/\tttaubarIaSfh} - 
         e^{-\Delta t_2/\tttaubarDepSfh}\right) \Big] ~.
\end{split}
\end{equation}

The new case is SNIa iron from stars that form before $t_c$
but explode after $t_c$.
For this third contribution the governing equation is
\begin{equation}
\label{eqn:ironc3}
\Mfedot + {\Mfe \over \tttaudep} = \mfeIa\bmdotstart
\end{equation}
with 
\begin{equation}
\label{eqn:c3int}
\bmdotstart = \mmdotstar \tauIa^{-1} \int_0^{t_c} e^{-{t'/\ttausfh}} 
              e^{-(t-t'-\td)/\tauIa}dt'~.
\end{equation}
(Compare to equation~\ref{eqn:app_expsfrint}; here we have assumed
$t > t_c+\td$ in setting the integration limit to $t_c$ rather than $t-\td$.)
Evaluating the integral gives
\begin{equation}
\label{eqn:bmdotstarc3}
\bmdotstart = \mmdotstar {\ttaubarIaSfh \over \tauIa} 
  \left[e^{t_c/\ttaubarIaSfh}-1\right] e^{-(t-\td)/\tauIa}~.
\end{equation}
With the substitution $t_2 = t-t_c$,
equation~(\ref{eqn:ironc3}) can now be written in the form
\begin{equation}
\label{eqn:ironc3a}
\Mfedot + {\Mfe \over \tttaudep} = K e^{-t_2/\tauIa}
\end{equation}
with
\begin{equation}
K = \mfeIa\mmdotstar {\ttaubarIaSfh \over \tauIa}
    \left[e^{t_c/\ttaubarIaSfh}-1\right] e^{-(t_c-\td)/\tauIa}~.
\end{equation}
This equation is solved by the usual technique with the boundary
condition that $\Mfe$ from this contribution starts from zero at $t=t_c$.
After some manipulation, and dividing by 
$M_g(t_2) = M_g(t_c) e^{-t_2/\tttausfh}$, one gets
\begin{equation}
\label{eqn:YfeIaEvolC3}
\begin{split}
Z_{\rm Fe,3}(t_2)
    = &Z_{\rm Fe,eq1}^{\rm Ia} \left[1-e^{-t_c/\ttaubarIaSfh}\right]
              e^{\td/\ttaubarIaSfh} \times \\
			  &{\tttaubarDepIa \over \ttaubarDepSfh}
	      \left[e^{-t_2/\tttaubarIaSfh}-e^{-t_2/\tttaubarDepSfh}\right]~,
\end{split}
\end{equation}
where $Z_{\rm Fe,eq1}^{\rm Ia}$ is the equilibrium abundance of
equation~(\ref{eqn:YfeIaeq2}) evaluated with the pre-$t_c$ parameters.
For small values of $t_2$, Taylor expansion in the final factor gives
\begin{equation}
\label{eqn:YfeIaEvolC3early}
\begin{split}
Z_{\rm Fe,3}(t_2)
   \approx & Z_{\rm Fe,eq1}^{\rm Ia} 
            \left[1-e^{-t_c/\ttaubarIaSfh}\right] \times \\
              & e^{\td/\ttaubarIaSfh}{t_2 \over \ttaubarDepSfh}~,
\end{split}
\end{equation}
which is linear in $t_2$ as expected.
The full solution for $t > t_c$ is
\begin{equation}
\label{eqn:YfeEvolChange}
\Yfe(t_2) = Z_{\rm Fe,1}(t_2)+Z_{\rm Fe,2}(t_2)+Z_{\rm Fe,3}(t_2)~.
\end{equation}

%%%%%%
\begin{figure*}
\centerline{\includegraphics[width=6.0truein]{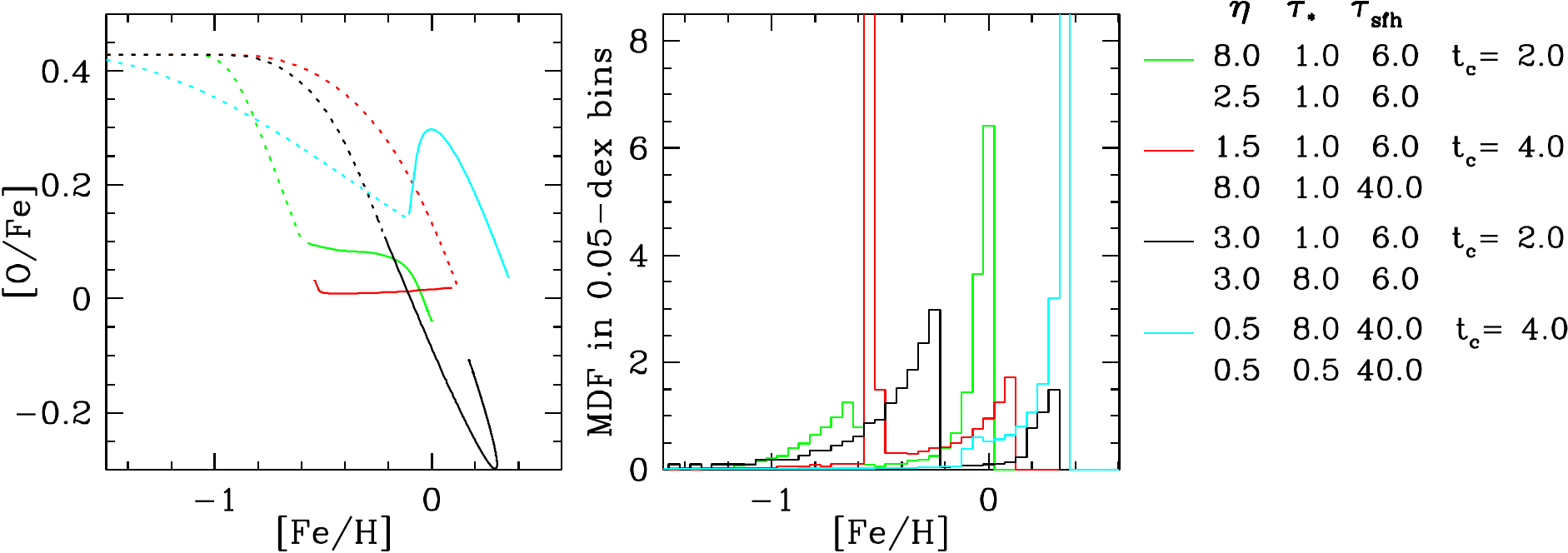}}
\caption{Tracks in $\ofe$ vs. $\feh$ (left) and iron MDFs (middle)
for models with sudden parameter changes, at time $t_c$.
Parameters before and after $t_c$ are listed in the legend at right.
In the left panel, dotted and solid lines denote pre-$t_c$ and 
post-$t_c$ evolution, respectively.
Green curves show a model with a sharp decrease in outflow efficiency
$\eta$ and corresponding increase in equilibrium abundance.
Red curves show the reverse case, an increase in $\eta$ that 
drives down the equilibrium abundance.  Black curves show a
model that transitions from high star formation efficiency
($\taustar=1\Gyr$) to low star formation efficiency ($\taustar=8\Gyr$).
Cyan curves show the reverse transition, from $\taustar=8\Gyr$
to $\taustar=0.5\Gyr$.
}
\label{fig:change1}
\end{figure*}

We can summarize these results as follows.
After a sudden change of parameters, the oxygen
abundance evolves from its value at $t_c$ to the new equilibrium
value on the timescale $\tttaubarDepSfh$.  The iron abundance
has three contributions: exponential decay of $\Yfe(t_c)$ on
the timescale $\tttaubarDepSfh$, growing iron with post-$t_c$
parameters that follows the usual behavior for exponential SFR
evolution with time variable $t_2$, and a contribution from
delayed pre-$t_c$ SNIa that grows linearly at small $t_2$ and
then decays exponentially as governed by equation~(\ref{eqn:YfeIaEvolC3}).
Different choices for parameter changes allow a rich variety
of behaviors.

Our assumption that the gas mass is continuous at $t_c$ implies
that the star formation rate changes discontinuously:
\begin{equation}
\label{eqn:sfrchange}
{\mdotstar(t_c+\epsilon)\over \mdotstar(t_c-\epsilon)} = 
{\ttaustar \over \tttaustar}~.
\end{equation}
If we instead assume that the gas supply is instantaneously
diluted by a factor $D$ at $t_c$, so that 
$M_g(t_c+\epsilon)/M_g(t_c-\epsilon)=D$, 
then the right hand side of this equation is multiplied by $D$.
In this case the contributions from pre-$t_c$ stars,
i.e., the terms $Z_{\rm O,1}$, $Z_{\rm Fe,1}$, and $Z_{\rm Fe,3}$,
are divided by $D$, while the contributions $Z_{O,2}$ and $Z_{\rm Fe,2}$ 
from post-$t_c$ stars are unchanged.

Figure~\ref{fig:change1} presents $\ofe-\feh$ tracks
and iron MDFs for several models
with sudden parameter changes.  Specific parameter values have
been chosen partly with regard to keeping the model curves
visually distinguishable.
Green curves show a model that transitions from high outflow
efficiency ($\eta=8$) to our fiducial outflow efficiency ($\eta=2.5$)
at $t_c=2\Gyr$.  The model track progresses rapidly to 
$\ofe=0.1$ at $\feh\approx -0.6$, but after the change 
of $\eta$ and the corresponding increase of equilibrium abundance,
it turns sharply towards higher $\feh$, maintaining approximately
constant $\ofe$.  The final downturn comes about $1.5\Gyr$
after the transition, when SNIa enrichment from post-$t_c$
star formation drives the model to its final equilibrium
abundances of $\ofe \approx \feh \approx 0$, the same
as in our fiducial model.  The majority of stars are
formed close to this final equilibrium.
As discussed by AWSJ and \citeauthor{nidever2014} (\citeyear{nidever2014},
their Fig.~16), a transition from high-$\eta$ to low-$\eta$
is one of the only ways to create a one-zone model that
exhibits substantial growth of $\feh$ at near-solar $\ofe$.

Red curves illustrate the reverse case of a model that approaches
equilibrium at low outflow efficiency ($\eta=1.5$) and slightly
super-solar $\feh$, then tracks backward to {\it low} $\feh$
at near-constant $\ofe$ because of an increase in outflow efficiency.
The final small uptick, again coming $\approx 1.5\Gyr$ after $t_c$,
arises because we have also increased $\tausfh$ from $6\Gyr$
to $40\Gyr$ (i.e., to near-constant SFR), and the latter case
leads to higher equilibrium $\ofe$.  If we maintained $\tausfh=6\Gyr$,
the curve would instead downtick to solar $\ofe$.
Progression from high metallicity to low metallicity goes
against conventional intuition about chemical evolution,
but it is perfectly reasonable if outflow efficiency changes
from low to high at late times, which could occur if decreasing
gas densities make it easier to drive outflows.
However, we have not found a model that produces steadily
decreasing $\feh$ together with steadily increasing $\ofe$,
resembling the $\afe$ locus for thin disk stars found by, e.g.,
\cite{adibekyan2012}, \cite{hayden2015}, and \cite{bertran2015}.

Black curves show a model in which $\eta$ and $\tausfh$ remain
constant but the star formation efficiency changes from
high ($\taustar = 1\Gyr$) to low ($\taustar=8\Gyr$).
After the transition, the model overshoots its original
equilibrium, evolving to sub-solar $\ofe \approx -0.25$ 
because oxygen enrichment at the low post-$t_c$ star formation efficiency
is too slow to balance the SNIa iron enrichment coming from
pre-$t_c$ stars.  Since the depletion of this iron is
also relatively slow, the model also reaches super-solar
$\feh \approx +0.25$.  The loop upward towards higher $\ofe$
commences a couple of Gyr after $t_c$, but it is a slow change
because of the relatively long depletion time,
$\taudep=\taustar/(1+\eta-r)=2.2\Gyr$.
In contrast to the other three models in Figure~\ref{fig:change1},
this model has not yet converged to its final (near-solar)
equilibrium abundances by the end of the calculation at $t=12.5\Gyr$.

Cyan curves show the opposite transition, from low star formation
efficiency ($\taustar=8\Gyr$, $\taudep=7.3\Gyr$) to high star
formation efficiency ($\taustar=0.5\Gyr$, $\taudep=0.45\Gyr$).
The pre-$t_c$ evolution has a slow decline in $\ofe$ characteristic
of long-$\taudep$ models.
Although the SFR is nearly constant ($\tausfh=40\Gyr$) before $t_c$ and
after $t_c$, there is a discontinuous factor of 16 jump in $\mdotstar$
at the transition (eq.~\ref{eqn:sfrchange}).
The result is a sudden upward boost of
$\ofe$ like those in the instantaneous burst calculations
of \S\ref{sec:bursts}.
This trend reverses within a Gyr because of the short 
post-$t_c$ depletion time, and the model quickly
evolves to the equilibrium abundances of its post-$t_c$
parameters.

The first three of these models produce bimodal MDFs, with
one peak corresponding to pre-$t_c$ parameters and one
to post-$t_c$ parameters.  Each of these peaks individually
has the characteristic form typically seen in Figure~\ref{fig:mdf1}.
For the second (red curve) model, the low metallicity peak
corresponds to stars formed {\it after} $t_c$ and the 
smaller, high metallicity peak to stars formed before
the transition when the equilibrium abundance was high.
For the final (cyan curve) model, the two peaks merge
into a single distribution because of the low pre-$t_c$
star formation efficiency.

\section{Extensions}
\label{sec:extensions}

There are a variety of ways that our analytic results can be extended
to accommodate interesting cases.

\subsection{Two-exponential SNIa Delay Time Distribution}
\label{sec:two-exp}

The exponential form of the SNIa DTD is essential to allowing the
analytic solutions for iron evolution derived in \S\ref{sec:evolution}.
While this form is reasonably consistent with existing empirical constraints
and with the predictions of population synthesis models,
it produces fewer supernovae at early times and late times compared to
the $t^{-1.1}$ power-law DTD favored by \cite{maoz2012a}.
Fortunately, it is trivial to generalize the solutions to a DTD
that is the sum of two exponential functions with different
normalizations and timescales.  One simply imagines (purely for
convenience) that these two exponentials correspond to two different 
populations of SNIa that produce distinguishable isotopes of iron:
the solution for each population individually is the same as before,
and one sums the iron from the two components to obtain the total.
Mathematically, the ability to do this follows from the linearity
of the governing differential equation.

%%%%%%
\begin{figure}
\centerline{\includegraphics[width=2.5truein]{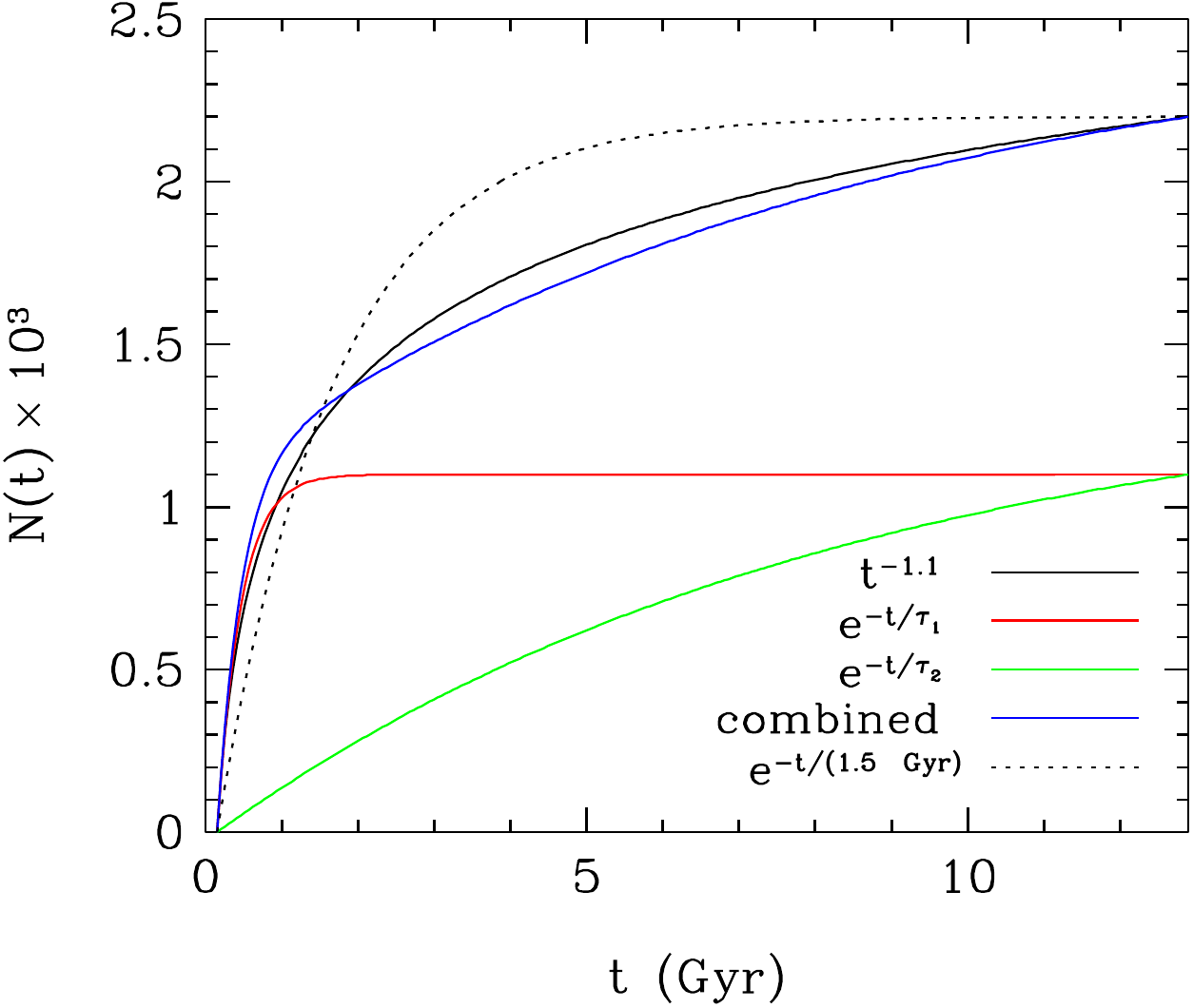}}
\caption{Cumulative number of SNIa per solar mass of star formation
(multiplied by $10^3$) as a function of time for our fiducial,
$1.5\Gyr$ exponential DTD (dotted black), for a $t^{-1.1}$ power
law DTD (solid black), and for the sum of two exponential DTDs
with timescales of 0.5 and $5.0\Gyr$ (blue; individual contributions
in red and green, respectively).  All cases have a minimum delay
time $\td=0.15\Gyr$ and are normalized to produce
$2.2 \times 10^{-3}$ SNIa per solar mass over $12.5\Gyr$.
}
\label{fig:SNIaNum}
\end{figure}
%%%%%%

%%%%%%
\begin{figure}
\centerline{\includegraphics[width=3.3truein]{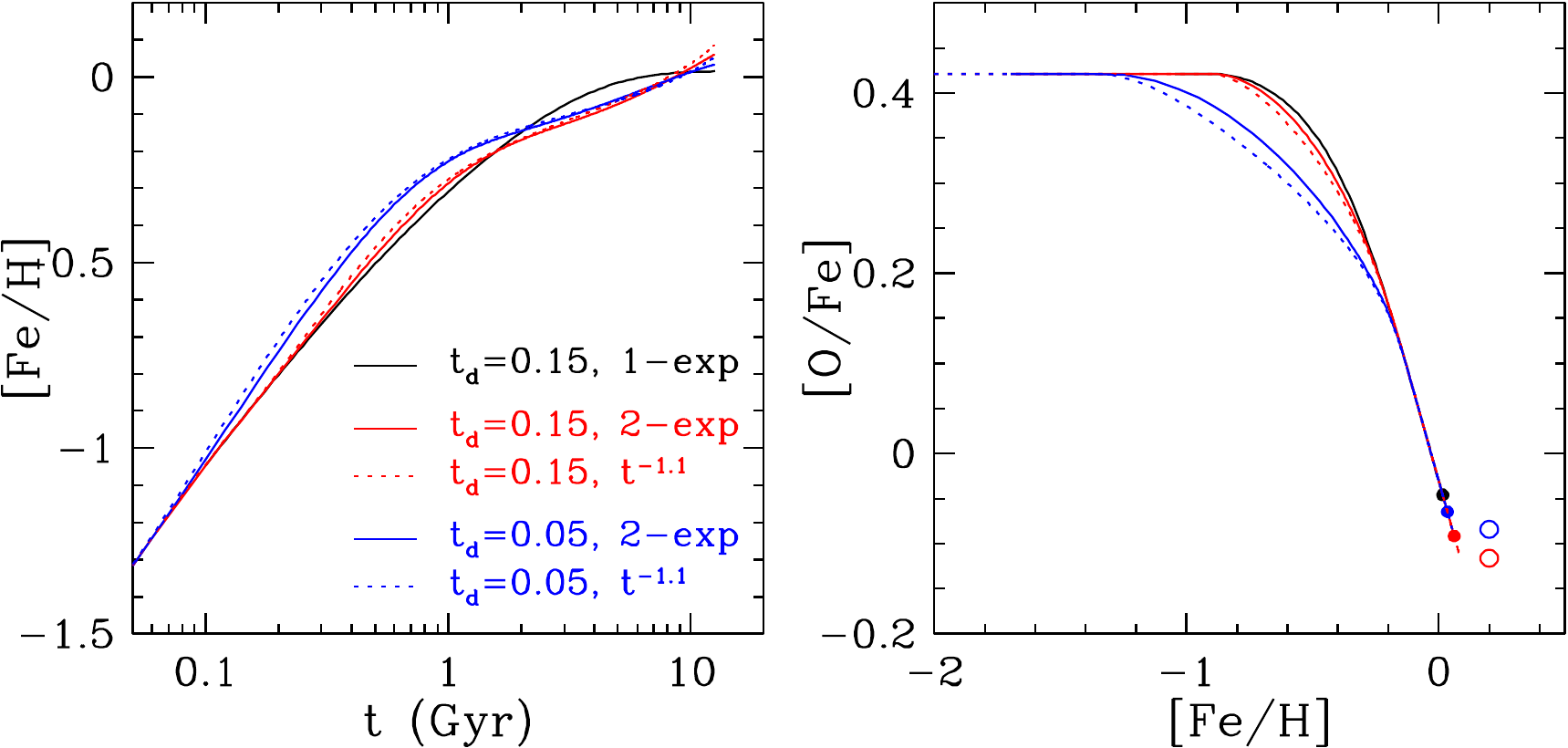}}
\caption{Evolution of \feh\ (left) and tracks in 
in \ofe\ vs. \feh\ (right) for our fiducial model (black),
which has an exponential SNIa DTD, 
for $t^{-1.1}$ power-law DTDs with minimum delay times
$\td=0.15\Gyr$ (red dotted) and $\td=0.05\Gyr$ (blue dotted),
and for the matched double-exponential DTDs described in the
text (red and blue solid).  Except for the DTD, all models have
the same parameters as the fiducial model.
Filled circles mark the final ($t=12.5\Gyr$) \ofe\ values for
the single- and double-exponential models, and horizontally offset open 
circles mark the final \ofe\ values for the power-law DTD models.
}
\label{fig:snmatch}
\end{figure}
%%%%%%

Figure~\ref{fig:SNIaNum} compares the cumulative number of SNIa
as a function of time from the $t^{-1.1}$ power-law and from a
sum of two exponentials with timescales of $0.5\Gyr$ and $5\Gyr,$
with a minimum delay time $t_d=0.15\Gyr$ in each case.
The power-law DTD is normalized to produce $2.2\times 10^{-3}$ SNIa
per solar mass of star formation over $12.5\Gyr$, and the two exponentials
are individually normalized to produce $1.1\times 10^{-3}$ SNIa per solar
mass over the same interval.  The two-exponential model approximates the 
result of the power-law model more closely than our single $1.5\Gyr$
exponential, though it still produces a larger fraction of its SNIa
at intermediate times and fewer at early and late times.  
For a minimum delay of $t_d=0.05\Gyr$ (not shown), we obtain 
comparable agreement by summing two exponentials with timescales
of $0.25\Gyr$ and $3.5\Gyr$, again normalized to produce equal
numbers of SNIa over $12.5\Gyr$.

Figure~\ref{fig:snmatch} compares $\feh$ enrichment histories and
$\ofe-\feh$ tracks for our
fiducial model to numerically calculated results for a 
$t^{-1.1}$ DTD and to analytic results for the matching two-exponential
DTDs described above.\footnote{The analytic results for $\YfeIa$
are computed using equation~(\ref{eqn:YfeIaEvolExp}) after multiplying
$\mfeIa$ by $(0.478,0.522)$ for $\tau=(0.5,5)\Gyr$ in the $t_d=0.15\Gyr$
case, and by $(0.493,0.507)$ for $\tau=(0.25,3.5)\Gyr$ in the
$t_d=0.05\Gyr$ case.  These multiplicative factors are not exactly
$(0.5,0.5)$ because we normalize the two exponentials to produce
the same number of SNIa over $12.5\Gyr$, while $\mfeIa$ is
defined out to $t=\infty$.}  
For a $0.15\Gyr$ minimum delay time, the $\feh$ enrichment is
nearly identical between the single-exponential and power-law DTD
at early times, with modest differences at $t > 2\Gyr$.  The most
noticeable difference in the $\ofe-\feh$ track is a decrease in
the final $\ofe$, from  $-0.05$ for the fiducial model to $-0.12$
for the power-law DTD.  The double-exponential DTD reproduces the
results of the power-law DTD quite accurately, with a maximum
difference in $\feh$ (and thus in $\ofe$, since CCSN enrichment
is unchanged) of about 0.02 dex.

With a single-exponential DTD, changing the minimum delay time
from $0.15\Gyr$ to $0.05\Gyr$ has only a small impact on the
$\ofe-\feh$ track, as shown previously in Figure~\ref{fig:history2}.
However, changing $\td$ has a much bigger impact for a 
$t^{-1.1}$ DTD because the early-time divergence of the SNIa
rate is only truncated by the minimum delay time.\footnote{Note
that $t^{-1.1}$ is critically different from $(\Delta t)^{-1.1}$,
which would imply an infinite number of SNIa for any $\td$.}
For $\td=0.05\Gyr$ 
the knee of the $\ofe-\feh$ track begins at lower $\feh$, and
the shape of the downturn is more gently curved.
The matched double-exponential DTD model again recovers the 
behavior of the power-law DTD model quite accurately, 
with a maximum difference of about 0.025 dex.
The final $\ofe$ value is again below that of the single-exponential
model, but it is above that of the $t_d=0.15\Gyr$ power-law model
because normalizing to the same number of SNIa over $12.5\Gyr$ 
implies fewer SNIa at late times for a smaller $t_d$.

\subsection{Enriched infall}
\label{sec:enriched}

We have previously assumed that infalling gas has primordial
composition, with no metals.  We can relax this assumption
using equation~(\ref{eqn:infall}) for the infall rate.
In the case of an exponential star formation history the 
result is particularly simple.  Here we have 
$\ddot{M}_* = -\mdotstar/\tausfh$, implying
\begin{equation}
\label{eqn:infall_exp}
\begin{split}
\dot{M}_{\rm inf} &= \mdotstar (1+\eta-r-\taustar/\tausfh) \\
                  &= \mdotstar \taustar/\taubarDepSfh ~.
\end{split}
\end{equation}
If the oxygen abundance of the infalling gas is $\Yoinf$, then
equation~(\ref{eqn:ModotTimedep}) for oxygen evolution can be 
revised to
\begin{equation}
\label{eqn:ModotEnriched}
\begin{split}
\Modot + {\Mo\over \taudep} &= \mocc\mdotstar + \Yoinf\dot{M}_{\rm inf} \\
  &= \mocc\mdotstar + \Yoinf\mdotstar \taustar/\taubarDepSfh \\
  &= \mocc\mdotstar (1+\Yoinf/\Yoeq)~,
\end{split}
\end{equation}
where the last equality substitutes $\Yoeq$ from equation~(\ref{eqn:Yoeq2}).
Thus, enriched infall has exactly the same effect as increasing the
CCSN oxygen yield by a factor of $1+\Yoinf/\Yoeq$.
The analogous argument for iron implies that enriched infall is
equivalent to increasing the CCSN iron yield (but not the SNIa yield)
by $1+\Yfeinf/\Yfecceq$.
In many circumstances we expect $Z_{\rm inf} \ll Z_{\rm eq}$,
in which case enriched infall has little impact.

For a linear-exponential SFH, enriched infall will be more important
at early times because the initial gas supply is low.
This case may also be solvable with our analytic approach, but
we have not investigated it.

Analytic solutions for a variety of enriched infall histories
are described by \cite{spitoni2015}, including solutions with
galactic fountains and exchange of metals between neighboring galaxies.
These analytic models adopt a ``linear Schmidt law,'' equivalent
to our assumption of constant $\taustar$.

\subsection{Oxygen Budget}
\label{sec:oxygen_budget}

\cite{peeples2014} present a global account of the metal budget
in and around low redshift star-forming galaxies (for earlier
work along similar lines see, e.g.,
\citealt{ferrara2005,gallazzi2008,zahid2012}).
Integral field (IFU) spectroscopic surveys
are beginning to allow spatially resolved accounting within
individual galaxies (e.g., \citealt{belfiore2015}).  It is useful
to connect our analytic results for oxygen MDFs (\S\ref{sec:mdf})
to this type of accounting.

With our assumptions, the total mass of oxygen produced by a system 
and returned to the ISM by time $t$ is
\begin{equation}
\label{eqn:Moprod}
\Moprod = \mocc\Mform
\end{equation}
where
\begin{equation}
\label{eqn:Mform}
\Mform = \int_0^t \mdotstart dt
\end{equation} 
is the total mass of stars that the system has formed.
The mass of these metals that is currently locked up in the
system's stars is
\begin{equation}
\label{eqn:Mostars}
\Mostars = \Yomean M_* = \Yomean\Mform (1-r)
\end{equation}
where 
\begin{equation}
\label{eqn:Yomean}
\Yomean = \int_0^{\infty} \Yo {dN\over d\Yo} d\Yo 
  \left[\int_0^{\infty} {dN \over d\Yo} d\Yo\right]^{-1}
\end{equation}
is the mean oxygen abundance of the stellar population.
If the abundance increases monotonically in time, then
the upper limit of the integral is, in practice, $\Yo(t)$.
If gas ejected by feedback is well mixed, so that it always has
the same oxygen abundance as the gas that is forming into stars,
then the mean metallicity of the ejected gas is
the same as the mean metallicity of the stars, making the ejected oxygen mass
\begin{equation}
\label{eqn:Moej}
\Moej = \eta\Yomean\Mform = \eta\Mostars /(1-r)~.
\end{equation}
Finally, the oxygen mass currently in the ISM depends on the 
current oxygen abundance rather than the mean abundance,
\begin{equation}
\label{eqn:Moism}
\Moism = \Yo(t) M_g(t) = \Yo(t) \taustar\mdotstart~.
\end{equation}

For a constant SFR, it is straightforward to evaluate the mean
of equation~(\ref{eqn:mdf1}) to find $\Yomean=\zomean \Yoeqc$ with
\begin{equation}
\label{eqn:YomeanCsfr}
\zomean = 1 - \zo(t)\taudep/t~,
\end{equation}
where $\zo(t)=1-e^{-t/\taudep}$ is the current scaled abundance.
Substituting in the above equations and using the 
definition~(\ref{eqn:YoeqCsfr}) of $\Yoeqc$ yields
\begin{equation}
\label{eqn:checksum}
\Mostars + \Moej + \Moism = \Moprod
\end{equation}
as expected.

For an exponential SFH
\begin{equation}
\label{eqn:MformExp}
\Mform = \int_0^t \mdotstaro e^{-t'/\tausfh} dt' = 
  \tausfh\mdotstaro\left(1-e^{-t/\tausfh}\right)~.
\end{equation}
Evaluating the mean of equation~(\ref{eqn:mdf2}) requires some manipulation,
with the end result
\begin{equation}
\label{eqn:YomeanExp}
\begin{split}
\zomean &= {\taudep\over\taubarDepSfh}\left[1-{\taubarDepSfh\over\tausfh}
  {e^{-t/\tausfh} \over 1-e^{-t/\tausfh}} \zo(t)\right] \\
  &= {\taudep\over\taubarDepSfh}\left[1-
    {\taubarDepSfh\mdotstart\over\Mform}\zo(t)\right]~.
\end{split}
\end{equation}
The sum of oxygen in stars plus oxygen ejected is
\begin{equation}
\label{eqn:sumExp}
\Mostars+\Moej=\Yoeq\zomean\Mform\left[(1-r) + \eta\right]~,
\end{equation}
and with the substitutions $\taudep (1+\eta-r)=\taustar$ and
$\Yoeq=\mocc\taubarDepSfh/\taustar$ one can again demonstrate
the closed accounting loop of equation~(\ref{eqn:checksum}).
Equation~(\ref{eqn:YomeanExp}) is useful in its own right,
relating the mean metallicity of a stellar population to the
current metallicity of the ISM under the assumptions of an
exponential SFH, constant $\taustar$, and instantaneous recycling
and enrichment.  In the limit of $\tausfh \gg t$ and $\tausfh \gg \taudep$,
equation~(\ref{eqn:YomeanExp}) approaches equation~(\ref{eqn:YomeanCsfr})
for a constant SFR.

\subsection{Intermediate Elements}
\label{sec:other_elements}

%%%%%%
\begin{figure}
\centerline{\includegraphics[width=2.5truein]{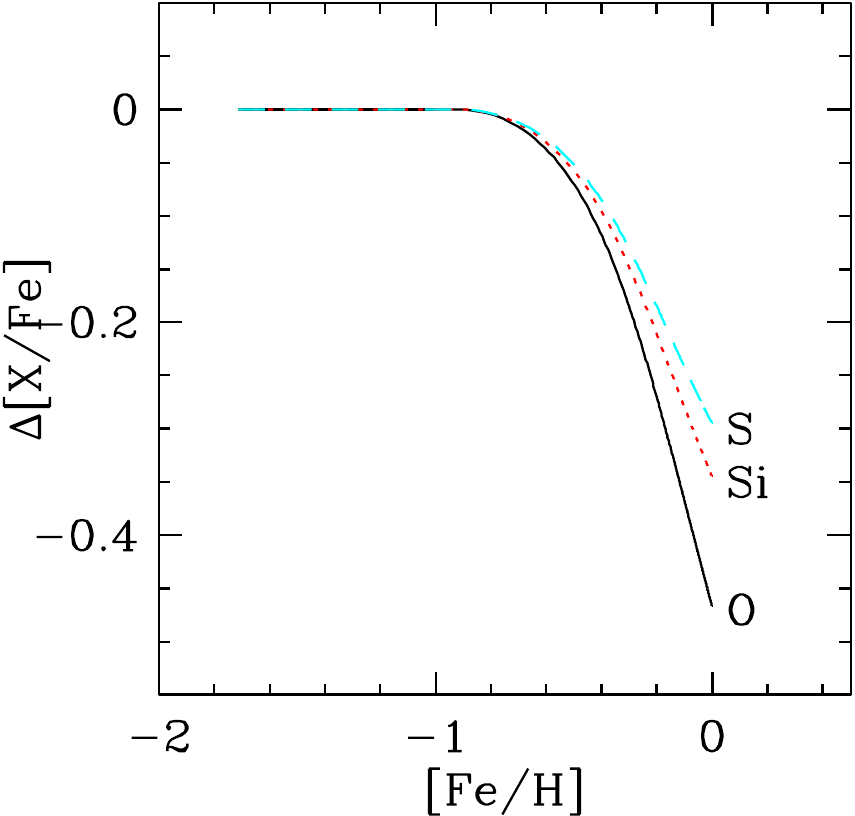}}
\caption{Tracks in the $\xfe-\feh$ plane for our fiducial model
model parameters ($\eta=2.5$, $\taustar=1\Gyr$, $\tausfh=6\Gyr$,
$\tauIa=1.5\Gyr$, and $\td=0.15\Gyr$) for oxygen (black solid),
silicon (red dotted), and sulfur (cyan dashed), based on 
equation~(\ref{eqn:Deltaxfe}).  
Because silicon and sulfur have increasingly significant
contributions from SNIa, they exhibit smaller drops in $\xfe$
between the CCSN plateau ($\Delta\xfe=0$ in this figure)
and the late-time equilibrium.
}
\label{fig:Deltaxfe}
\end{figure}
%%%%%%

In our calculations, we treat oxygen as an ``idealized'' $\alpha$-element,
produced only by CCSNe, with a metallicity independent yield.
Based on comparison to the more complete calculations in AWSJ, this
idealization is good at the $\approx 0.05$ dex level.
For a given star formation history, the evolution of $\ofe$ relative
to the early-time, CCSN plateau is governed by 
equation~(\ref{eqn:Deltaofe}) and thus by the ratio $\YfeIa(t)/\Yfecc(t)$.
The value of $\ofe$ at the plateau depends on CCSN yields
and solar abundances as described by equation~(\ref{eqn:plateau}).
As noted in \S\ref{sec:ev_csfr}, any other idealized $\alpha$-element 
should follow exactly the same track relative to its own early-time plateau.
In particular, with the yields adopted by AWSJ, Mg production is also 
dominated almost entirely by CCSNe with metallicity-independent yield,
so it should follow a nearly identical track to oxygen.

Moving up the periodic table, Si and S are $\alpha$-elements
with increasingly significant predicted contributions from SNIa,
though their production is still dominated by CCSN.  For metallicity
independent yields $\mxcc$ and $\mxIa$ of an element X,
the evolution of the relative abundance can be expressed
in terms of the iron evolution and the relative yields:
\begin{equation}
\label{eqn:xratio}
\begin{split}
{\Yx(t)\over\Yfe(t)} &=
 {\Yfecc(t)\mxcc/\mfecc + \YfeIa(t)\mxIa/\mfeIa \over
  \Yfecc(t) + \YfeIa(t)} \\
 &= {\mxcc \over \mfecc} \times 
   {1+{\mxIa/\mxcc \over \mfeIa/\mfecc}\YfeIa(t)/\Yfecc(t) \over
    1+\YfeIa(t)/\Yfecc(t)}~.\\
\end{split}
\end{equation}
The plateau value of the abundance ratio is just $\mxcc/\mfecc$, so
in combination with equation~(\ref{eqn:Deltaofe}) we can write
\begin{equation}
\label{eqn:Deltaxfe}
\Delta\xfe = \Delta\ofe + \log_{\rm 10}
  \left[1+{\mxIa/\mxcc \over \mfeIa/\mfecc}{\YfeIa(t) \over \Yfecc(t)}\right]~,
\end{equation}
which as expected yields $\Delta\xfe=\Delta\ofe$ for any element
with $\mxIa=0$, and yields $\Deltaxfe=0$ for iron.
Figure~\ref{fig:Deltaxfe} shows relative abundance
tracks for Si and S compared to the oxygen track, for the parameters
of our fiducial model.  Based on the IMF-integrated yields computed
by AWSJ and the W70 model yield of \cite{iwamoto1999}, we have adopted
$\msicc=0.0013$ and $\msiIa=0.24\msicc=0.00031$ for silicon and
$\mscc=0.00056$ and $\msIa=0.36\mscc=0.00020$ for sulfur.
Because of the increasing contribution of SNIa, the drops
between the plateau and the equilibrium are successively smaller.
For our adopted yields and the solar photospheric
abundance values of \cite{lodders2003},
the implied values of the plateau are
$\ofe_{\rm plateau}=+0.43$, $[{\rm Si}/{\rm Fe}]_{\rm plateau}=+0.26$,
and $[{\rm S}/{\rm Fe}]_{\rm plateau}=+0.21$.
If we used the \cite{lodders2003} recommended proto-solar
abundances (corrected for diffusion) rather than photospheric abundances,
then all of these values would drop by 0.07 dex.

Our prediction of similar $\Delta\xfe$ for oxygen and magnesium
and progressively lower values for Si and S is not
in good agreement with observations.
For example, the plateau values of $[{\rm Mg}/{\rm Fe}]$ and
$[{\rm Si}/{\rm Fe}]$ found for solar neighborhood stars by
\cite{adibekyan2012} are approximately $+0.3$ and $+0.2$, but
analyzing the same sample \cite{bertran2015} find $\ofe$ values
for thick disk stars as high as $+0.5-0.8$.
The compilation of $[{\rm S}/{\rm Fe}]$ 
values by \cite{kacharov2015} (their Fig.~6), for individual stars
and star clusters, suggests a plateau at $+0.4$.
These discrepancies could reflect systematic errors in the observational
abundances, which are difficult to place on a consistent scale across
a wide range of $\feh$.  Alternatively they could indicate a breakdown
of our assumptions, particularly the assumption that the IMF-averaged
yields are independent of metallicity.  The yields of different elements
are strongly dependent on stellar mass, so even if the yield at fixed
stellar mass is metallicity independent, the IMF-averaged yield could
vary if the IMF itself depends on metallicity or if the mass range
of stars that explode as CCSNe changes with metallicity.
Homogeneous analyses of large data sets such as APOGEE, Gaia-ESO, and GALAH
should clarify whether there is indeed a discrepancy with observations
to be explained.

\subsection{Time-Dependent $\tau_*$}
\label{sec:timedeptaustar}

Perhaps the least desirable restriction of our analytic solutions
is the requirement of constant SFE timescale $\tau_*$.  
While observations are consistent with a constant
SFE for molecular gas in typical galaxies \citep{leroy2008},
the non-linear form of the Kennicutt-Schmidt law 
\citep{schmidt1959,kennicutt1998} implies that the SFE decreases
with decreasing total (atomic + molecular) gas surface density.
For one-zone models with declining star formation histories, therefore,
a $\taustar$ that increases with time would represent this
situation more accurately.

Unfortunately constant $\taustar$, or more precisely constant
$\taudep$, has a special place in allowing
analytic solutions to our evolution equations. These generically
have the form
\begin{equation}
\label{eqn:evol_generic}
\dot{M}(t) + {M(t)\over \taudep} = m F(t)~,
\end{equation}
where the forcing function is $F(t)=\mdotstart$ for CCSN products
and $F(t)=\bmdotstart$ for SNIa products.  Our analytic solution 
methods require analytic expressions for
$\mu(t) = \exp[\int dt/\taudep]$ and for $\int \mu(t) F(t) dt$.
For constant $\taudep$, $\int\mu(t)F(t)dt = \int e^{t/\taudep} F(t)dt$ 
is analytic for interesting choices of $F(t)$, including ones involving
an exponential DTD for SNIa.  If $\taudep$ is a function of $t$,
then the integral is analytic only in special cases.

One such case is 
\begin{equation}
\label{eqn:taustaro}
\taustar(t) = \taustaro(1+t/\taudepo)
\end{equation}
and thus
\begin{equation}
\label{eqn:taudepo}
\taudep(t) = \taudepo(1+t/\taudepo)~,
\end{equation}
a depletion timescale that grows linearly from a starting value $\taudepo$,
reaching double its initial value after $\taudepo$ and approaching
$\taudep(t) \approx t$ at late times.
This case yields $\mu(t) = 1+t/\taudepo$ and allows analytic solutions
for both the oxygen mass and the iron mass as a function of time.
We assume constant $\eta$ and $r$, so the gas mass at a given time is
$M_g(t) = \mdotstart\taustar(t) = \mdotstart\taustaro(1+t/\taudepo)$.
For an exponential star formation history, our usual solution methods
yield, after some calculation,
\begin{equation}
\label{eqn:YoEvolTimedepTau}
\begin{split}
\Yo(t) = & {\mocc\over 1+\eta-r} \left({\tausfh\over\taudepo}\right) 
           \left(1+{t\over \taudepo}\right)^{-2} e^{t/\tausfh} \times \\
  & \left[\left(1+{\tausfh\over\taudepo}\right)\left(1-e^{-t/\tausfh}\right) -
    {t\over \taudepo}e^{-t/\tausfh}\right]~,
\end{split}
\end{equation}
and
\begin{equation}
\label{eqn:YfeIaEvolTimedepTau}
\begin{split}
\YfeIa(t) = & {\mfeIa\over 1+\eta-r} \left({\taubarIaSfh \over \tauIa}\right)
           \left(1+{t\over \taudepo}\right)^{-2} e^{t/\tausfh} \times \\
  & {1\over\taudepo} \Bigg[\left(1+{t\over\taudepo}\right)
   \left(\tauIa e^{-\Delta t/\tauIa} - \tausfh e^{-\Delta t/\tausfh}\right) + \\
  & {1\over \taudepo}\left(\tauIa^2 e^{-\Delta t/\tauIa} - 
                           \tausfh^2 e^{-\Delta t/\tausfh}\right) - \\
  & {1\over \taudepo}(\tauIa - \tausfh)(\td+\taudepo+\tauIa+\tausfh) \Bigg]~,
\end{split}
\end{equation}
where the latter expression is for $t>\td$ only and $\YfeIa=0$ at $t\leq\td$.
These expressions are not particularly intuitive, and they do
not asymptotically approach an equilibrium value like our 
expressions for constant $\taudep$.  Nonetheless, the behavior
for realistic parameter values is not radically different from that 
of our previous solutions.

%%%%%%
\begin{figure}
\centerline{\includegraphics[width=3.5truein]{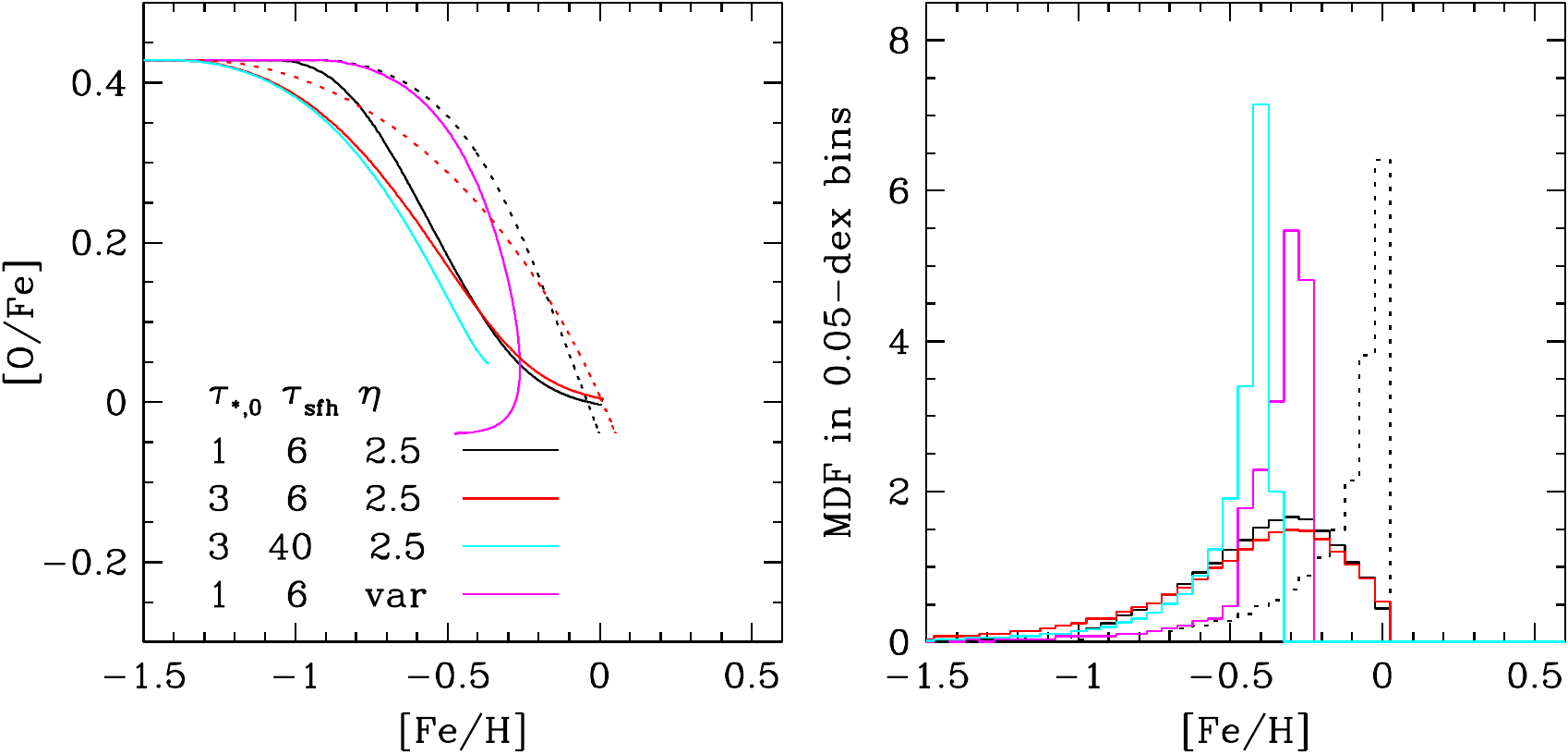}}
\caption{Tracks in $\ofe-\feh$ (left) and iron MDFs (right) for models
with time-dependent SFE timescale $\taustar$.  Solid black and red
curves show models with $\taustar = \taustaro(1+t/\taudepo)$
for $\taustaro=1$ and $3\Gyr$, respectively, and other parameters
equal to those of our fiducial model.  Dotted curves in the left panel
show two corresponding models with constant $\taustar=\taustaro$,
and the dotted histogram in the right panel shows the MDF of
the fiducial ($\taustar=1\Gyr$) model.
Cyan curves show a model with $\taustaro=3\Gyr$ and a long
$\tausfh$ that implies a nearly constant SFR.
Magenta curves show a model in which $\eta(t)$ varies
simultaneously with $\taustar(t)$ in a way that keeps
$\taudep$ constant, following equations~(\ref{eqn:taustart})
and~(\ref{eqn:etat}) with $q=2$ and $H(t)$ rising linearly
from zero to one over 12.5 Gyr.
}
\label{fig:timedeptau}
\end{figure}

Figure~\ref{fig:timedeptau} shows $\ofe-\feh$ tracks and iron MDFs
for three example cases.  The first (black curves) has our fiducial
model parameters but the time-dependent $\taudep$ of 
equation~(\ref{eqn:taudepo}), with $\taudepo=\taustaro/(1+\eta-r)=0.323\Gyr$.
Compared to the constant $\taustar$ case (dotted curves) the
$\ofe$ abundance ratio turns down at lower $\feh$ because $\taustar$
has already grown by a factor $\sim 5$ by the time SNIa enrichment 
becomes important.  At late times, the model track turns rightward to 
increasing $\feh$ at nearly solar $\ofe$ because the depletion
timescale starts to exceed the SFH timescale.  The abundances
at $t=12.5\Gyr$ are similar to those of the fiducial model.
Increasing $\taustaro$ from $1\Gyr$ to $3\Gyr$ (red curves) 
shifts the knee to lower $\feh$ but makes little difference
at late times, when the model has again approached $\taudep \approx t$.
The cyan curve shows the same case but with a nearly constant SFR
($\tausfh=40\Gyr$), which eliminates the rightward turn in the
$\ofe-\feh$ track.  Note that the SFE timescales of these models
at $t=12.5\Gyr$ are $\taustar \approx t(1+\eta-r) \approx 40\Gyr$,
so the SFE at late times is extremely low.

MDFs for the two $\tausfh=6\Gyr$ models qualitatively resemble the
closed/leaky box form, which we found previously for ``gas starved''
models in which $\tausfh$ is comparable to $\taudep$.
In both classes of models, the star formation rate declines 
substantially over a time interval in which the metallicity
continues to grow, producing a smooth turnover rather than a
sharp cutoff in $\feh$.  The $\tausfh=40\Gyr$ model, on the
other hand, has a sharply peaked MDF, similar in form to
that of our fiducial (constant $\taudep$) model but shifted
down in $\feh$ by 0.5-dex.

Because it is $\taudep$ that enters our mass evolution equations,
not $\taustar$ directly, another way to allow time-dependent $\taustar$
is to introduce a time-dependent $\eta$ that compensates to keep
$\taudep$ constant.  For example, we can take
\begin{equation}
\label{eqn:taustart}
\taustar(t) = \taustaro[1+q H(t)]~,
\end{equation}
where $q$ is a constant and $H(t)$ is a function that goes from 0 to 1
with an arbitrary time-dependence.  If we simultaneously require
\begin{equation}
\label{eqn:etat}
\begin{split}
\eta(t)&=[1+qH(t)]{\taustaro/\taudep}-1+r \\
       &= \eta_0+qH(t)(1+\eta_0-r)
\end{split}
\end{equation}
then $\taudep$ is constant and our standard solutions for
$\Mo(t)$, $\Mfecc(t)$, and $\MfeIa(t)$ apply.
Compared to these standard solutions, however, the gas supply 
$M_g(t)=\taustar(t)\mdotstart$ at a given time is larger by a 
factor $\taustar(t)/\taustaro=1+qH(t)$ and the abundances
are lower by the same factor, with the abundance ratios unchanged.
This behavior is pleasantly simple and holds for arbitrary $H(t)$,
but it must be seen as the combined effect of raising $\taustar$
and raising $\eta$, not either alone.
Magenta curves in Figure~\ref{fig:timedeptau} show a case in
which $\taustar$ rises linearly from $1\Gyr$ to $3\Gyr$ over the
$12.5\Gyr$ span of the calculation.  The $\ofe-\feh$ track loops
back to low $\feh$ after approaching solar $\ofe$, similar to 
the model in Figure~\ref{fig:change1} for which a sudden increase
in $\eta$ drives down the equilibrium abundance, though the increase
here is steady rather than sudden.  This model produces a 
sharply peaked MDF like that of our typical constant $\taustar$
models, though the shape depends on the adopted $qH(t)$.

There may be other combinations of $\taudep(t)$ and $\mdotstart$
that yield analytic results for interesting illustrative cases,
at least for oxgyen.  Achieving analytic results for iron with
a realistic SNIa DTD is harder because the $\mu(t)F(t)$ integral 
must still be analytic when the forcing function is $\bmdotstart$.

\subsection{Generic sum of star formation histories}
\label{sec:sfhsum}

Returning to models with constant $\taustar$ and $\eta$,
we can combine results we have derived previously for constant,
exponential, or linear-exponential star formation histories
to derive solutions for more complex histories.
Suppose that we have a solution to the generic evolution 
equation (\ref{eqn:evol_generic}) for the forcing functions
associated with a star formation history $\mdotstarone(t)$
yielding a solution $M_1(t)$, where $M_1$ could refer to the oxygen,
CCSN iron, or SNIa iron mass.
Now consider a second solution $M_2(t)$ for a second 
star formation history $\mdotstartwo(t)$ with the same $\taustar$
and $\eta$ (both constant in time).  Linearity tells us that for
the star formation history $\mdotstar(t)=\mdotstarone(t)+\mdotstartwo(t)$
we can add the two forcing terms on the right hand side of
equations (\ref{eqn:evol_generic}) and get a solution with
$M(t)=M_1(t)+M_2(t)$.  
Our solutions have the generic form $Z(t)=Z_{\rm eq} G(t)$
(e.g., equations~\ref{eqn:YoEvolExp}, \ref{eqn:YfeIaEvolExp}, 
\ref{eqn:YoEvolLinExp}, \ref{eqn:YfeIaEvolLinExp}),
implying $M(t)=Z_{\rm eq} G(t)\mdotstart\taustar$.
To get the abundance for the combined star formation history
we must add the two element masses and divide by the total
gas mass $M_g(t)=\taustar[\mdotstarone(t)+\mdotstartwo(t)]$, obtaining
\begin{equation}
\label{eqn:sfhsum}
Z(t) = {\mdotstarone(t)Z_{\rm eq,1}G_1(t)+\mdotstartwo(t)Z_{\rm eq,2}G_2(t)
        \over \mdotstarone(t)+\mdotstartwo(t)}~.
\end{equation}
The solution for the combined star formation history is thus an average
of the two individual solutions weighted by their contribution
to the current star formation rate.

%%%%%%
\begin{figure}
\centerline{\includegraphics[width=3.5truein]{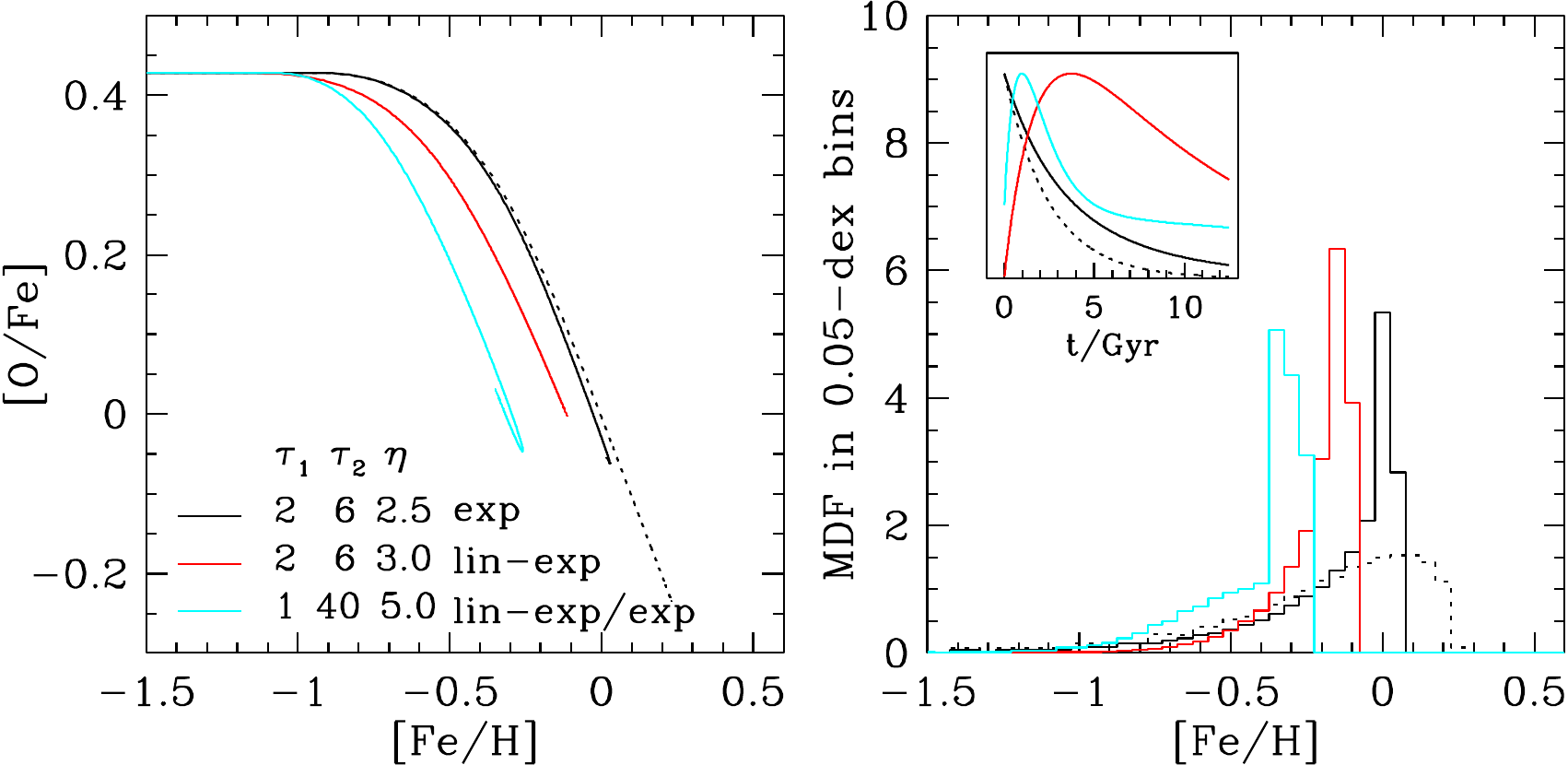}}
\caption{Tracks in $\ofe-\feh$ (left) and iron MDFs (right) for models
in which the star formation history (SFH) is a sum of two components.
All models have $\taustar=1\Gyr,$ $\tauIa=1.5\Gyr$, $\td=0.15\Gyr$.
Solid black curves show an SFH that is the sum of two exponentials
with timescales $\tausfh=2\Gyr$ and $6\Gyr$.  Red curves show
an SFH with the same two timescales but a linear-exponential form.
Cyan curves show the sum of a $1\Gyr$ linear-exponential and a
nearly constant ($\tausfh=40\Gyr$) exponential.  Dotted black curves
show the track and resulting MDF for a pure $2\Gyr$ exponential.
The inset in the right panel shows the star formation histories.
Values of $\eta$, indicated in the legend, have been chosen
to maintain visual clarity.
}
\label{fig:sumsfh}
\end{figure}

Figure~\ref{fig:sumsfh} shows $\ofe-\feh$ tracks and iron MDFs
for three examples of this class.  
In contrast to the models in
Figure~\ref{fig:change1}, all parameters are fixed throughout the
evolution of each of these models, even though the star formation
history is the sum of components with different timescales.
The star formation histories themselves are shown in the 
inset of the right hand panel.  In all cases we have chosen 
normalizations so that the integral (over $12.5\Gyr$) of the
long timescale component is twice that of the short timescale
component.  

We have previously seen that for smooth star formation histories
the shapes of tracks depend mainly on the outflow mass-loading
and SFE timescale ($\eta$ and $\taustar$),
with little dependence on $\mdotstart$,
but the star formation history has a stronger impact on the MDF.
Solid black curves in Figure~\ref{fig:sumsfh} show a star formation history 
that is the sum of two exponentials with $\tausfh=2\Gyr$ and $6\Gyr$,
while the dotted curves show results for a single $\tausfh=2\Gyr$
exponential for reference.  The two models follow nearly identical
tracks at early times, but the combined model stops at the
near-solar equilibrium abundances characteristic of $\tausfh=6\Gyr$
while the $\tausfh=2\Gyr$ model continues on to sub-solar $\ofe$
and super-solar $\feh$.
The MDF of the $\tausfh=2\Gyr$ model has the smooth cutoff 
characteristic of gas-starved models, expected because $\tausfh < 2\tauIa$
(see \S\ref{sec:traditional}).  The combined model, on the other hand,
has a sharply peaked MDF because of its longer $\tausfh$ at late times.

Red curves show a similar case, but for linear-exponential histories
instead of exponential histories.  The overall behavior is similar
to that of the combined exponential model, but the abundances are
slightly lower because of the slower approach to equilibrium.
(We have also increased $\eta$ from 2.5 to 3.0 to keep better
visual separation of the models; results for $\eta=2.5$ are
shifted slightly towards those of the combined exponential model.)
Cyan curves show a more extreme example with a star formation
history that begins with a sharply peaked, $\tausfh=1\Gyr$, 
linear-exponential burst followed by a nearly constant SFR
($\tausfh=40\Gyr$ exponential).  Here we have adopted $\eta=5$
to keep visual separation from the other two models.
The rapid early burst produces fast enrichment, and after
reaching slightly sub-solar $\ofe$ the model track loops back
to the equilibrium abundances of the constant SFR model.
The combination of two components with very different timescales
makes a clear imprint on the MDF, which has a linear rise
followed by a sharp peak.

\section{Conclusions}
\label{sec:conclusions}

Our main analytic results apply to one-zone models with 
metallicity-independent stellar yields and constant values
of the governing parameters, in particular the star formation
efficiency timescale $\taustar$ and the outflow efficiency $\eta$.
Crucially, our incorporation of a realistic DTD for Type Ia supernovae
(either exponential or a sum of exponentials approximating a power law)
allows us to compute the separate evolution of $\alpha$ and iron-peak
elements.
Realistic galaxies are more complex than these one-zone models,
but our calculations provide a number of insights that 
are useful to understanding more general chemical evolution scenarios.
The descriptions below continue to use oxygen as a representative
$\alpha$ element, but our conclusions about oxygen also apply to other
elements whose production is dominated by CCSNe and
whose yields have weak metallicity dependence.
We list our principal conclusions in several broad categories,
then briefly discuss future applications of our results.

\smallskip
\noindent
{\it Equilibrium abundances}
\smallskip

1.\ Under fairly general conditions, the abundances in a one-zone
model with constant parameters
evolve to an equilibrium in which the production of new
metals is balanced by the combination of dilution and 
loss of metals to star formation and
outflows.  If the star formation rate is approximately constant,
then oxygen approaches equilibrium on the gas depletion timescale
$\taudep=\taustar/(1+\eta-r)$, while iron approaches equilibrium
on the gas depletion or SNIa timescale ($\tauIa \approx 1.5\Gyr$),
whichever is longer.

2.\ At early times ($t \ll \tauIa$) the $\ofe$ ratio is determined
by CCSN yields, $\Yo/\Yfe = \mocc/\mfecc$.  For constant SFR, the
equilibrium abundances at late times are
$\Yo=\mocc/(1+\eta-r)$ and $\Yfe = (\mfecc+\mfeIa)/(1+\eta-r)$,
with strong dependence on $\eta$.  The equilibrium abundance ratio is
$\Yo/\Yfe=\mocc/(\mfecc+\mfeIa)$, depending only on yields.
Declining star formation histories 
lead to higher equilibrium abundances and lower equilibrium $\ofe$.

3.\ Elevated $\ofe$ ratios are a sign that the iron abundance, at least,
has not yet reached equilibrium.  Conversely, models with approximately
solar or sub-solar $\ofe$ are usually close to equilibrium in $\feh$.
It is difficult to construct a model with constant parameters that
shows significant increase in $\feh$ after reaching near-solar $\ofe$;
an extremely long gas depletion timescale is required.  

4.\ A population can be low metallicity either because it has not
had time to evolve to equilibrium or because the equilibrium
abundance itself is low.  Different factors may dominate in
different situations, and both may be important in some cases.
For example, gas rich dwarf galaxies may be metal poor because of
low star formation efficiencies (large $\taustar$), so that
they remain below equilibrium, and dwarf spheroidal galaxies
may have had their star formation truncated before reaching
equilibrium.  However, when star formation is vigorous the 
timescale for achieving equilibrium can be short, and many
high-redshift galaxies may be low metallicity not because they
are young but because they have high outflow rates that keep
their equilibrium abundances low.  Metallicity gradients in
galaxies could arise from slower star formation (departure
from equilibrium) or higher outflow efficiency (low equilibrium abundance)
at larger radii.
Mixing processes, not included in our models, may also play an
important role in regulating gradients
(e.g., \citealt{schoenrich2009,bilitewski2012,pezzulli2016}).

\smallskip
\noindent
{\it Effect of star formation history}
\smallskip

5.\ The behavior of one-zone models can be significantly different
if the star formation rate is declining on a timescale $\tausfh$
that is comparable to the depletion timescale $\taudep$ or, for iron,
to the SNIa timescale $\tauIa$.  
A value of $\tausfh \approx \taudep$ arises when
the system is ``gas starved,'' i.e., when the rate of gas accretion
is much lower than the rate of gas depletion.
In these cases the equilibrium abundances become large
because metals are deposited in a rapidly declining gas supply,
but the timescale to reach equilibrium becomes long.

6.\ Models with exponential ($\mdotstar \propto e^{-t/\tausfh}$)
and linear-exponential ($\mdotstar \propto t e^{-t/\tausfh}$) star
formation histories have the same equilibrium abundances, but
the early-time abundances are a factor of two lower for
linear-exponential histories, and the approach to equilibrium
is considerably slower.  
In either case, a more rapidly declining star formation
history (shorter $\tausfh$) leads to lower $\ofe$ at equilibrium
because the delayed SNIa enrichment, which comes from an earlier time when
star formation was more rapid, is more important compared to 
ongoing CCSN enrichment.

%7.\ While SNIa enrichment begins after the minimum delay time $\td$,
%the location of the knee in $\ofe-\feh$ tracks depends mainly on
%the value of $\tauIa$, or more generally on the time required for SNIa
%to produce a substantial fraction of their total metals.

\smallskip
\noindent
{\it Metallicity distribution functions}
\smallskip

7.\ When $\tausfh \gg \taudep$ and $\tausfh \gg \tauIa$, a
generic situation for a system with ongoing gas accretion,
metallicity distribution functions (MDFs) are sharply peaked
near the equilibrium abundance.  The limiting case of constant
SFR has $dN/d\Yo \propto (1-\Yo/\Yoeq)^{-1}$ up to the
current abundance $\Yo(t)$.
For oxygen, $\tausfh = 2\taudep$ is a critical case with
constant $dN/d\Yo$, and shorter $\tausfh$ produces
declining $dN/d\Yo$.  The traditional ``closed box'' or
``leaky box'' scenarios represent the limiting case of
no gas accretion, yielding $\tausfh=\taudep$, 
and this produces an exponential MDF 
$dN/d\Yo \propto \exp[-\Yo(1+\eta)/\mocc]$.
However, these scenarios do {\it not} capture the
typical behavior for star-forming systems with continuing accretion.

8.\ Similar considerations apply to iron MDFs, but here the
relevant comparison timescale is usually $\tauIa$ rather
than $\taudep$.  A system with rapid accretion {\it and}
rapid depletion can produce a rising $dN/d\Yo$ and a declining
$dN/d\Yfe$ if the timescales are such that 
$2\taudep \la \tausfh \la 2\tauIa$.  While the MDF shapes
of $\alpha$-elements and iron-peak elements should typically
be similar in form, cases where they differ can provide a
distinctive diagnostic of enrichment timescales.

9.\ At early times ($t \ll \taudep$ and $t \ll \tauIa$) and low metallicities,
exponential star formation histories produce constant $dN/dZ$ and
linear-exponential star formation histories produce $dN/dZ\propto Z$,
given our assumptions of constant $\taustar$ and metallicity-independent 
yields.  This behavior is sensitive to our assumption of instantaneous
recycling of CCSN products.

\smallskip
\noindent
{\it Starbursts and sudden changes}
\smallskip

10.\ Bursts of star formation, on the scale of an entire galaxy
or of an individual molecular cloud or star-forming region, boost
the rate of CCSN enrichment relative to SNIa enrichment.
If the CCSN products are retained by the system, then a burst that
converts a significant fraction of the available gas into
stars can easily boost $\ofe$ by 0.1-0.3 dex.
The low observed scatter in $\ofe$ at fixed $\feh$ (along either
the ``thick disk'' or ``thin disk'' sequences) sets limits
on the importance of this time-varying enrichment effect.
In typical molecular clouds, it is probably small because
of low conversion efficiency, and perhaps because the
CCSN metals are lost before they can be incorporated into new stars.
However, molecular clouds vary in their properties, and 
occasional systems that form stars efficiently and retain
their metals could be a source of rare $\alpha$-enhanced
stars at intermediate ages \citep{martig2014,chiappini2015,haywood2015}.
Bursty star formation histories in dwarf galaxies could have
a significant impact on their $\ofe$ distributions, relative to
smooth star formation histories with the same time-averaged behavior
\citep{gilmore1991}.

11.\ A rapid change from efficient star formation (short $\taustar$) to
inefficient star formation (long $\taustar$) can lead to a substantial
drop in $\ofe$, as iron deposition exceeds oxygen deposition.  Conversely,
a rapid change from low efficiency to high efficiency leads to a 
temporary boost in $\ofe$.

12.  One can produce an evolutionary sequence that has increasing $\feh$ at
low $\ofe$ by decreasing $\eta$, and thus raising the equilibrium
abundance itself,
after the population has already evolved to an initial equilibrium.
Alternatively,
by increasing the outflow efficiency at late times, one can construct
an evolutionary track that reaches solar $\ofe$ and $\feh$ and
then moves backward to lower $\feh$ at constant $\ofe$ because
of the reduction in equilibrium abundance.
This backward evolution scenario allows a single evolutionary
track to produce a bimodal distribution in the 
$\ofe$-$\feh$ plane.

%11.\ For an exponential star formation history, the impact of enriched
%infall with abundance $\Yinf$ is equivalent to increasing the CCSN
%yield of the same element by a factor $(1+Z/Z_{\rm eq}^{\rm cc})$.
%This effect will typically be small, though it could be significant
%in low metallicity galaxies or the outer regions of disks.

\smallskip
\noindent
{\it Intermediate elements}
\smallskip

13.\ In any one-zone model,
all ``idealized'' $\alpha$ elements, by which we mean
elements produced entirely by CCSNe with a metallicity-independent yield,
follow the same track in $[{\rm X}/{\rm Fe}]$ vs. $\feh$, 
described by equation~(\ref{eqn:Deltaofe}).
In particular, the drop between the CCSN plateau and the
eventual equilibrium $[{\rm X}/{\rm Fe}]$ is the same
for all such elements.
When an observed element does not show this behavior, it indicates
that there is another significant source (SNIa or AGB production),
or that the IMF-averaged yield is metallicity dependent,
or that the observational estimates of the abundance are
systematically biased; any one of these would be an interesting conclusion.
Elements like sulfur that have a subdominant but non-negligible
contribution from SNIa should show intermediate behavior,
as described by equation~(\ref{eqn:Deltaxfe}).
Detailed comparisons of the $[{\rm X}/{\rm Fe}]$ tracks of
different $\alpha$-elements should provide significant insights
on their production mechanisms.  Even when yields have weak metallicity
dependence at a fixed stellar mass, the IMF-averaged yield could
change with metallicity either because the IMF itself changes
or because the mass ranges of stars that explode as CCSNe
(instead of collapsing to black holes) change with metallicity.

\smallskip
\noindent
{\it Applications}
\smallskip

Beyond these insights, we expect our analytic solutions to have
significant practical utility for modeling observations.
The key equations are~(\ref{eqn:YoEvolExp}), (\ref{eqn:YfeccEvolExp}),
(\ref{eqn:YfeIaEvolExp}) for oxygen, CCSN iron, and SNIa iron
evolution with an exponential star formation history 
and the corresponding equations~(\ref{eqn:YoEvolLinExp}),
(\ref{eqn:YfeccEvolLinExp}), and~(\ref{eqn:YfeIaEvolLinExp}) for
linear-exponential star formation histories.
Section~\ref{sec:changes} and \S\ref{sec:extensions} describe
a variety of ways to extend these results, including more
complex star formation histories, double-exponential DTDs for SNIa,
models with discontinuous parameter changes, and $\alpha$-elements
with significant SNIa contributions.
Analytic solutions enable rapid explorations of parameter space,
zeroing in on regions of observational or physical interest
that merit detailed numerical modeling.
They are useful for characterizing degeneracies among parameters,
to better understand whether a fit to data is unique or one among many.
In more quantitative terms, their speed of calculation makes them
useful for statistical modeling of data sets via likelihood
methods, Markov Chain Monte Carlo sampling, or related techniques 
(e.g., \citealt{kirby2013}).  
While most systems of interest are more complicated than a one-zone
model, some may be usefully approximated by mixtures of one-zone
models, or by a one-zone model whose geometry and kinematics
change over time to represent heating or contraction.

These models are aimed principally at the interpretation of resolved
stellar populations with star-by-star abundance measurements.
They may also be useful for modeling gas phase abundances and
connecting them to underlying stellar populations, an application 
of growing importance in the era of large IFU surveys such as
CALIFA \citep{sanchez2012}, SAMI \citep{croom2012}, and
MaNGA \citep{bundy2015}.  They can also be
incorporated into population synthesis models of galactic spectral
energy distributions, tying a distribution of stellar metallicities
to an inferred history of star formation.  They can be used as an 
approximate tool for post-processing numerical simulations to
make chemical evolution predictions in simulations that do not
explicitly track multi-element enrichment.

The high-fiber revolution that transformed the study of 
large scale structure is now transforming our knowledge of
the multi-element distributions of stellar populations in 
the Milky Way and its neighbors.  These rich data sets,
often augmented by phase space information from astrometry
and age information from asteroseismology, offer many
clues to the history of our Galaxy.  One of the challenges
in interpreting these clues is evaluating the uniqueness
of successful models, especially in light of systematic
uncertainties in element yields and the observed abundances
themselves.  Flexible approximate models can play a valuable
role in mapping out the variety of routes to a given final
state and identifying the observational features that may best 
distinguish competing scenarios.

Appendix~\ref{appx:userguide} provides a step-by-step guide
to using our analytic results for computing enrichment histories,
$\afe$ tracks, and MDFs.

\bigskip\noindent
We are grateful to Ralph Sch\"onrich and Jennifer Johnson for
several years worth of invaluable education about chemical evolution
modeling.  We thank Evan Kirby, Molly Peeples, Ralph Sch\"onrich,
and Philipp Kempski for numerous helpful comments on an earlier version
of the manuscript.  This work was supported by NSF grant AST-1211853.

\appendix

\section{A. SNIa Enrichment}
\label{appx:SNIa}

Define the SNIa rate $R(\nu)$ such that $R(\nu)d\nu$ is the number
of SNIa per unit mass of stars formed that explode in the time interval
$\nu \rightarrow \nu + d\nu$ following the formation of a population
at time $\nu = 0$.
The units of $R(\nu)$ are $M_\odot^{-1}\yr^{-1}$.  We use the variable $\nu$
to avoid confusion with time $t$ in a continuously forming stellar population.

As already noted in equation~(\ref{eqn:mfeIa}), we define the population
averaged SNIa iron yield to be
\begin{equation}
\label{eqn:app_mfeIa}
\mfeIa \equiv \KfeIa \int_0^\infty R(\nu)d\nu~,
\end{equation}
where $\KfeIa$ is the mean mass of iron ejected per SNIa.
This expression implicitly assumes that the integral converges to a
finite value, which it does for an exponential DTD.
For a form of the DTD for which the integral does not converge,
such as $R(\nu) \propto \nu^{-1}$, one can simply cut off the integral
at some time larger than the age of the universe, with no loss of generality
in our equations.

For a star formation history $\mdotstar(t)$, 
the rate at which SNIa inject iron to the ISM is
\begin{equation}
\label{eqn:app_injection}
\MfeIadot(t) = \KfeIa\int_0^t \mdotstar(t')R(t-t')dt' = \mfeIa\bmdotstart
\end{equation}
where $\bmdotstart$ is given by equation~(\ref{eqn:app_bmdotstart}).
For a DTD that is zero prior to $\td$ and exponential thereafter,
$R(\nu) = R_0 e^{-(\nu-\td)/\tauIa}$, implying
\begin{equation}
\label{eqn:app_norm}
\int_0^\infty R(\nu)d\nu = R_0 \tauIa
\end{equation}
and
\begin{equation}
\label{eqn:app_bmdotstart}
\bmdotstart = \tauIa^{-1} \int_0^{t-\td}
\mdotstar(t') e^{-(t-t'-\td)/\tauIa} dt'~,
\end{equation}
where the upper limit is set to $t-\td$ because the SNIa rate is zero
for more recent star formation.
For the case of a constant SFR, evaluating this integral yields
\begin{equation}
\label{eqn:app_csfr}
\bmdotstart = \mdotstar \left[1-e^{-\Delta t/\tauIa}\right]~,
  \quad {\rm constant~SFR}~,
\end{equation}
with $\Delta t \equiv t-\td$.
%{\bf Check not missing an $e^{\td/\tauIa}$ factor.}

For an exponentially declining star formation history,
$\mdotstar(t) = \dot{M}_{*,0}e^{-t/\tausfh}$,
\begin{equation}
\label{eqn:app_expsfrint}
\bmdotstart = \dot{M}_{*,0} \tauIa^{-1} \int_0^{t-\td} e^{-t'/\tausfh}
  e^{-(\Delta t -t')/\tauIa} dt'~.
\end{equation}
Since $e^{-\Delta t/\tauIa}$ is independent of $t'$ it can be factored
out of the integral, yielding
\begin{equation}
\label{eqn:app_expsfrint2}
\bmdotstart = \dot{M}_{*,0} \tauIa^{-1} e^{-\Delta t/\tauIa} 
  \int_0^{t-\td} e^{-t'/\tausfh} e^{t'/\tauIa} dt'~.
\end{equation}
Using the notation~(\ref{eqn:taubar}) allows the integrand to
be written $e^{t'/\taubarIaSfh}$, and evaluating the integral yields
\begin{eqnarray}
\label{eqn:app_expsfr}
\bmdotstart &=& \dot{M}_{*,0} 
  \left({\taubarIaSfh\over\tauIa}\right) e^{-\Delta t/\tauIa}
  \left[e^{\Delta t/\taubarIaSfh}-1\right] \\
  &=& \dot{M}_{*,0} \left({\taubarIaSfh\over\tauIa}\right) 
      \left[e^{-\Delta t/\tausfh}-e^{-\Delta t/\tauIa}\right]~,
\end{eqnarray}
where the second equality uses $\tauIa^{-1}-\taubarIaSfh^{-1}=\tausfh^{-1}$.
From here we can factor out $e^{-\Delta t/\tausfh}$ and use the
substitution $\dot{M}_{*,0} e^{-\Delta t/\tausfh} = \mdotstart e^{\td/\tausfh}$
to write
\begin{equation}
\label{eqn:app_expsfr2}
\bmdotstart = \mdotstart e^{\td/\tausfh} {\taubarIaSfh\over \tauIa}
  \left[1-e^{-\Delta t/\taubarIaSfh}\right]~,
  \quad {\rm exponential~SFH}~,
\end{equation}
which in turn leads to equation~(\ref{eqn:mdotstar_ratio}).

This example illustrates several of the features that arise
throughout our calculations, in particular the appearance and behavior of
harmonic difference timescales.
While $\taubarIaSfh$ can be positive or negative, the factor in $[...]$
always has the same sign as $\taubarIaSfh$, forcing $\bmdotstart$ to be 
positive.
Note that for $\tauIa \approx \tausfh$, and thus large $\taubarIaSfh$,
Taylor-expanding the exponential
yields
\begin{equation}
\label{eqn:app_expsfrapprox}
\bmdotstart = \mdotstart e^{\td/\tausfh} {\Delta t \over \tauIa}~.
\end{equation}
Thus, the result does not diverge even as $\tauIa \rightarrow \tausfh$
and $\taubarIaSfh \rightarrow \infty$.

For $\mdotstar(t) \propto t e^{-t/\tausfh}$ the procedure is
similar, but the integral that enters is $\int xe^x dx$ rather than 
$\int e^x dx$, leading to combinations of linear and exponential terms.
The result can be expressed
\begin{equation}
\label{eqn:app_linexp}
{\bmdotstar \over \mdotstar} = e^{\td/\tausfh} 
  \left({\taubarIaSfh\over\tauIa}\right){\taubarIaSfh \over t}
  \left[{\Delta t \over \taubarIaSfh}+e^{-\Delta t/\taubarIaSfh} -1 \right]~,
  \quad \hbox{{\rm linear-exponential~SFH}}.
\end{equation}

\section{B. A User's Guide}
\label{appx:userguide}

Our analytic results provide a flexible tool for computing
$\ofe-\feh$ tracks, age-metallicity and age-$\ofe$ relations,
and MDFs for comparisons to observational data or as inputs for
population synthesis or theoretical models.
Here we provide a step-by-step guide for such calculations,
which can be implemented easily in a plotting package such as {\tt sm},
or any programming language.  We begin with the most straightforward
case of a single-exponential SNIa DTD and model parameters $\eta$, 
$\taustar$, and $\tausfh$ that are fixed throughout the evolution.
We then summarize how to implement the two-exponential DTDs of 
\S\ref{sec:two-exp}, which approximate a $t^{-1.1}$ power-law,
and how to implement models with sudden parameter changes,
like those of \S\ref{sec:sudden}.

\medskip\noindent
{\it Set the physical parameters}

\begin{itemize}
\item
Choose values for the supernova yield parameters $\mocc$, $\mfecc$,
and $\mfeIa$.  The default values listed in Table~\ref{tbl:variables} are
reasonable choices, based on the assumptions described in 
\S\ref{sec:eq_equations}.

\item
Choose a value of the mass recycling parameter $r$.  Our fiducial
value $r=0.4$ is a good choice for a Kroupa IMF, and results are only weakly
sensitive to this parameter.

\item
Choose values for the solar oxygen and iron abundance.  
To convert from the conventional shifted number density scale
$x_i \equiv 12+\log(X_i/H)$ to mass fractions use
\begin{eqnarray}
\log Z_{{\rm O},\odot} &=& 
  (x_{\rm O}-12)+\log(16)+\log(0.71) = -2.25+(x_{\rm O}-8.69) ~, 
  \label{eqn:solarO} \\
\log Z_{{\rm Fe},\odot} &=& 
  (x_{\rm Fe}-12)+\log(55.85)+\log(0.71) = -2.93+(x_{\rm Fe}-7.47) ~,
  \label{eqn:solarFe}
\end{eqnarray}
where we have adopted a solar hydrogen mass fraction of 0.71 and 
an atomic weight of 55.85 for iron.

\item
Choose values for the SNIa DTD timescale $\tauIa$ and the minimum
delay time $\td$.  Our fiducial values are $\tauIa=1.5\Gyr$ and 
$\td=0.15\Gyr$.  
\end{itemize}

\noindent
{\it Choose model parameters and compute equilibrium abundances}

\begin{itemize}
\item
Specify values of the mass-loading parameter $\eta$, the
star formation efficiency timescale $\taustar$, and the $e$-folding
timescale for the star formation rate, $\tausfh$.  

\item
Calculate the depletion timescale $\taudep=\taustar/(1+\eta-r)$.
Calculate the harmonic difference timescales $\taubarDepSfh$,
$\taubarDepIa$, and $\taubarIaSfh$ from equation~(\ref{eqn:taubar}).
If you will be exploring many parameter values, it may be useful to 
multiply $\taustar$, $\tausfh$, and $\tauIa$ by numbers that are 
very slightly different from one to avoid exact equalities that 
lead to undefined values of the harmonic difference timescales.
As discussed in the text, abundance results are convergent and physical
even near limits where one of these timescales diverges.

\item
Compute the equilibrium abundances $\Yoeq$, $\Yfecceq$,
and $\YfeIaeq$ from equations~(\ref{eqn:Yoeq2})-(\ref{eqn:YfeIaeq2}).

\end{itemize}

\noindent
{\it Compute time evolution and MDF}

\begin{itemize}
\item
Choose an exponential or linear-exponential star formation history.
By setting $\tausfh$ to a large value, one can also use these cases to 
approximate a constant SFR or linearly rising SFR, respectively.

\item
Compute $\Yo(t)$ and $\Yfe(t) = \Yfecc(t)+\YfeIa(t)$ using
equations~(\ref{eqn:YoEvolExp}), (\ref{eqn:YfeccEvolExp}), and 
(\ref{eqn:YfeIaEvolExp}) for an exponential SFH or
equations~(\ref{eqn:YoEvolLinExp}), (\ref{eqn:YfeccEvolLinExp}), and 
(\ref{eqn:YfeIaEvolLinExp}) for a linear-exponential SFH.
These can be converted to $\ohh=\log(\Yo/Z_{{\rm O},\odot})$,
$\feh=\log(\Yfe/Z_{{\rm Fe},\odot})$, and $\ofe=\ohh-\feh$.

\item
To compute an iron MDF, first choose bins for $\feh$.
Then compute $\Yfe(t)$ with a constant time spacing 
and add a quantity proportional to $\mdotstart$,
and hence to the number of stars formed during the time interval,
to the $\feh$ bin in which $\Yfe(t)$ falls.
The resulting histogram can be multiplied by a constant to normalize
it to unit integral.  Short time spacings (e.g., $0.002\Gyr$) may
be necessary to get accurate and smooth results at low metallicity.

\item Distributions of $\ofe$ and $\ohh$ can be computed in the
same way as the iron MDF.  Equations~(\ref{eqn:mdf1}) and~(\ref{eqn:mdf2}) 
provide analytic forms for the oxygen MDF for a constant
or exponential SFH, respectively.

\item
To model a more general star formation history that is a sum of exponentials
and/or linear-exponentials, follow equation~(\ref{eqn:sfhsum}).

\end{itemize}

\medskip\noindent
{\it Two-exponential DTD}
\smallskip

To approximate a $t^{-1.1}$ DTD with a minimum delay time of $\td=0.15\Gyr$,
we recommend a sum of two exponential DTDs with timescales of
$\tau=0.5\Gyr$ and $5\Gyr$.  Proceed as before, but compute and add 
two values of $\YfeIa(t)$ to get the total SNIa iron contribution, 
multiplying the value of $\mfeIa$ by $(0.478,0.522)$ for $\tau=(0.5,5)$
so that each exponential is normalized to produce half of the SNIa iron
over $12.5\Gyr$ (see footnote in \S\ref{sec:two-exp}).
Note that the harmonic difference timescales and $\YfeIaeq$ must
be computed separately for the two exponentials.

For a minimum delay time of $\td=0.05\Gyr$, we recommend a sum of two
exponential DTDs with $\tau=(0.25,3.5)\Gyr$ and $\mfeIa$ multiplied
by $(0.493,0.507)$.  Experimentation along the lines illustrated
in Figure~\ref{fig:SNIaNum} can be carried out to find exponential
combinations for other minimum delay times or other 
functional forms of the DTD.

\medskip\noindent
{\it Sudden Parameter Changes}
\smallskip

Section~\ref{sec:sudden} describes how to calculate a model in which
parameters ($\taustar$, $\eta$, $\tausfh$) change from one value to 
another at a transition time $t_c$.  For times $t < t_c$, proceed as
before, and record the values of $\Yo(t_c)$ and $\Yfe(t_c)$ for 
subsequent use.  For times $t>t_c$, compute quantities $\tttaudep$,
$\tttaubarIaSfh$, $\tttaubarDepSfh$, and $\tttaubarDepIa$ using
the the post-$t_c$ parameter values.  Compute the corresponding
equilibrium abundances.  Defining $t_2 = t-t_c$, compute the contributions
$Z_{\rm O,1}(t_2)$ and $Z_{\rm O,2}(t_2)$ from 
equations~(\ref{eqn:Yo1}) and~(\ref{eqn:YoEvolExp2}) and 
sum them to get $\Yo(t_2)$.
For iron, compute and sum the values of
$Z_{\rm Fe,1}(t_2)$, $Z_{\rm Fe,2}(t_2)$, and $Z_{\rm Fe,3}(t_2)$
from equations~(\ref{eqn:Yfe1}), (\ref{eqn:YfeEvolExp2}), 
and (\ref{eqn:YfeIaEvolC3}), respectively.

%------------------------------------------------------------------------------

\bibliographystyle{apj} 
\bibliography{equilibrium}

%\begin{thebibliography}{% number of papers cited here}
% contents of .bbl file
%\end{thebibliography}

\end{document}